\documentclass{JFM-FLM_Au}
\usepackage{subcaption}  
\usepackage{comment}
\usepackage{graphicx}
\usepackage[normalem]{ulem}

\lefttitle{Author 1 et al.}
\righttitle{Journal of Fluid Mechanics}

\title{Symmetry-Constrained Exact Coherent Structures in Plane Poiseuille Flow}

\author{Akshit Nanda \and Ritabrata Thakur}

\affiliation{Department of Applied Mechanics, Indian Institute of Technology Delhi, New Delhi, India}

\corresau{Ritabrata Thakur, ritabrata@iitd.ac.in}

\begin{document}
\maketitle

\begin{abstract}
Turbulence in wall-bounded shear flows is increasingly understood through exact coherent structures (ECS)---invariant solutions of the
Navier--Stokes equations that act as organising centres in the
high-dimensional state space. Here we report five new ECS of plane Poiseuille
flow: two relative periodic orbits (RPOs) and three travelling waves
(TWs), that are computed in four distinct symmetry-invariant subspaces using a
Newton--Krylov--hookstep solver initialised from direct numerical simulations. We trace each state through one-parameter continuations in both the
Reynolds number $Re$ and the spanwise period $L_z$. All five states are
organised around counter-rotating rolls sustaining streamwise velocity
streaks, yet they exhibit qualitatively different stability properties: the
two RPOs are linearly stable within their symmetry subspace, while the three
TWs are saddle-type solutions whose instabilities are, respectively, mixed
oscillatory-and-monotone, purely monotone and purely oscillatory. The
continuation diagrams reveal their bifurcation geometry, from simple
saddle-node folds for the RPOs to a pronounced S-shaped multi-branch
structure with three coexisting dissipation levels for one of the travelling
waves. Along each branch, the roll--streak topology is preserved across the
folds, with the different branches distinguished by their amplitude and
gradient intensity rather than by a change in spatial organisation. Floquet
spectra evaluated at multiple points along each branch shows that folds generically act as sites of reduced instability for the
travelling waves, while upper branches develop stronger and sometimes
qualitatively different unstable modes. These include the emergence of linearly
stable segments on the intermediate and upper branches of the multi-branch
travelling wave, despite the solution being unstable at its reference
parameters. The combined structural and spectral analysis across the
$(Re, L_z)$ parameter space provides a detailed picture of how these
coherent states persist, deform and change their dynamical character, giving
a set of state-space landmarks for understanding the organisation of
turbulent dynamics in channel flow.
\end{abstract}

\begin{keywords}
Authors should not enter keywords on the manuscript, as these must be chosen by the author during the online submission process and will then be added during the typesetting process (see \href{https://www.cambridge.org/core/services/aop-file-manager/file/61436b61ff7f3cfab749ce3a/JFM-Keywords-Sept-2021.pdf.}{Keyword PDF} for the full list).  Other classifications will be added at the same time.
\end{keywords}


\section{Introduction}
\label{sec:Introduction}

The onset and maintenance of turbulence in wall-bounded shear flows remain
central questions in fluid mechanics. Turbulent shear flows display persistent
spatial and temporal patterns, often referred to as coherent structures
\citep{hussain1986coherent,robinson1991coherent}, which play a key role in
turbulent transport processes and energy transfer from the mean shear.
Following the pioneering ideas of \cite{landau1944problem} and the subsequent
emergence of exact coherent structures (ECS) in canonical shear flows, a
dynamical-systems perspective has become a powerful framework for understanding
subcritical transition and sustained turbulence. In this view, the dynamics of
the incompressible Navier--Stokes equations is represented as a trajectory in a
high-dimensional state space, where equilibria, travelling waves and (relative)
periodic orbits appear as invariant sets that organise the surrounding dynamics
\citep{Gibson2008,graham2021exact}. Although the Navier--Stokes system is
formally infinite-dimensional, viscous dissipation confines the long-time
dynamics to a finite-dimensional manifold in state space
\citep{hopf1948mathematical,doering1995applied}, and turbulence can be regarded
as a chaotic attractor whose skeleton is formed by these invariant solutions
and the homo- and heteroclinic connections between them
\citep{graham2021exact}. The identification of such invariant solutions has
provided insight into the building blocks of turbulence and the
self-sustaining processes (SSP) in shear flows
\citep{hamilton1995regeneration,waleffe1997self,waleffe2001exact,jimenez1999autonomous,graham2021exact}.

Over the last three decades, numerous invariant solutions have been computed in
plane Couette, plane Poiseuille and pipe flows, including equilibria,
travelling waves (TWs) and relative periodic orbits (RPOs)
\citep{nagata1990three,clever1992three,faisst2003traveling,wedin2004exact,
gibson2009equilibrium,park2015exact}. The first ECS were discovered by
\cite{nagata1990three} in plane Couette flow, arising through a saddle-node
bifurcation. Also, \cite{clever1992three,clever1997tertiary} independently found
the same solutions by continuing Rayleigh--B\'{e}nard--Couette flow to a zero
Rayleigh number. Building on ideas about the SSP, \cite{waleffe1998three,
waleffe2001exact,waleffe2003homotopy} used body-force continuation to find
travelling waves in both Couette and Poiseuille flows, coining the
term ``exact coherent structures.'' Travelling wave solutions with analogous
near-wall structure were subsequently found in pipe flow
\citep{faisst2003traveling,wedin2004exact}, where they originate from
saddle-node bifurcations at Reynolds numbers ($Re$) as low as  $1250$. The
presence of these finite-amplitude solutions across different geometries
points to a generic self-sustaining mechanism in wall-bounded shear flows
\citep{waleffe1997self,jimenez1999autonomous,graham2021exact}. This mechanism
has been further clarified through the resolvent framework of
\cite{mckeon2010critical}, which shows that many ECS can be efficiently
represented by a small number of resolvent modes derived from the linearised
Navier--Stokes operator \citep{sharma2013coherent,mckeon2017engine}.

In plane Poiseuille flow in particular, early studies identified nonlinear
travelling waves closely related to the near-wall streak-vortex cycle
\citep{waleffe2001exact}, while later work revealed extensive sets of
invariant solutions organised by discrete symmetries and domain parameters such
as spanwise and streamwise lengths
\citep{gibson2014spanwise,zammert2015crisis,Aghor_Gibson_2025}.
\cite{park2015exact} found several channel flow ECS and showed that
the turbulent attractor is organised around the upper-branch
solution of a set they denoted ``P4.'' The turbulent trajectory
intermittently visits the vicinity of the corresponding lower-branch
solution, strengthening the connection between ECS and turbulent
dynamics. The dynamical significance of these states lies in their typically
low-dimensional unstable manifolds, which act as gateways between laminar and
turbulent dynamics and provide a deterministic skeleton for the observed
spatio-temporal complexity.

Beyond travelling waves, periodic and relative periodic orbits play an essential
role in organising the chaotic dynamics: close to any chaotic attractor there
exists an infinity of unstable periodic orbits \citep{holmes2015dynamical}.
\cite{kawahara2001periodic} discovered two periodic orbits in plane Couette
flow. One displays a full regeneration cycle of streamwise vortices and
low-speed streaks; the other consists of a weaker standing-wave motion with
milder spatiotemporal variation. They show that turbulence wanders around the former while intermittently visiting
the latter, with heteroclinic connections serving as the pathways for bursting
trajectories. This picture has been broadly confirmed in subsequent studies
\citep{toh2003periodic,viswanath2007recurrent,halcrow2009heteroclinic}.
\cite{budanur2017relative} found that relative periodic orbits collectively
fill out the turbulent region of state space in pipe flow. A turbulent
trajectory shadows individual RPOs for finite time intervals when it visits
their neighbourhoods. In plane Couette flow, \cite{van2011homoclinic}
established a homoclinic connection of a periodic orbit on the
laminar-turbulent boundary that exhibits the essential features of turbulent
bursting. Subsequently, \cite{lustro2019onset} directly demonstrated a
homoclinic tangency, i.e.\ the first non-transverse intersection of the stable
and unstable manifolds of the weaker periodic orbit of
\cite{kawahara2001periodic}, thereby confirming the onset of transient chaos.
More broadly, the stable and unstable
manifolds of ECS organise the transport of trajectories in state space,
dictating how the turbulent state moves from one invariant solution to another
\citep{Gibson2008,budanur2017relative,park2018bursting}. Recently,
\cite{suri2024predictive} showed that invariant manifolds of unstable periodic
orbits can also explain and predict sudden large-scale flow reversals and
excursions in turbulence, further demonstrating the organising power of
nonchaotic invariant solutions in high-dimensional flows.

A closely related concept is that of edge states. These are invariant solutions residing on the laminar-turbulent boundary, possessing only one unstable direction that leads either to laminar flow or to (possibly transient) turbulence. These states represent the marginal dynamical behaviour separating the two regimes and have been studied in a variety of flows
\citep{skufca2006edge,schneider2008laminar,schneider2010localized,
duguet2009localized,schneider2009edge,willis2009turbulent,xi2012dynamics,
kerswell2018nonlinear}. \cite{kerswell2018nonlinear} provided a comprehensive
review of nonlinear nonmodal stability theory, clarifying how the basin
boundary between laminar and turbulent attractors is structured by such edge
states and connecting their study to optimal perturbation and transition
problems. The existence of these solutions on the basin boundary further
motivates a thorough exploration of invariant solutions across different
symmetry subspaces, since edge states may belong to symmetry-restricted
subsets of state space that are not accessible by generic searches.

A critical methodological advance enabling the systematic computation of these
invariant solutions was the Newton--Krylov--hookstep algorithm introduced by
\cite{viswanath2007recurrent}. This method combines a matrix-free
Newton--Krylov iteration with a locally constrained optimal hookstep, making it
possible to converge onto highly unstable solutions directly from turbulent
initial conditions without requiring a bifurcation-continuation scenario.
\citeauthor{viswanath2007recurrent} applied it to compute five new periodic and
relative periodic orbits in plane Couette turbulence, each demonstrating the
breakup and re-formation of near-wall coherent structures. They showed that
enforcing each independent second-order symmetry during the Krylov iteration
roughly halves the effective search-space dimension, yielding a corresponding
speedup in convergence. A tutorial on the practical aspects of computing ECS is
provided by \cite{willis2019computing}. This observation connects directly to
the role of symmetry subspaces in organising invariant solutions: by projecting
onto a symmetric subspace prior to each time integration, one confines the
Newton search to a dynamically invariant subset of state space. This
simultaneously reduces computational cost and eliminates redundant solution
branches related by coordinate transformations.
\cite{viswanath2009critical} subsequently
extended this framework to study the critical layer structure and stable
manifolds of lower-branch travelling waves in pipe flow. Disturbances of the
laminar solution that evolve onto the stable manifold of a given lower-branch
state provide a direct dynamical pathway from laminar to turbulent flow,
thereby linking the computation of invariant solutions to the transition
problem.

The role of symmetry has received renewed attention in this context.
\cite{Aghor_Gibson_2025} showed that enforcing reflection and shift symmetries
when searching for invariant solutions in plane Poiseuille flow not only
improves numerical efficiency but also reveals previously unseen ECS. Plane
Poiseuille flow possesses several discrete symmetries, including reflections
about the wall-normal and spanwise mid-planes and discrete translations in the
streamwise and spanwise directions. Restricting the dynamics to invariant
subspaces associated with these symmetries yields dynamically self-contained
systems that can host distinct branches of equilibria, travelling waves and
RPOs. Understanding how these symmetry-related subspaces are populated, and how
branches within a given subspace are created and destroyed by bifurcations, is
therefore crucial for mapping the organisation of invariant solutions. The
systematic exploitation of symmetries has also revealed spatially localised
invariant solutions in extended domains
\citep{schneider2010snakes,brand2014doubly}, including doubly localised
spot-like equilibria in plane Couette flow that closely resemble transitional
turbulent spots. This demonstrates that localised turbulent patterns can also
be understood as exact solutions of the Navier--Stokes equations.

In the present work we compute and analyse five new exact coherent
states of the incompressible Navier--Stokes equations in plane Poiseuille flow:
two relative periodic orbits and three travelling waves.
All solutions are obtained using a Newton--Krylov solver initialised from direct
numerical simulations and are subsequently continued in Reynolds number $Re$
and spanwise domain length $L_z$. Their parametric dependence is summarised
through bifurcation diagrams plotted as suitable dissipation-based measures
versus $Re$ and versus $L_z$.

The continuation diagrams reveal folds and symmetry-constrained bifurcations
along each of the five branches, clarifying how the corresponding coherent
states are created, how their dissipation levels evolve with $Re$ and $L_z$,
and where they undergo qualitative changes in stability or symmetry within
their respective subspaces. Linear stability analysis shows that the two RPOs
are stable within their symmetry subspace, aside from neutral directions
associated with continuous symmetries (time shifts and streamwise/spanwise
translations), whereas each travelling wave possesses a small number of
unstable directions. In all cases the number of unstable modes remains low,
so that these ECS can exert a strong organising influence on nearby dynamics in
state space, in line with observations for other travelling waves and RPOs in
wall-bounded flows
\citep{viswanath2007recurrent,gibson2009equilibrium,park2015exact,graham2021exact}.
To trace how stability changes along each solution branch, we also compute
Floquet spectra at selected points along the continuation curves, including
at fold/turning points and on opposite branches. This spectral information
reveals where multipliers cross the unit circle and how the number and
character of unstable directions evolve with $Re$ and $L_z$.

Taken together, the two RPOs and three TWs extend the known
catalogue of invariant solutions in plane Poiseuille flow and provide new
evidence for the central role of discrete symmetries in structuring the
existence, stability and bifurcation landscape of exact coherent structures in
wall-bounded shear flows.

The remainder of the paper is organised as follows.
Section~\ref{sec:theory} summarises the governing equations, symmetry
operations and the construction of invariant subspaces.
Section~\ref{sec:numerics} describes the numerical formulation, Newton--Krylov
solver and continuation procedures.
Section~\ref{sec:results} presents the five new exact coherent structures,
their basic properties and representative flow fields along with their Floquet
spectra and discusses the dynamical role of the new ECS in state space.
Section~\ref{sec:continuation} analyses their continuation in Reynolds number
and spanwise domain length, with emphasis on folds, symmetry-constrained
bifurcations in dissipation-based diagrams, and the evolution of linear
stability along each branch.
Finally, Section~\ref{sec:conclusions} summarises the main findings and
outlines directions for future work.

\section{Theoretical background}
\label{sec:theory}

Plane Poiseuille flow, the motion of a viscous incompressible fluid between two
parallel plates driven by a constant pressure gradient or an equivalent
bulk--velocity constraint, serves as a canonical system for studying
wall--bounded turbulence and exact coherent structures
\citep{kim1987turbulence,graham2021exact,Aghor_Gibson_2025}. We consider the
flow in a doubly periodic domain with streamwise and spanwise periodic lengths
$L_x$ and $L_z$, respectively, following the formulation of
\cite{Aghor_Gibson_2025}. The dynamics admit several discrete and continuous
symmetries that strongly influence the organisation of invariant solutions. The
set of all symmetry transformations that leave the equations of motion
invariant defines the \emph{plane--Poiseuille symmetry group}, denoted by
$G_{PPF}$, which we describe below together with the class of invariant
solutions of interest.

\subsection{Governing equations and invariant solutions}

We focus on plane Poiseuille flow driven by a prescribed constant bulk velocity
$U_{\mathrm{bulk}}$ in the streamwise direction and zero mean velocity in the
spanwise direction. The Reynolds number is defined as
\begin{equation}
    Re = \frac{U_c\, h}{\nu},
\end{equation}
where $h$ is the channel half-height, $\nu$ is the kinematic viscosity, and
$U_c = \tfrac{3}{2}U_{\mathrm{bulk}}$ is the centreline velocity of the
parabolic laminar profile carrying the same streamwise bulk flux. After
non-dimensionalisation using $h$ and $U_c$, the incompressible Navier--Stokes
equations read
\begin{equation}
    \frac{\partial \mathbf{u}_{\mathrm{tot}}}{\partial t}
    + \mathbf{u}_{\mathrm{tot}}\cdot\nabla\mathbf{u}_{\mathrm{tot}}
    = -\nabla p_{\mathrm{tot}}
    + \frac{1}{Re}\nabla^2 \mathbf{u}_{\mathrm{tot}},
    \qquad
    \nabla\cdot\mathbf{u}_{\mathrm{tot}} = 0,
    \label{eq:NSfull}
\end{equation}
posed on the domain
\begin{equation}
    \Omega = [0,L_x)\times[-1,1]\times[0,L_z),
\end{equation}
with periodic boundary conditions in $x$ (streamwise) and $z$ (spanwise), and
no-slip conditions $\mathbf{u}_{\mathrm{tot}}=\mathbf{0}$ at the walls
$y = \pm 1$. 

The total velocity and pressure are decomposed into a steady laminar base flow
and fluctuation fields,
\begin{equation}
    \mathbf{u}_{\mathrm{tot}} = \mathbf{U}(y) + \mathbf{u},
    \qquad
    p_{\mathrm{tot}} = P + p,
\end{equation}
where the laminar base velocity is
\begin{equation}
    \mathbf{U}(y) = (1 - y^2)\,\mathbf{e}_x,
    \label{eq:laminar}
\end{equation}
for which $U_c = 1$ and $U_{\mathrm{bulk}} = 2/3$ in non-dimensional units.
The base pressure takes the form $P(t) = x\,P_x(t) + z\,P_z(t)$, so that the
total pressure gradient is
\begin{equation}
    \nabla p_{\mathrm{tot}}(\mathbf{x},t)
    = P_x(t)\,\mathbf{e}_x + P_z(t)\,\mathbf{e}_z + \nabla p(\mathbf{x},t),
\end{equation}
where $\mathbf{e}_x$ and $\mathbf{e}_z$ are unit vectors in the streamwise and
spanwise directions. The mean pressure-gradient components $P_x(t)$ and
$P_z(t)$ are not prescribed but instead adjust dynamically at each instant to
enforce the bulk-velocity constraints: constant streamwise bulk flux and zero
spanwise mean flow. The fluctuation pressure $p$ is constrained to have zero
spatial mean, $\langle\nabla p\rangle_\Omega = 0$, which uniquely determines
$P_x(t)$ and $P_z(t)$ \citep{Aghor_Gibson_2025}.

Substituting the decomposition into \eqref{eq:NSfull} and using the fact that
$\mathbf{U}(y)$ satisfies the base-flow equations, one obtains the governing
equation for the velocity fluctuation
$\mathbf{u}(\mathbf{x},t) = [u,v,w](x,y,z,t)$:
\begin{equation}
    \frac{\partial \mathbf{u}}{\partial t}
    + \mathbf{u}\cdot\nabla\mathbf{u}
    = -\nabla p
    + \frac{1}{Re}\nabla^2\mathbf{u}
    - \left(\frac{2}{Re} + P_x\right)\mathbf{e}_x
    - P_z\,\mathbf{e}_z
    + 2y\,v\,\mathbf{e}_x
    - (1-y^2)\,\frac{\partial\mathbf{u}}{\partial x},
    \qquad
    \nabla\cdot\mathbf{u} = 0,
    \label{eq:NSfluct}
\end{equation}
where $v$ is the wall-normal component of $\mathbf{u}$. The term
$-2y\,v\,\mathbf{e}_x$ arises from the wall-normal advection of base-flow
momentum, i.e.\ from the product $(\mathbf{u}\cdot\nabla)\mathbf{U}
= v\,U'(y)\,\mathbf{e}_x = -2yv\,\mathbf{e}_x$ upon substitution of
$\mathbf{U}'(y) = -2y\,\mathbf{e}_x$; the term
$-(1-y^2)\,\partial_x\mathbf{u}$ accounts for the streamwise advection of the
fluctuation by the base flow, $(\mathbf{U}\cdot\nabla)\mathbf{u}
= U(y)\,\partial_x\mathbf{u}$. The mean pressure-gradient terms
$P_x(t)$ and $P_z(t)$, together with the viscous correction
$\tfrac{2}{Re}\,\mathbf{e}_x$, act as spatially uniform body forces determined
instantaneously as Lagrange multipliers enforcing incompressibility and the
bulk-velocity constraints; they are therefore functions of $\mathbf{u}$ alone.
The entire right-hand side of the momentum equation in \eqref{eq:NSfluct} thus defines a well-determined
function $f(\mathbf{u})$ of the fluctuation field, rendering the system
autonomous \citep{Aghor_Gibson_2025}.

The natural function space for the fluctuation field is
\begin{equation}
    \mathbb{U}
    = \bigl\{\mathbf{u}\in L^2(\Omega) :
      \nabla\cdot\mathbf{u} = 0,\;
      \mathbf{u}\big|_{y=\pm 1} = \mathbf{0}
    \bigr\},
\end{equation}
with $L^2$ norm defined by $\|\mathbf{u}\|^2 = \langle\mathbf{u}\cdot\mathbf{u}\rangle_\Omega$.
Since $\mathbb{U}$ is a vector space, any linear combination of states in
$\mathbb{U}$ remains in $\mathbb{U}$. The evolution is represented compactly as
\begin{equation}
    \frac{\partial\mathbf{u}}{\partial t} = f(\mathbf{u}),
    \qquad
    \mathbf{u}(t) = \phi^t\!\left(\mathbf{u}(0)\right),
    \label{eq:dynamics}
\end{equation}
where $f:\mathbb{U}\to\mathbb{U}$ denotes the right-hand side of the momentum equation in
\eqref{eq:NSfluct} and $\phi^t$ is the corresponding nonlinear time-evolution
operator.

An \emph{invariant solution} is a velocity field $\mathbf{u}\in\mathbb{U}$
satisfying
\begin{equation}
    \sigma\,\phi^t(\mathbf{u}) - \mathbf{u} = \mathbf{0},
    \label{eq:invariant}
\end{equation}
for some time $t > 0$ and some symmetry $\sigma\in G_{PPF}$
\citep{Aghor_Gibson_2025}. Different choices of $\sigma$ and the temporal
condition yield distinct solution types:
\begin{itemize}
    \item \textbf{Equilibria} satisfy \eqref{eq:invariant} for all $t > 0$
    with $\sigma = 1$; they are time-independent solutions of
    \eqref{eq:dynamics}.

    \item \textbf{Travelling waves} satisfy \eqref{eq:invariant} for all
    $t > 0$ with $\sigma(t) = \tau(c_x t,\, c_z t)$, a phase shift that
    grows linearly in time at fixed wavespeeds $c_x$ and $c_z$ (one or both
    non-zero), where $\tau(a,b)$ denotes a spatial translation by $(a,b)$ in
    the $(x,z)$-plane (see \S\,\ref{sec:symmetry}).

    \item \textbf{Periodic orbits} satisfy \eqref{eq:invariant} with
    $\sigma = 1$ and a fixed period $T > 0$, with $\phi^t(\mathbf{u})\neq
    \mathbf{u}$ for $0 < t < T$.

    \item \textbf{Relative periodic orbits} satisfy \eqref{eq:invariant} with
    a fixed period $T > 0$ and a non-trivial symmetry $\sigma \neq 1$.
\end{itemize}

In practice, invariant solutions are computed by solving the discretised form
of \eqref{eq:invariant} from initial guesses $\hat{\mathbf{u}}$ (and, where
applicable, $\hat{\sigma}$ and $\hat{T}$) that approximately satisfy the
equations. For doubly periodic rectangular domains at low to moderate Reynolds
numbers, the velocity field is represented spectrally and $\phi^t$ is advanced
with a finite-difference time-stepping scheme, yielding a nonlinear system with
$\mathcal{O}({10^4})$--$\mathcal{O}({10^6})$ unknowns. The resulting system is solved efficiently using
trust-region Newton--Krylov methods \citep{viswanath2007recurrent}, with $T$
and any relevant phase parameters treated as additional unknowns.

\subsection{Symmetries, equivariance and invariant subspaces}
\label{sec:symmetry}

The admissible symmetries $\sigma$ in~\eqref{eq:invariant} are those that leave
the Navier--Stokes equations~\eqref{eq:NSfull} invariant. The symmetry group $G_{PPF}$ is
generated by two discrete reflections and two continuous translations
\citep{Aghor_Gibson_2025}:
\begin{align}
\sigma_y[u,v,w](x,y,z)&=[u,-v,w](x,-y,z), \\
\sigma_z[u,v,w](x,y,z)&=[u,v,-w](x,y,-z), \\
\tau(a,b)[u,v,w](x,y,z)&=[u,v,w](x+a,y,z+b),
\end{align}
where $(a,b)\in\mathbb{R}^2$ denote phase shifts in the streamwise and
spanwise directions, taken modulo $(L_x,L_z)$ for the periodic domain. The
full symmetry group can thus be written as
\begin{equation}
G_{PPF}=\langle \sigma_y,\sigma_z,\{\tau(a,b):a,b\in\mathbb{R}\}\rangle.
\end{equation}
For later convenience, we define the half--domain translations
\begin{equation}
\tau_x=\tau(L_x/2,0),\qquad
\tau_z=\tau(0,L_z/2),\qquad
\tau_{xz}=\tau(L_x/2,L_z/2),
\end{equation}
which generate discrete shift subgroups that play a central role in identification of many known ECS \citep{gibson2014spanwise,Aghor_Gibson_2025}.

The Navier--Stokes operator $f$ defined above and its flow map $\phi^t$ are equivariant with
respect to $G_{PPF}$:
\begin{equation}
f(\gamma\mathbf{u})=\gamma f(\mathbf{u}),
\qquad
\phi^t(\gamma\mathbf{u})=\gamma\phi^t(\mathbf{u}),
\label{eq:equivariance}
\end{equation}
for all $\gamma\in G_{PPF}$ and all $t$. Thus, if $\mathbf{u}(t)$ is a
solution, then $\gamma\mathbf{u}(t)$ is also a solution for every
$\gamma\in G_{PPF}$. If the initial condition is invariant under a subgroup
$H\subset G_{PPF}$, i.e.\ $\mathbf{u}(0)=h\mathbf{u}(0)$ for all $h\in H$,
then this symmetry is preserved for all time and the trajectory remains within
the corresponding invariant subspace
\begin{equation}
\mathbb{U}_H=\{\mathbf{u}\in\mathbb{U}:\mathbf{u}=h\mathbf{u}\;\forall\,h\in H\}.
\end{equation}

In this study we work within several such invariant subspaces. The two RPOs
are computed in the subspace defined by combined reflections
$(\sigma_y,\sigma_z)$ together with the half--box shift $\tau_{xz}$, i.e.,\
the subgroup
\begin{equation}
H_{\text{RPO}} = \langle \sigma_y, \sigma_z, \tau_{xz} \rangle \subset G_{PPF}.
\end{equation}
The three travelling waves are computed in three distinct subspaces:
\begin{equation}
H_{\text{TW1}} = \langle \sigma_y, \tau_x, \tau_z \rangle,\qquad
H_{\text{TW2}} = \langle \sigma_y, \tau_{xz} \rangle,\qquad
H_{\text{TW3}} = \langle \sigma_y, \sigma_z, \tau_{xz} \rangle.
\end{equation}
Each of these subgroups defines a dynamically self--contained system that can
host its own branches of equilibria, travelling waves and RPOs, and the
symmetry constraints strongly shape the properties of the solutions that
reside within them \citep{Aghor_Gibson_2025}.

\subsection{Leveraging symmetry in the computation of invariant solutions}

The interplay between symmetry and invariant solutions has both theoretical and
computational implications. On the theoretical side, the choice of subgroup
$H\subset G_{PPF}$ determines which classes of equilibria, travelling waves and
RPOs can exist within the corresponding invariant subspace $\mathbb{U}_H$.
For example, enforcing $\sigma_z$ symmetry prohibits solutions with net
spanwise drift, while including $\tau_{xz}$ symmetry constrains the admissible
streamwise--spanwise phase relations. Different subspaces can therefore support
qualitatively different families of ECS even at the same $(Re,L_x,L_z)$.

From a computational perspective, enforcing symmetry constraints reduces the
effective dimensionality of the search space explored by the Newton--Krylov
solver. If $n$ independent second--order symmetries are imposed, the cost of
the search typically decreases by a factor of order $2^n$, as demonstrated in
detail by \cite{Aghor_Gibson_2025}. Working in the subspaces
$H_{\text{RPO}}$, $H_{\text{TW1}}$, $H_{\text{TW2}}$ and $H_{\text{TW3}}$ thus
reduces the number of degrees of freedom substantially while isolating
dynamically distinct recurrent motions.

ECS reported in this study are computed using symmetry
projection operators and time--stepping/Newton routines implemented in
\texttt{Channelflow~2.0} \citep{Gibson2008,park2015exact}. The
RPOs are obtained as symmetry--constrained relative periodic orbits in
$\mathbb{U}_{H_{\text{RPO}}}$, while the three travelling waves are computed as
relative equilibria in $\mathbb{U}_{H_{\text{TW1}}}$,
$\mathbb{U}_{H_{\text{TW2}}}$ and $\mathbb{U}_{H_{\text{TW3}}}$, respectively.
This symmetry--aware formulation provides an efficient route to
discovering new branches of exact coherent structures in plane Poiseuille flow.

\section{Numerical methods and computational setup}
\label{sec:numerics}

\texttt{Channelflow~2.0} uses a
Fourier--Chebyshev spectral--Galerkin discretisation of the incompressible
Navier--Stokes equations in doubly periodic channel domains. In this framework we compute five
exact coherent structures (ECS) of plane Poiseuille flow: two relative periodic
orbits (RPO1, RPO2) and three travelling waves (TW1--TW3). The RPOs are
restricted to the symmetry-invariant subspace
$\langle\sigma_y,\sigma_z,\tau_{xz}\rangle$, while the travelling waves are
computed in three distinct symmetry subspaces generated by
$\langle\sigma_y,\tau_x,\tau_z\rangle$, $\langle\sigma_y,\tau_{xz}\rangle$ and
$\langle\sigma_y,\sigma_z,\tau_{xz}\rangle$, respectively (see
\S\ref{sec:symmetry}). In each case we use direct numerical simulation (DNS) [at $Re$ = (1000, 1500, 1900 and 2000) ]
to obtain initial guesses, followed by Newton--Krylov
refinement of the appropriate invariance condition.

\subsection{Formulation and numerical computation of invariant solutions}

The invariant solutions sought here are equilibria, travelling waves and
relative periodic orbits (RPOs) of the fluctuation field $\mathbf{u}$ evolving
under \eqref{eq:dynamics}. As described in \S\ref{sec:theory}, these are
characterised by
\begin{equation}
\phi^{T}(\mathbf{u}) = \sigma\,\mathbf{u},
\label{eq:inv_num}
\end{equation}
for some period $T>0$ and symmetry $\sigma\in G_{PPF}$. Equilibria correspond
to $\sigma=1$ and time--independent $\mathbf{u}$, travelling waves to
continuous translations $\sigma=\tau(c_x T,c_z T)$ with constant wave speed
$(c_x,c_z)$, and RPOs to discrete shifts $\sigma\neq1$ involving reflections
and/or discrete translations.

In practice, the solver represents the flow in a comoving frame
and solves a discretised version of \eqref{eq:inv_num} using a
Newton--Krylov--hookstep method
\citep{viswanath2007recurrent,Gibson2008,park2015exact}. For RPOs,
the unknowns are the state $\mathbf{u}$, the period $T$ and the discrete shift
parameters (e.g.,\ the streamwise drift $a_x$); for travelling waves, the
unknowns are $\mathbf{u}$ and the wavespeeds $(c_x,c_z)$ in the moving frame.
At each Newton step, the time--$T$ map $\phi^T(\mathbf{u})$ (or its appropriate
specialisation for travelling waves) is evaluated by DNS using the spectral
solver. Convergence is declared when the normalised residual
\begin{equation}
\frac{\|\phi^{T}(\mathbf{u})-\sigma\,\mathbf{u}\|_2}{\|\mathbf{u}\|_2}
\le 10^{-13},
\end{equation}
and the updates in $(T,a_x,c_x,c_z)$ are consistent with this tolerance. The
resulting systems involve $\mathcal{O}(10^5$--$10^6)$ spectral coefficients and
a comparable number of nonlinear equations.

Once a single ECS is converged at a given $(Re,L_z)$, it is tracked in the
parameter space by gradually varying either the Reynolds number $Re$ or the
spanwise domain length $L_z$, using the converged solution at one parameter
value as the initial guess at the next. This yields the bifurcation and
continuation curves presented in later sections.

\subsection{Domains, parameters and spectral discretisation}
\label{subsec:domain}

All computations employ the non--dimensionalisation introduced in
\S\ref{sec:theory}, with channel half--height $h=1$ and laminar centreline
velocity $U_c=1$, so that the Reynolds number is $Re=U_c h/\nu$ and the
laminar bulk velocity is $U_{\text{bulk}}=2/3$. The wall--normal extent of the
domain is fixed at $L_y=2$, while the streamwise and spanwise lengths $(L_x,L_z)$
are chosen to accommodate the different coherent structures.

Three domain configurations are used:

\begin{itemize}
  \item \textbf{Configuration~A} (RPO1, RPO2, TW2):
  \[
  L_x\times L_y\times L_z = \pi\times2\times\frac{\pi}{2},
  \qquad
  N_x\times N_y\times N_z = 48\times81\times48.
  \]
  This ``minimal'' domain supports the two relative periodic orbits at
  $Re=1000$ (RPO1) and $Re=1500$ (RPO2), both in the subspace
  $\langle\sigma_y,\sigma_z,\tau_{xz}\rangle$, as well as a travelling wave
  TW2 at $Re=2000$ in the subspace $\langle\sigma_y,\tau_{xz}\rangle$.

  \item \textbf{Configuration~B} (TW1):
  \[
  L_x\times L_y\times L_z = 4\pi\times2\times\pi,
  \qquad
  N_x\times N_y\times N_z = 48\times81\times49.
  \]
  This extended streamwise and spanwise domain is used for TW1 at $Re=1900$
  in the subspace $\langle\sigma_y,\tau_x,\tau_z\rangle$, allowing longer--wavelength
  structures and additional spanwise variation.

  \item \textbf{Configuration~C} (TW3):
  \[
  L_x\times L_y\times L_z = \pi\times2\times\pi,
  \qquad
  N_x\times N_y\times N_z = 48\times81\times48.
  \]
  This square planform domain supports TW3 at $Re=2000$ in the subspace
  $\langle\sigma_y,\sigma_z,\tau_{xz}\rangle$.
\end{itemize}

In all cases, the velocity field is expanded in Fourier modes in the periodic
directions $x$ and $z$ and in Chebyshev polynomials in $y$, with the $3/2$
dealiasing rule applied in $x$ and $z$.

Time advancement of the DNS used to evaluate $\phi^t$ is performed by a
semi--implicit scheme in which viscous terms are treated implicitly and
nonlinear terms explicitly, with adaptive substepping to satisfy a prescribed
CFL constraint. This time--stepping configuration is used consistently for the
long DNS runs that generate initial guesses and for the Newton--Krylov
refinement of both RPOs and travelling waves.

\subsection{Identification of recurrent and traveling states}
\label{sec:recurrent}

Initial guesses for the ECS are obtained from direct numerical simulations of
plane Poiseuille flow within each symmetry-invariant subspace. Each DNS is
initialised with a smooth, divergence-free velocity field whose Fourier
coefficients decay rapidly at high wavenumber, and is then integrated for 5000 timesteps . We analyse the
resulting trajectories using a combination of recurrence diagnostics, energy
signals and phase-plane projections to identify candidate RPOs and travelling
waves. For TW2, a different strategy is employed: the initial guess is
obtained via edge tracking on the laminar--turbulent boundary, as described
below.

\paragraph*{Recurrence analysis for RPOs.}

For the RPOs, a primary diagnostic is the recurrence function
\begin{equation}
R(t,T)=\|\mathbf{u}(t+T)-\mathbf{u}(t)\|_2,
\end{equation}
evaluated over a grid of base times $t$ and time lags $T$. The corresponding
recurrence plots reveal intervals where the trajectory nearly retraces itself
after a delay $T$. Diagonal bands or ridges of low $R(t,T)$ indicate repeated
near-recurrence of full velocity fields and thus suggest candidate periods.
Local minima of $R(t,T)$ along such bands provide both an approximate period
$T$ and a candidate state $\mathbf{u}(t)$ for Newton--Krylov refinement of an
underlying RPO.

To validate each candidate, we complement $R(t,T)$ with the kinetic energy
\begin{equation}
E(t)=\frac{1}{2}\,\langle|\mathbf{u}(t)|^2\rangle,
\end{equation}
where $\langle\cdot\rangle$ denotes a volume average over $\Omega$. For each
candidate period $T$, the segment $E(t)$ over $[t,t+T]$ is examined for
approximately symmetric modulation about its midpoint, which helps
distinguish genuine near-periodicity from accidental close returns.

\paragraph*{Energy-balance diagnostics in the $(D,I)$ plane.}

A further diagnostic, used for both RPOs and travelling waves, is based on the
instantaneous energy input $I(t)$ and dissipation $D(t)$, defined as
\begin{align}
I(t) &= \frac{1}{2L_z}\int_0^{L_z}\int_{-1}^{1}
\Big(p_{\text{tot}}u_{\text{tot}}\big|_{x=0}
     -p_{\text{tot}}u_{\text{tot}}\big|_{x=L_x}\Big)\,dy\,dz, \\
D(t) &= \frac{1}{2L_xL_z}\int_0^{L_z}\int_{-1}^{1}\int_0^{L_x}
\big(|\nabla u_{\text{tot}}|^2+|\nabla v_{\text{tot}}|^2+|\nabla w_{\text{tot}}|^2\big)
\,dx\,dy\,dz.
\end{align}
In statistical equilibrium one expects $D(t)$ and $I(t)$ to be comparable on
average, and exact invariant solutions correspond to precisely balanced energy
input and dissipation in an appropriate frame.

For RPO candidates, we examine trajectories in the $(D,I)$ plane over
intervals $[t,t+T]$ suggested by the recurrence analysis. A clean relative
periodic orbit generates a closed or nearly closed loop in the $(D,I)$ plane
when observed over one period: the flow returns to the same combination of
energy input and dissipation. Candidate initial conditions are therefore
required to exhibit near-closure of the loop, with the end point
$(D(t+T),I(t+T))$ lying close to $(D(t),I(t))$.

For travelling-wave candidates (TW1 and TW3), the same $(D,I)$ diagnostics
provide a way to locate states that are nearly stationary in an
appropriate comoving frame. We search along trajectories for times $t_*$ at
which the instantaneous energy balance is nearly satisfied,
\begin{equation}
|D(t_*)-I(t_*)| < 2\times10^{-3},
\end{equation}
and at which the temporal fluctuations of $D$ and $I$ are locally small. Such
points correspond to segments of the trajectory that linger near a
quasi-steady balance of energy input and dissipation. When projected onto the
desired symmetry subspace, these states serve as dynamically informed initial
guesses for travelling waves.

\paragraph*{Edge tracking for TW2.}

The initial guess for TW2 is obtained not from turbulent DNS but from edge
tracking
\citep{skufca2006edge,schneider2008laminar,kerswell2018nonlinear}. It is a
bisection procedure that identifies states on the laminar--turbulent boundary
in state space. Starting from a turbulent snapshot, the perturbation
amplitude relative to the laminar profile is iteratively bisected and at each
step, the flow is integrated forward in time, and the amplitude is adjusted
upward if the trajectory decays toward laminar flow or downward if it
evolves toward turbulence. Successive bisections confine the trajectory to
the basin boundary, where it converges toward an edge state possessing a
single unstable direction that separates the laminar and turbulent basins of
attraction
\citep{schneider2008laminar,kerswell2018nonlinear}. In our case, the
edge-tracking procedure is carried out within the
$\langle\sigma_y\rangle$-invariant subspace at $Re = 2000$. The trajectory
on the basin boundary settles into a nearly steady state with weak velocity
fluctuations (peak $|u| \approx 0.13$, substantially smaller than TW1 or
TW3), and the cross-flow energy signal confirms that the dynamics approach a
fixed point in the comoving frame. This near-steady state, projected onto
the $\langle\sigma_y,\tau_{xz}\rangle$ subspace, provides the initial guess
from which TW2 is converged via the Newton--Krylov--hookstep solver. The
edge-tracking time series ($\|\mathbf{u}\|$ and cross-flow energy as
functions of time) are shown in the supplementary material (figure~S1).

\paragraph*{Selection of initial guesses and symmetry projection.}

For each symmetry subspace, we combine the above diagnostics to select a
small number of candidate initial guesses:
\begin{itemize}
  \item For RPO1 and RPO2 in the subspace
  $\langle\sigma_y,\sigma_z,\tau_{xz}\rangle$, we require:
  (i)~pronounced recurrence ridges in $R(t,T)$,
  (ii)~approximately symmetric modulation of $E(t)$ over one candidate
  period, and (iii)~near-closure of the loop in the $(D,I)$ plane.

  \item For TW1 in $\langle\sigma_y,\tau_x,\tau_z\rangle$ and TW3 in
  $\langle\sigma_y,\sigma_z,\tau_{xz}\rangle$, we select times $t_*$ with
  $|D(t_*)-I(t_*)|<2\times10^{-3}$ and small local variation in $(D,I)$,
  indicating near-steady behaviour, and then project the corresponding
  fields onto the desired symmetry subspace.

  \item For TW2 in $\langle\sigma_y,\tau_{xz}\rangle$, the initial guess
  is the near-steady state obtained from edge tracking, projected onto the
  target subspace as described above.
\end{itemize}

Each selected state is projected by averaging over the generators of the
appropriate subgroup $H$ (see \S\ref{sec:symmetry}), ensuring that the
subsequent Newton--Krylov iterations remain in $\mathbb{U}_H$. In this way we
obtain dynamically consistent initial conditions for both relative periodic
orbits and travelling waves, from which the fully converged ECS reported in
later sections are computed.

\section{Results}
\label{sec:results}

In this section we present the complete set of five new exact coherent structures
(ECS) of plane Poiseuille flow computed in this study: two relative periodic
orbits (RPOs) and three travelling waves (TWs), each residing in a specified
symmetry-invariant subspace. The two RPOs are found at $Re=1000$ and $Re=1500$
in the subspace $\langle \sigma_y,\sigma_z,\tau_{xz}\rangle$ with
$\tau_{xz}=\tau(L_x/2,L_z/2)$. The three TWs are obtained at higher Reynolds
numbers in distinct symmetry subspaces: TW1 at $Re=1900$ in
$\langle \sigma_y,\tau_x,\tau_z\rangle$, TW2 at $Re=2000$ in
$\langle \sigma_y,\tau_{xz}\rangle$, and TW3 at $Re=2000$ in
$\langle \sigma_y,\sigma_z,\tau_{xz}\rangle$. All states are refined to machine
precision using a Newton--Krylov--hookstep method for invariant solutions \citep{viswanath2007recurrent} starting from
near-recurrent (RPOs) or near-steady-in-a-comoving-frame (TWs) episodes observed
in DNS. For the TWs, the DNS used to obtain
initial guesses were performed in closely related symmetry-invariant subspaces;
in particular, the converged TWs reported above may possess additional discrete
shift symmetries beyond those enforced during the exploratory DNS. We mention
this here only in passing, and return to it in the dedicated TW subsections.

To compactly visualise candidate recurrences and to document the raw DNS
behaviour from which these ECS are extracted, we use instantaneous power input
$I(t)$ and viscous dissipation $D(t)$ and consider their projection in the
$(D,I)$ plane. Close approaches to the identity line $D=I$ (equivalently, small
imbalance $|D(t)-I(t)|$) and near-closure of loops in $(D,I)$ provide a robust,
model-independent indicator of time windows suitable for Newton--Krylov
initialisation. For the RPOs, these diagnostics are complemented by explicit
temporal recurrence checks (energy modulation and overlap of $T$-shifted
segments). For the TWs, the same $(D,I)$ diagnostics identify intervals where
the dynamics approach a steady state in an appropriate comoving frame; for TW2
we additionally use the cross-flow energy signal from edge tracking to
highlight the approach to the edge state and the refinement window.

\begin{figure}[htbp]
\centering
  \begin{subfigure}[htbp]{0.48\textwidth}
  \centering
    \includegraphics[width=\linewidth]{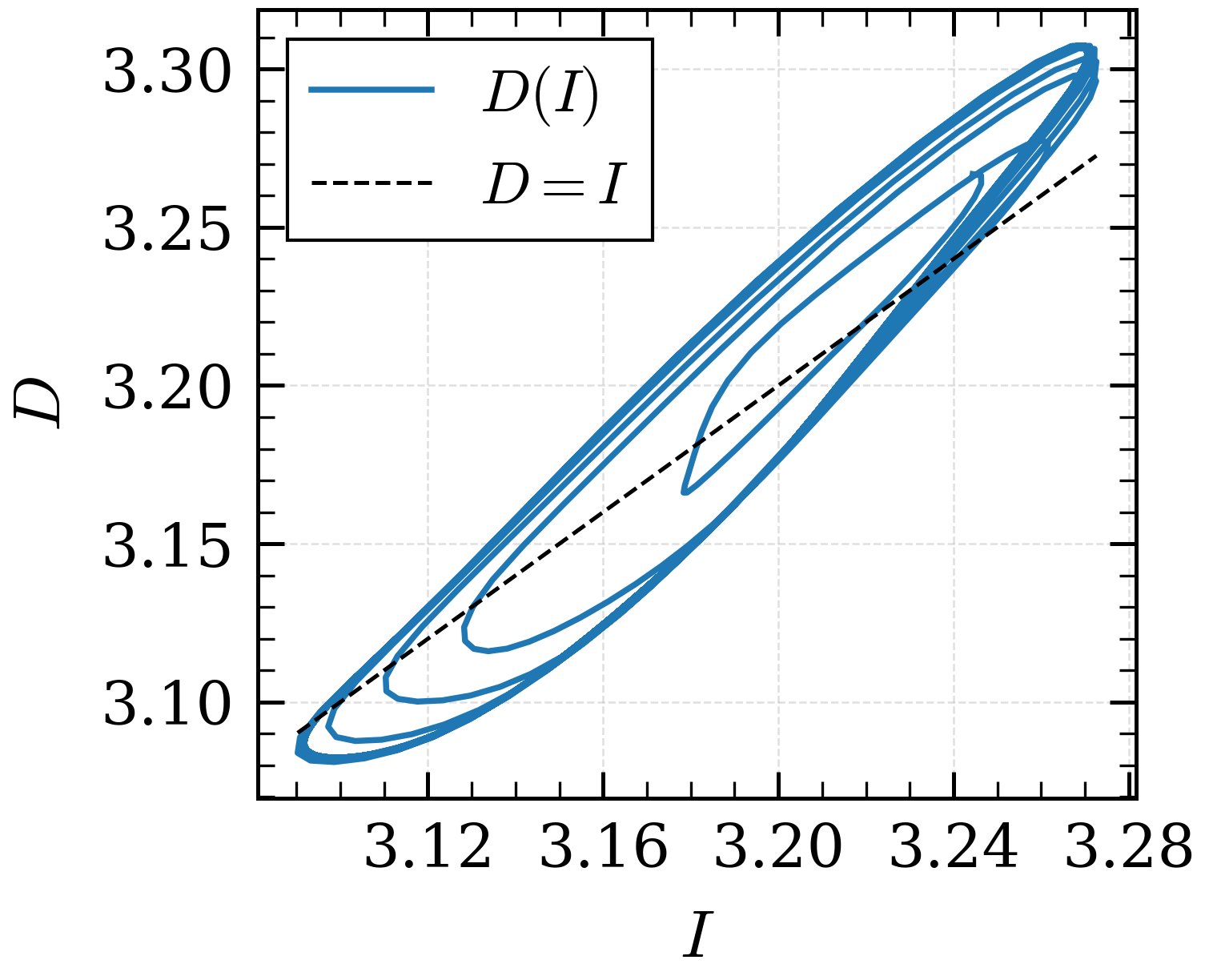}
    \caption{$Re=1000$ (RPO1 candidate DNS).}
    \label{fig:DI_RPO1_raw}
  \end{subfigure}
  \centering
  \begin{subfigure}[htbp]{0.48\textwidth} 
    \includegraphics[width=\linewidth]{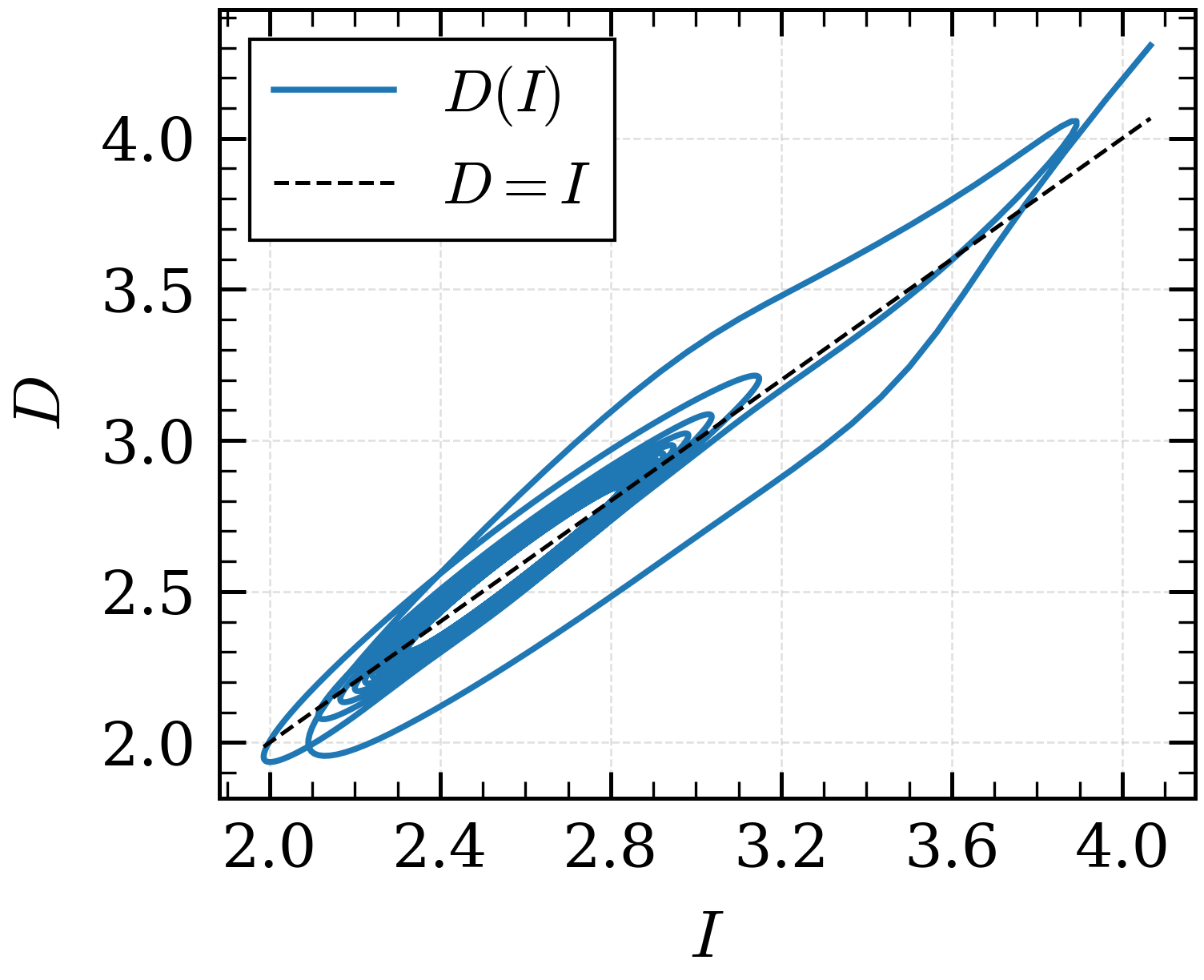}
    \caption{$Re=1500$ (RPO2 candidate DNS).}
    \label{fig:DI_RPO2_raw}
  \end{subfigure}
  \vspace{2mm}
  \begin{subfigure}[htbp]{0.48\textwidth}
  \centering
    \includegraphics[width=\linewidth]{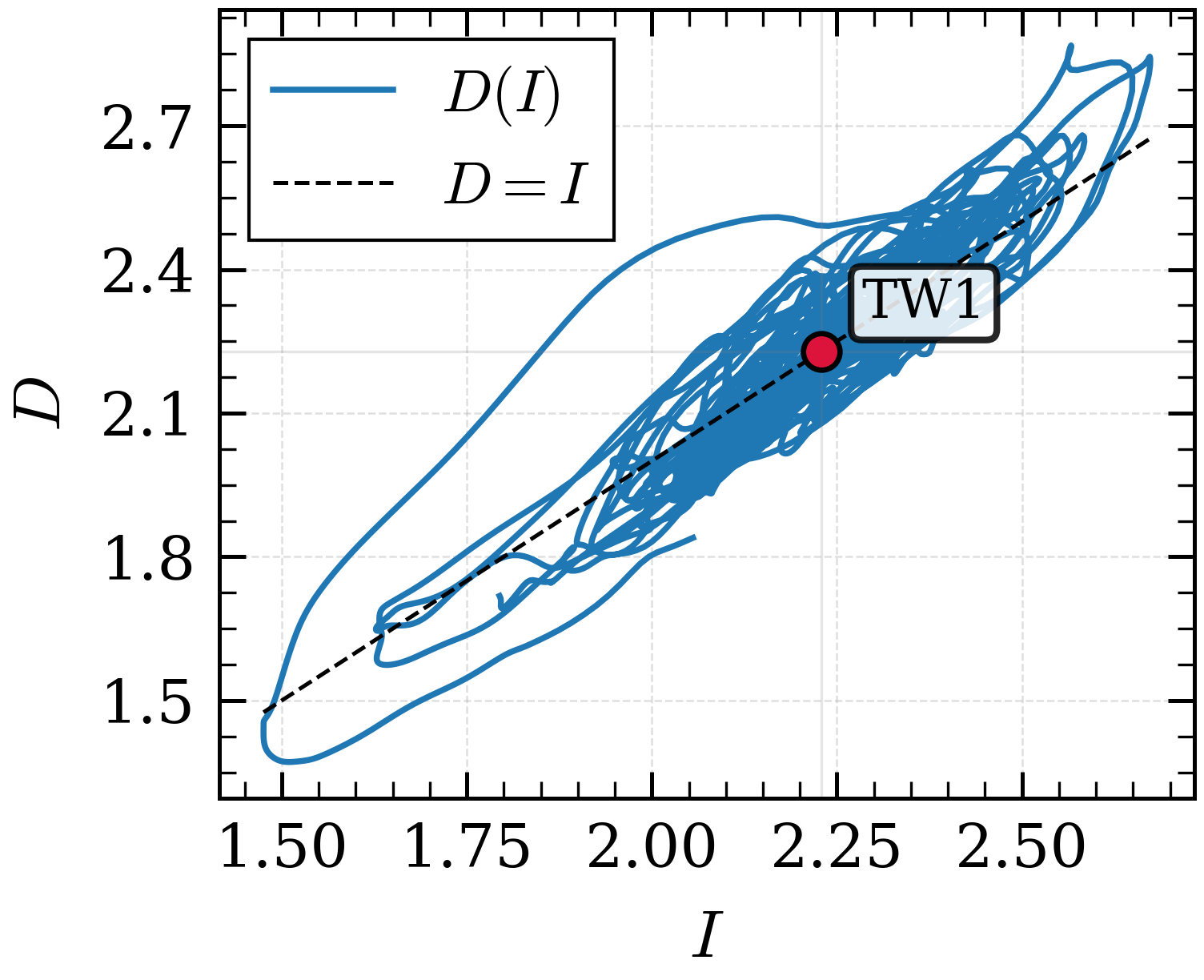}
    \caption{$Re=1900$ (TW1 candidate DNS).}
    \label{fig:DI_TW1_raw}
  \end{subfigure}
  \begin{subfigure}[htbp]{0.48\textwidth}
    \includegraphics[width=\linewidth]{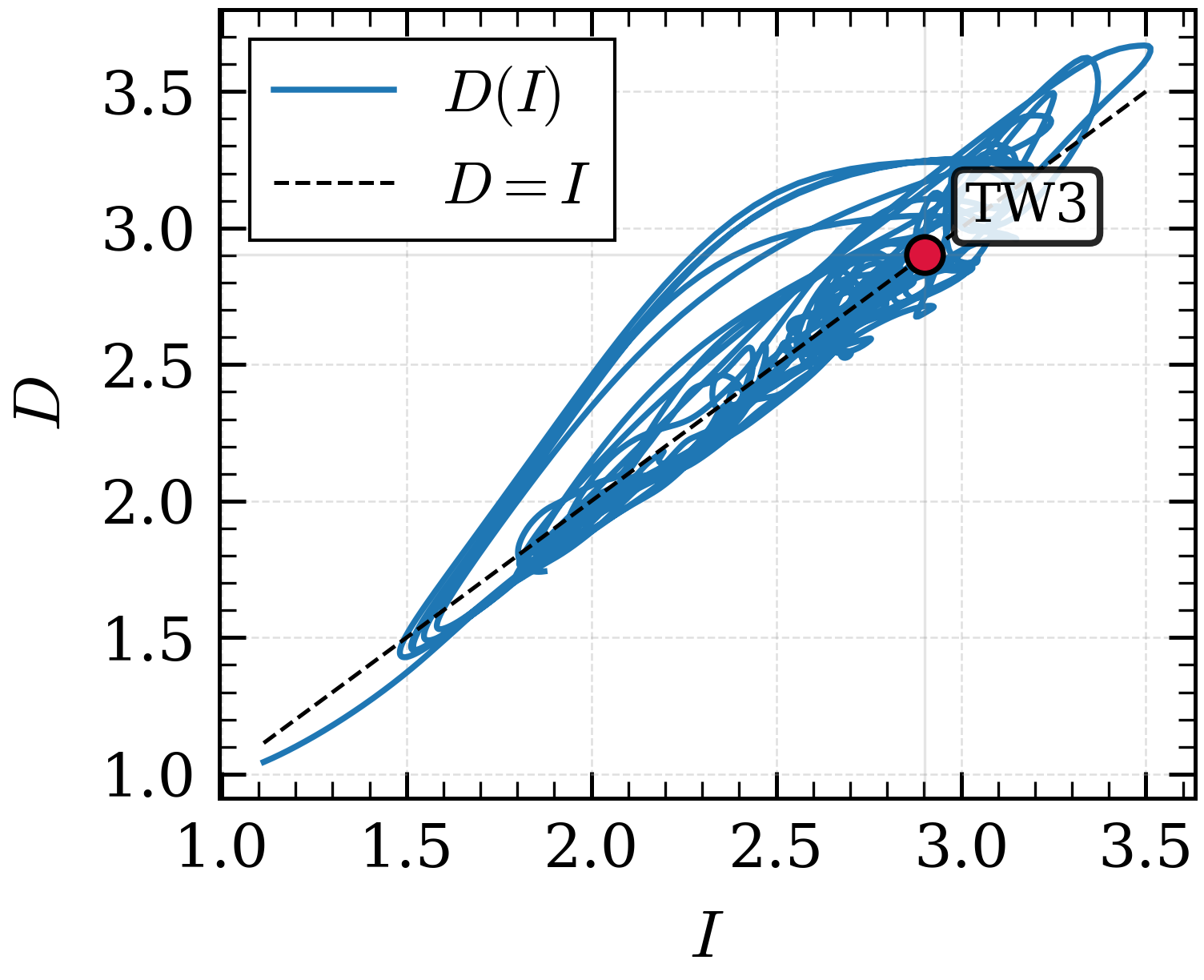}
    \caption{$Re=2000$ (TW3 candidate DNS).}
    \label{fig:DI_TW3_raw}
  \end{subfigure}
  \caption{Raw $(D,I)$ (dissipation--input) projections of DNS trajectories used
  to initialise the Newton--Krylov searches for the ECS reported in this paper.
  Near approaches to the identity line $D=I$ (small $|D-I|$) and near-closure of
  loops in the $(D,I)$ plane provide a compact signature of candidate recurrence
  windows (RPOs) or near-steady episodes in a comoving frame (TWs).}
  \label{fig:DI_all_raw}
\end{figure}

Figure~\ref{fig:DI_all_raw} summarises the raw $(D,I)$ projections for the four
DNS data sets that directly supplied the initial guesses for RPO1, RPO2, TW1
and TW3. The remaining TW (TW2) is obtained from edge tracking at $Re=2000$;
for this case we show the corresponding $\|u\|$ (the $L^2$-norm of the velocity
field) and cross-flow energy as functions of time in
figure~S1 of the supplementary material. The subsequent subsections analyse each ECS in
turn (RPO1, RPO2, then TW1--TW3), reporting their defining parameters (domain,
resolution, drift/wavespeed where applicable), representative structure
visualisations, and linear stability as quantified by Floquet spectra.

\subsection{RPO1 at \texorpdfstring{$Re=1000$}{Re=1000}}
\label{sec:rpo1}

\subsubsection{Overview and defining parameters}

\begin{figure}[htbp]
  \centering
  \begin{subfigure}[htbp]{0.49\textwidth}
    \centering
    \includegraphics[width=\linewidth]{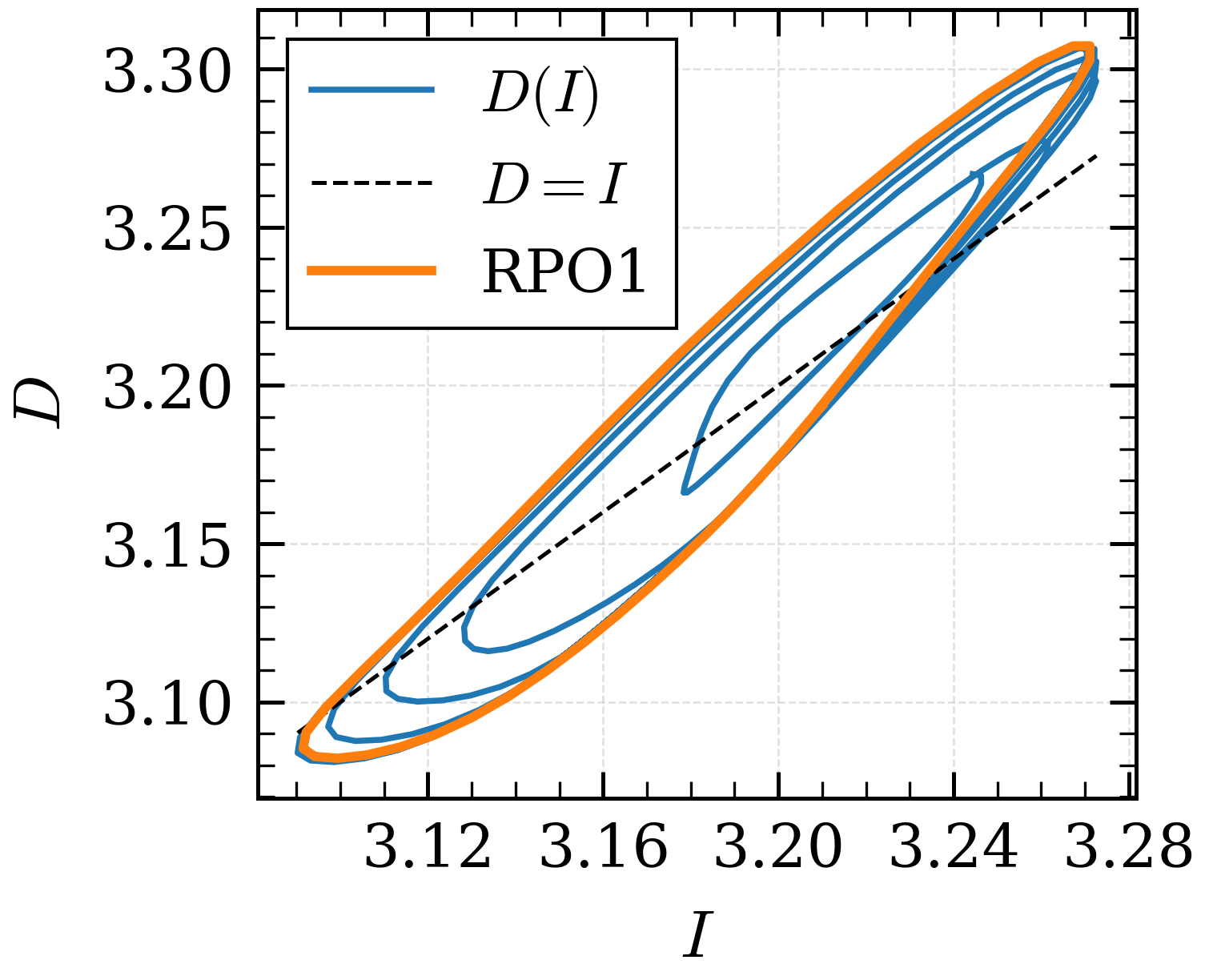}
    \caption{Converged RPO demarcated on the phase plot.}
    \label{fig:rpo1_phase_shifted}
  \end{subfigure}\hfill
  \begin{subfigure}[htbp]{0.49\textwidth}
    \centering
    \includegraphics[width=\linewidth]{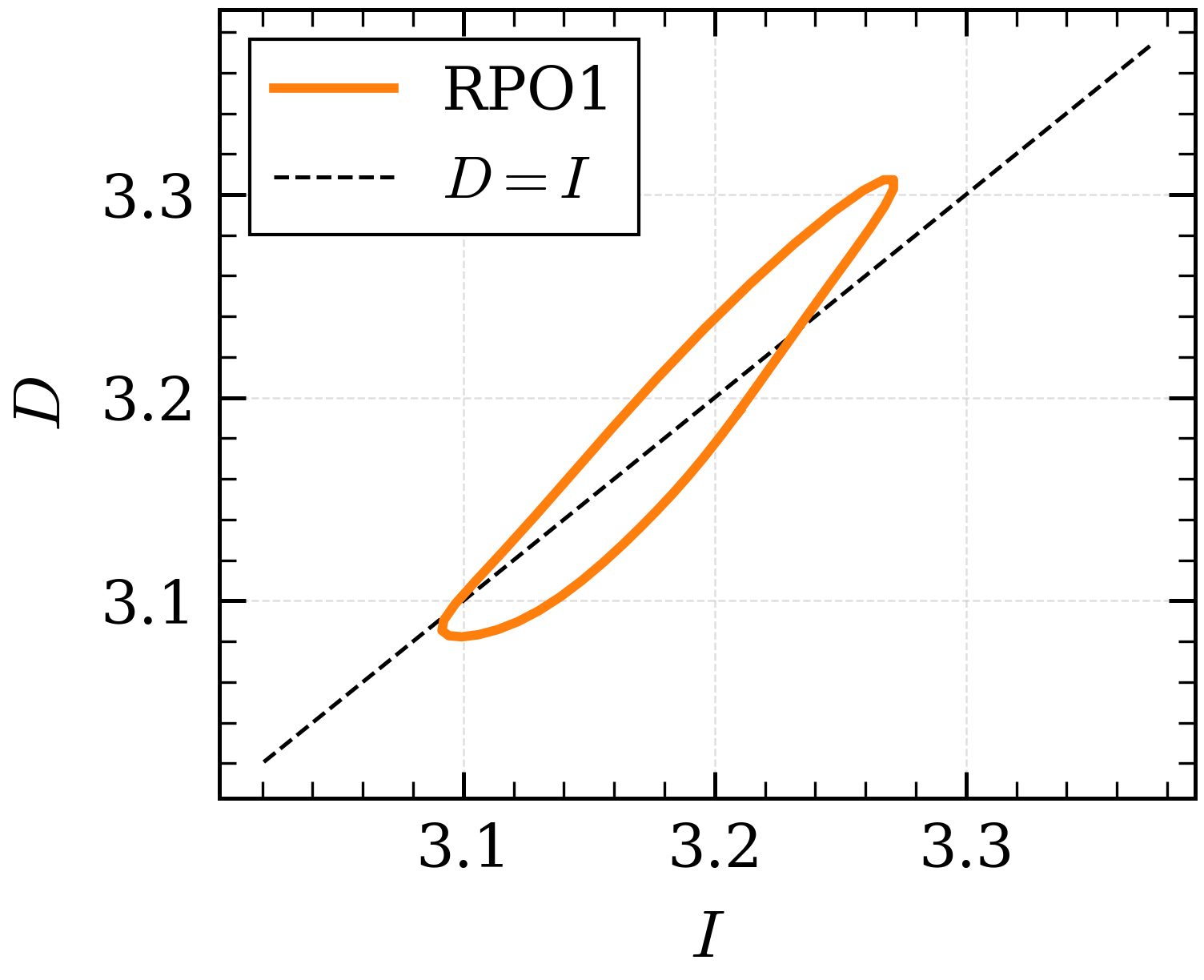}
    \caption{Single-period segment in $(D,I)$.}
    \label{fig:rpo1_phase_segment}
  \end{subfigure}
  \caption{Phase-plane diagnostics for RPO1 at $Re=1000$. Repeated overlap of
  $T$-shifted loops indicates a robust relative recurrence consistent with a
  drift $(a_x,a_z)$ over one period.}
  \label{fig:rpo1_phase}
\end{figure}

RPO1 is a relative periodic orbit of plane Poiseuille flow at $Re=1000$,
computed in the symmetry-invariant subspace
$\langle \sigma_y,\sigma_z,\tau_{xz}\rangle$, where $\tau_{xz}=\tau(L_x/2,L_z/2)$.
It satisfies the relative-periodicity condition
\[
  \tau(a_x,a_z)\,\phi^{T}(\mathbf{u})=\mathbf{u},
\]
with the numerically converged parameters
\[
  T \approx 41,\qquad
  a_x \approx -0.0604,\qquad
  a_z \approx 0.
\]
In the $(D,I)$ representation, RPO1 appears as a closed loop that is nearly
retraced by successive $T$-shifted segments of the DNS trajectory, providing a
clear near-recurrence window for Newton--Krylov--hookstep refinement.
Figure~\ref{fig:rpo1_phase} shows (i) the converged RPO demarcated on the
phase plot, and (ii) a representative single-period segment used to illustrate
the closure.

\subsubsection{Structure}

\begin{figure}[htbp]
  \centering
  \begin{subfigure}[t]{0.48\textwidth}
    \centering
    \includegraphics[width=\linewidth]{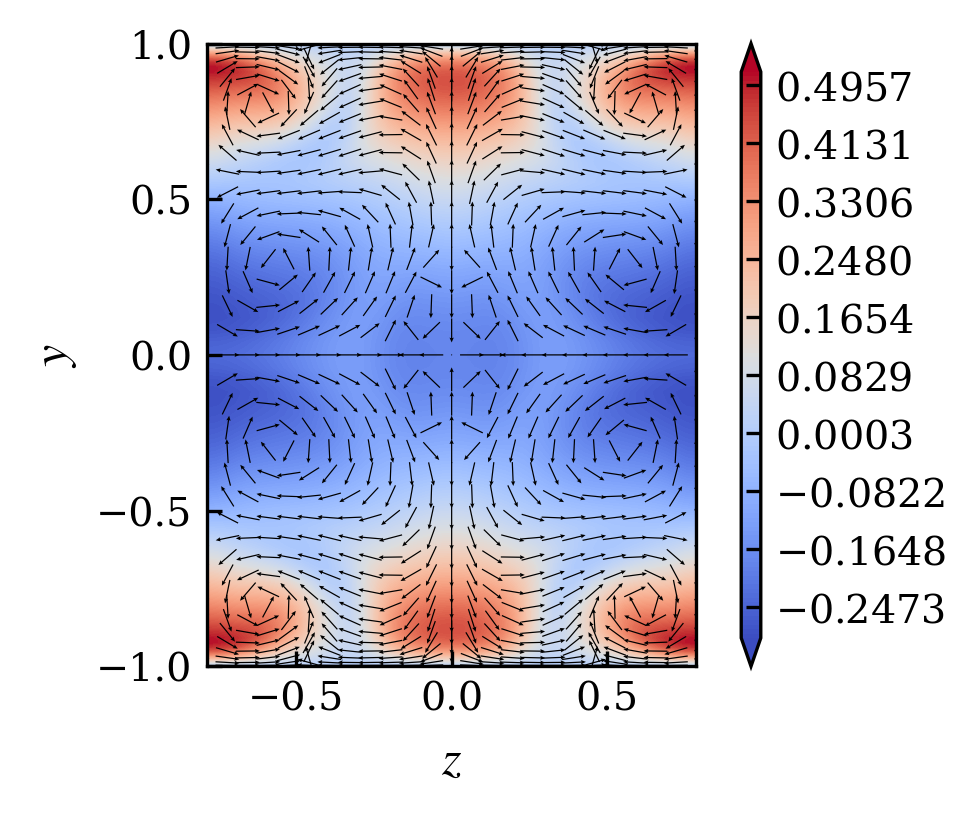}
    \caption{$(y,z)$ slice: crossflow quiver with $u$ contours.}
    \label{fig:rpo1_yz}
  \end{subfigure}
  \begin{subfigure}[t]{0.48\textwidth}
    \centering
    \includegraphics[width=\linewidth]{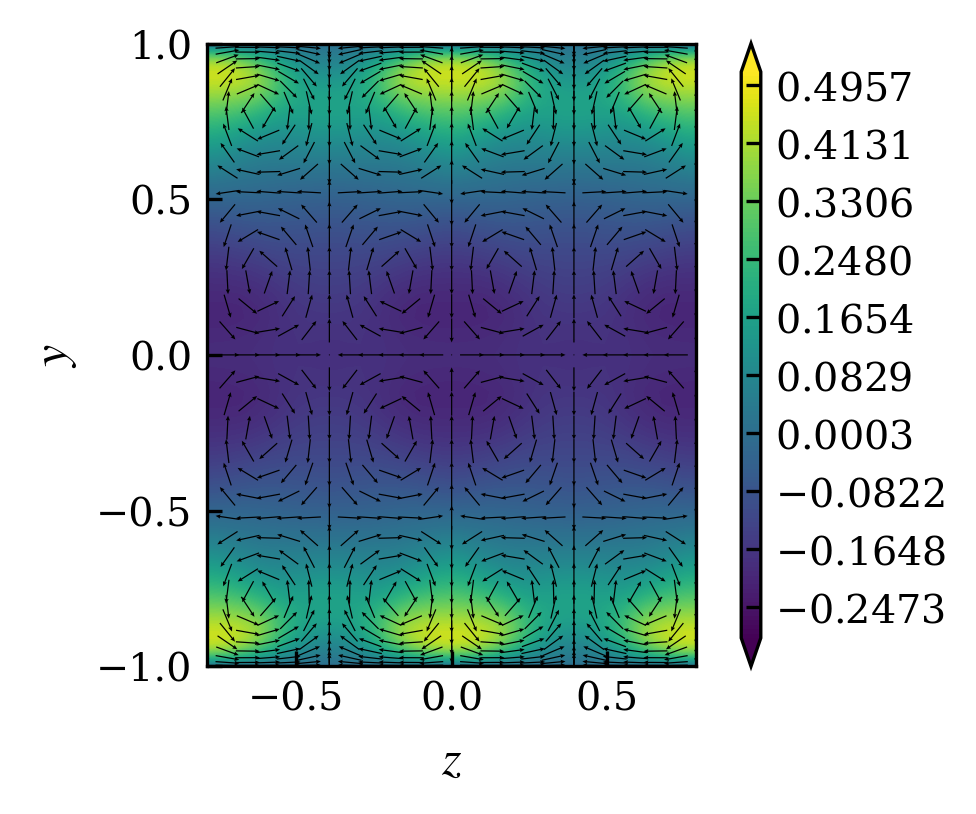}
    \caption{$(y,z)$ slice: crossflow quiver with $\langle u\rangle_x$ contours.}
    \label{fig:rpo1_yz_xavg}
  \end{subfigure}
  \begin{subfigure}[t]{\textwidth}
    \centering
    \includegraphics[width=0.72\linewidth]{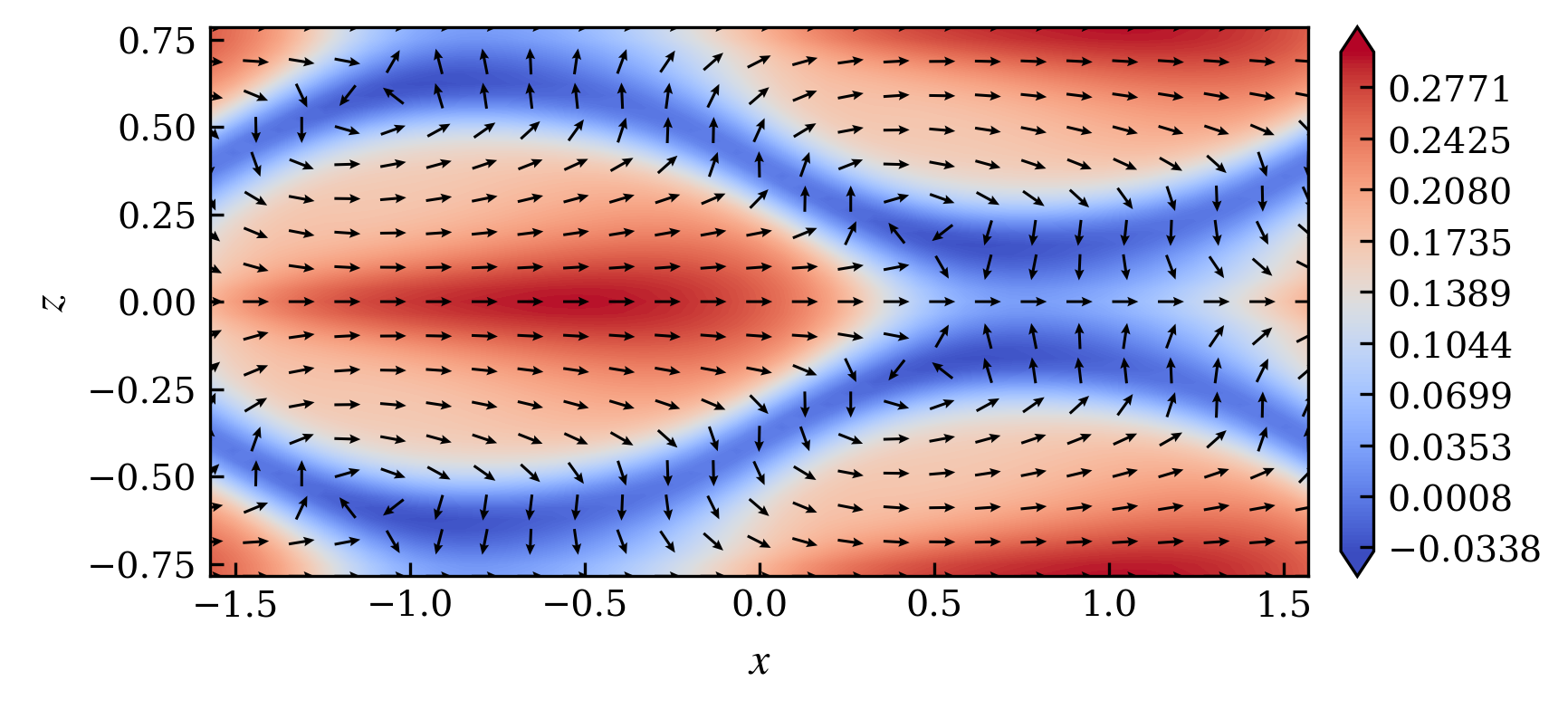}
    \caption{$(x,z)$ slice at $y=0.5$: in-plane quiver with $u$ contours.}
    \label{fig:rpo1_xz}
  \end{subfigure}
  \caption{Structure of RPO1 at $Re=1000$ shown via streak--roll projections.
  Background colour: streamwise velocity $u$ (or $\langle u\rangle_x$ in the
  middle panel). Arrows: in-plane velocity components in the plotted plane.}
  \label{fig:rpo1_quiver}
\end{figure}

Figure~\ref{fig:rpo1_quiver} summarises the spatial organisation of RPO1
through three complementary two-dimensional views, with the background colour
showing the streamwise velocity $u$ (or its streamwise average
$\langle u\rangle_x$) and the arrows indicating in-plane velocity components.
The cross-plane $(y,z)$ slice (figure~\ref{fig:rpo1_yz}) reveals an array of
counter-rotating roll cells flanking alternating high- and low-speed streaks:
high-speed regions concentrate near the walls, while lower-speed regions occupy
a broader portion of the channel interior, consistent with roll-induced
redistribution of streamwise momentum. The pattern respects the imposed
reflections $\sigma_y$ (about the channel midplane) and $\sigma_z$ (about the
spanwise midplane). When the background is replaced by the
streamwise-averaged field $\langle u\rangle_x$
(figure~\ref{fig:rpo1_yz_xavg}), the spanwise organisation of the streaks is
isolated: multiple streak bands appear in $z$, with the strongest modulation
near the walls and weaker contrast near the centreline, an arrangement
consistent with the half-box shift $\tau_{xz}$ that constrains the admissible
streak/roll tiling. The streamwise--spanwise $(x,z)$ slice at $y=0.5$
(figure~\ref{fig:rpo1_xz}) shows long streamwise-coherent streak segments with
clear spanwise modulation and an alternating pattern in $x$; the in-plane
arrows indicate a spanwise transport that directs flow from high-$u$ regions
toward low-$u$ regions, a hallmark of streak redistribution by crossflow
motions.

\subsubsection{Floquet spectrum and linear stability}

The linear stability of a relative periodic orbit is determined by its Floquet
spectrum. To obtain this spectrum, one considers the evolution of infinitesimal
perturbations to the orbit over one period $T$: the eigenvalues of the
resulting linearised map are the Floquet multipliers $\Lambda$. An RPO is
linearly stable if all multipliers satisfy $|\Lambda| \leq 1$, with the
equality $|\Lambda|=1$ reserved for neutral directions arising from continuous
symmetries of the system (e.g.\ time translation along the orbit, or spatial
translations in homogeneous directions). The corresponding Floquet exponents,
$\lambda = (1/T)\log\Lambda$, provide growth rates ($\Re(\lambda)$) and
frequencies ($\Im(\lambda)$). Stability requires $\Re(\lambda) \leq 0$ for all
non-neutral modes.

\begin{figure}[htbp]
  \centering
  \includegraphics[width=\linewidth]{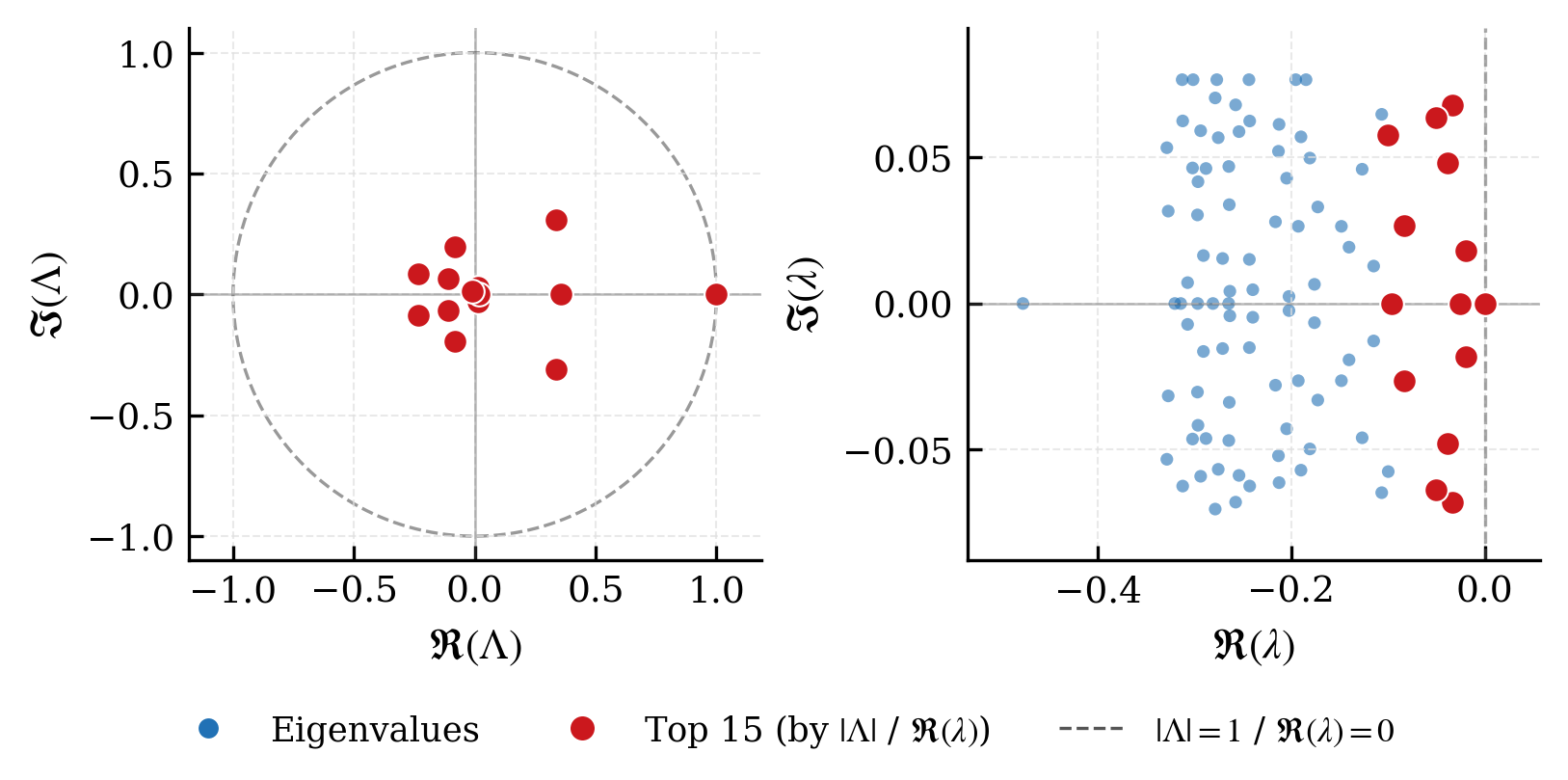}
  \caption{Floquet spectrum of RPO1 at $Re=1000$. (a)~Multipliers $\Lambda$ in
  the complex plane; the dashed circle marks $|\Lambda|=1$.
  (b)~Corresponding exponents $\lambda=(1/T)\log\Lambda$; the dashed vertical
  line marks $\Re(\lambda)=0$. In both panels, blue markers show all computed
  eigenvalues and red markers highlight the 15 eigenvalues with largest
  $|\Lambda|$ (left) or largest $\Re(\lambda)$ (right).}
  \label{fig:rpo1_floquet}
\end{figure}

Figure~\ref{fig:rpo1_floquet} shows the computed Floquet spectrum of RPO1.
In both panels, blue markers denote the full set of computed eigenvalues, while
the red markers highlight the 15 eigenvalues of largest magnitude $|\Lambda|$
(left panel) or largest real part $\Re(\lambda)$ (right panel). The left panel
displays the multipliers in the complex plane together with the unit circle
$|\Lambda|=1$. All multipliers lie on or inside the unit circle, with two real
multipliers very close to unity ($\Lambda \approx 1.000017$ and
$\Lambda \approx 1.000005$). These correspond to the neutral directions
associated with continuous symmetries. Specifically, time translation along
the orbit and streamwise translation and their marginal departure from
exactly $|\Lambda|=1$ reflects the finite numerical precision of the
computation. All remaining multipliers are well inside the unit circle,
indicating that perturbations along every non-neutral direction decay over one orbit period.

The right panel shows the corresponding exponents $\lambda$ in the complex
plane, with the dashed vertical line marking $\Re(\lambda)=0$. Consistent with
the multiplier picture, the two near-neutral modes appear with
$\Re(\lambda)\approx 0$, while all other exponents have strictly negative real
parts, including several damped oscillatory pairs with nonzero
$\Im(\lambda)$. Taken together, these results confirm that RPO1 is linearly
stable within the symmetry subspace
$\langle \sigma_y,\sigma_z,\tau_{xz}\rangle$, aside from the neutral
directions generated by continuous symmetries.

\subsection{RPO2 at \texorpdfstring{$Re=1500$}{Re=1500}}
\label{sec:rpo2}

\subsubsection{Overview and defining parameters}

\begin{figure}[htbp]
  \centering
  \begin{subfigure}[t]{0.48\textwidth}
    \centering
    \includegraphics[width=\linewidth]{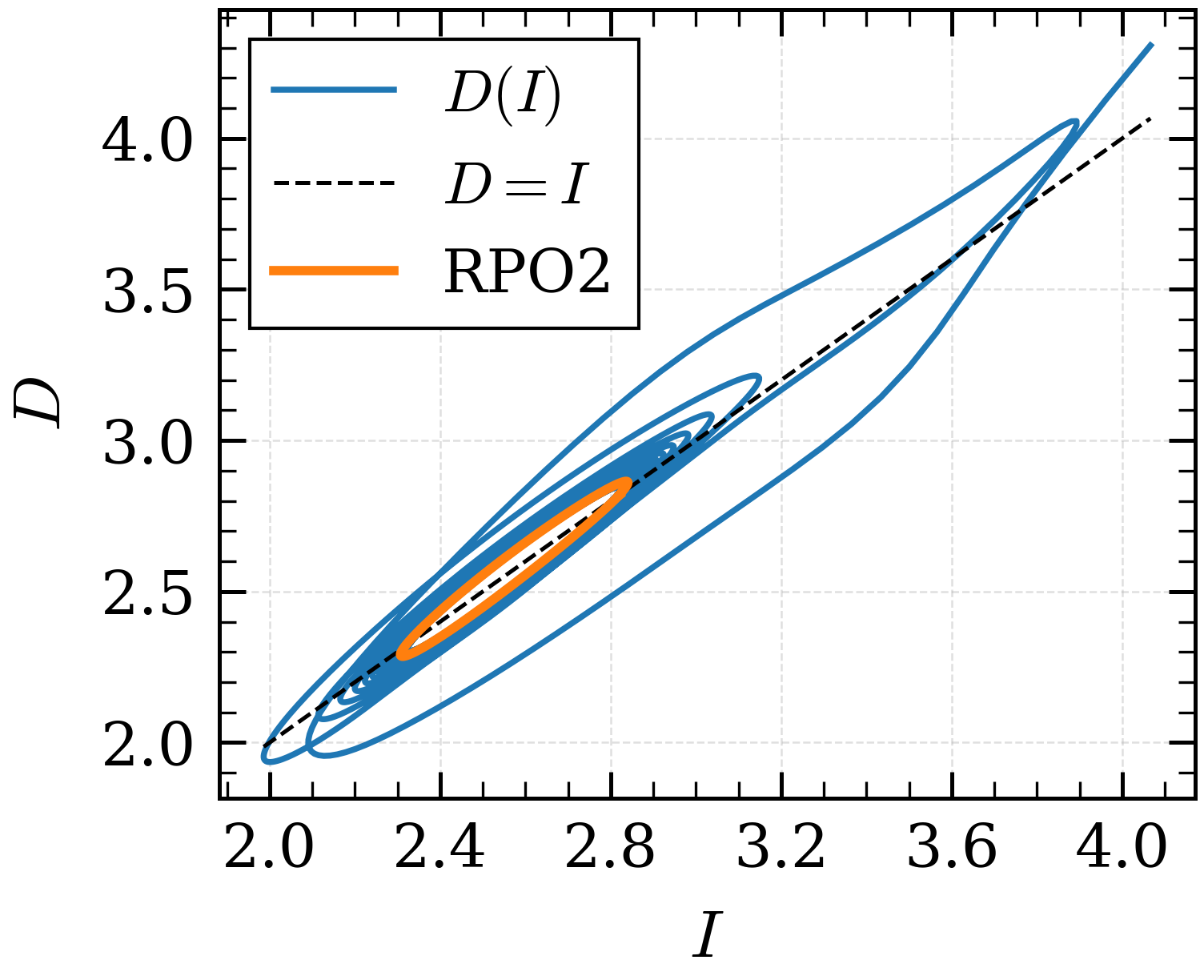}
    \caption{Converged RPO demarcated on the phase plot.}
    \label{fig:rpo2_phase_shifted}
  \end{subfigure}
  \hfill
  \begin{subfigure}[t]{0.48\textwidth}
    \centering
    \includegraphics[width=\linewidth]{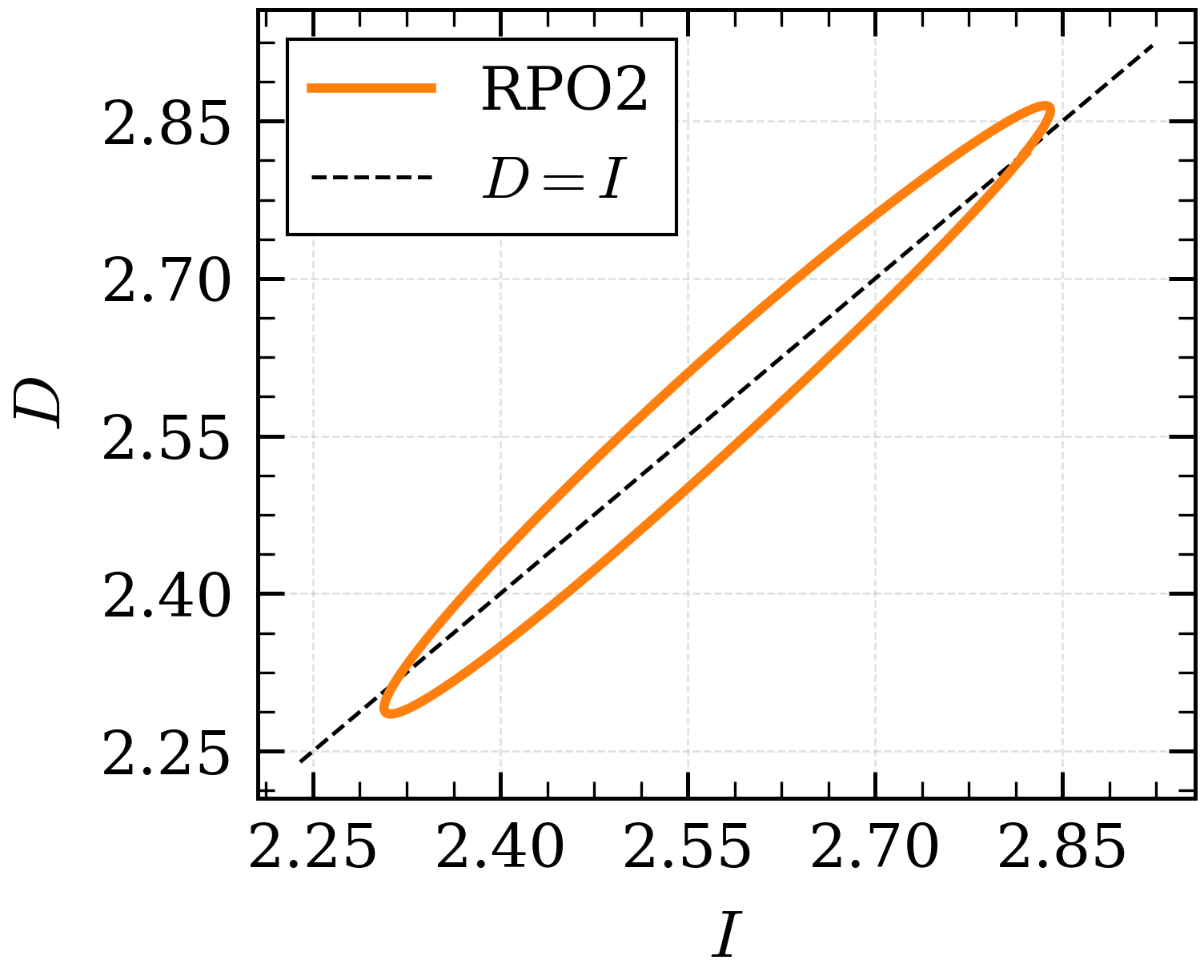}
    \caption{Single-period segment in $(D,I)$.}
    \label{fig:rpo2_converged}
  \end{subfigure}
  \caption{Phase-plane diagnostics for RPO2 at $Re=1500$. Near-closure of the
  loop in the $(D,I)$ plane indicates a robust relative recurrence consistent
  with a drift $(a_x,a_z)$ over one period.}
  \label{fig:RPO2_phase}
\end{figure}

RPO2 is a relative periodic orbit of plane Poiseuille flow at $Re=1500$,
extracted from a turbulent DNS trajectory and refined in the
symmetry-invariant subspace
$\langle \sigma_y,\,\sigma_z,\,\tau_{xz}\rangle$.
It satisfies the relative-periodicity condition
\begin{equation}
\tau_x(a_x)\,\tau_z(a_z)\,\phi^{T}(\mathbf{u}) \;=\; \mathbf{u},
\label{eq:rpo2_invariance}
\end{equation}
with the numerically converged parameters
\[
T\approx 153,\qquad
a_x \approx -0.037,\qquad
a_z \approx 0
\]
In the $(D,I)$ representation, the parent DNS trajectory exhibits a long
near-recurrent episode whose loop closes closely after one period (up to the
drift), providing a robust initial guess for Newton--Krylov--hookstep
refinement. Figure~\ref{fig:RPO2_phase} shows (i) the converged RPO
demarcated on the phase plot, and (ii) a representative single-period segment
used to illustrate the closure.

\subsubsection{Structure}
\label{sec:rpo2_structure}

\begin{figure}[htbp]
  \centering
  \begin{subfigure}[t]{0.48\textwidth}
    \centering
    \includegraphics[width=\linewidth]{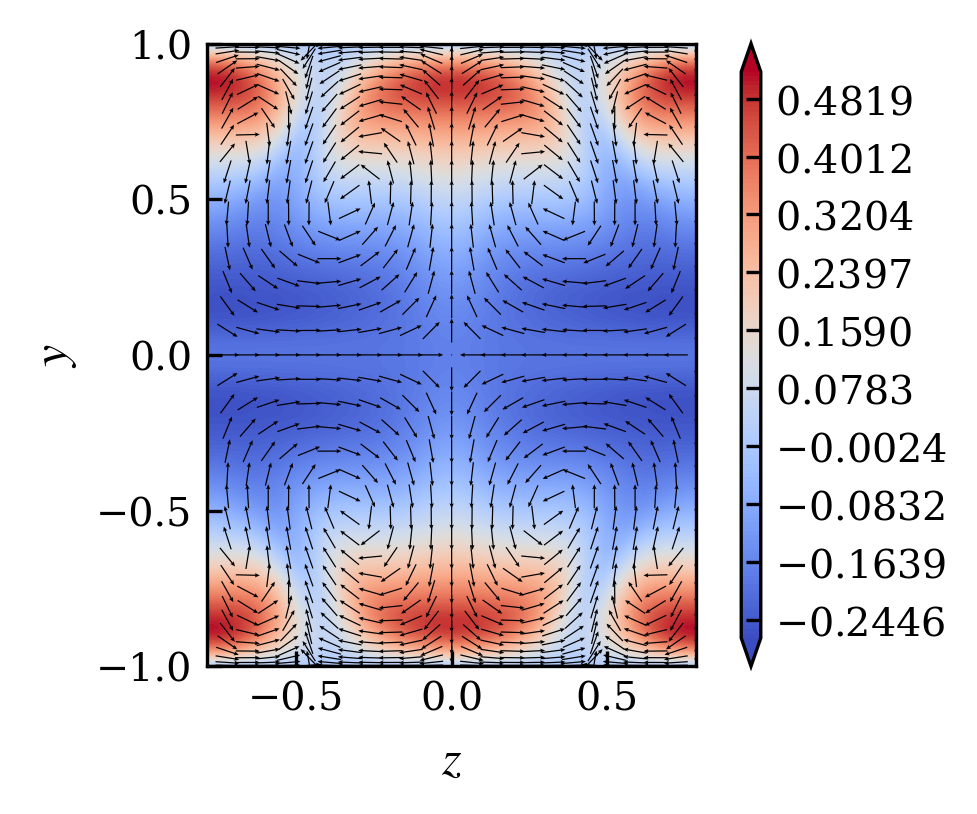}
    \caption{$(y,z)$ slice: $(v,w)$ quiver with $u$ contours.}
    \label{fig:rpo2_yz}
  \end{subfigure}
  \begin{subfigure}[t]{0.48\textwidth}
    \centering
    \includegraphics[width=\linewidth]{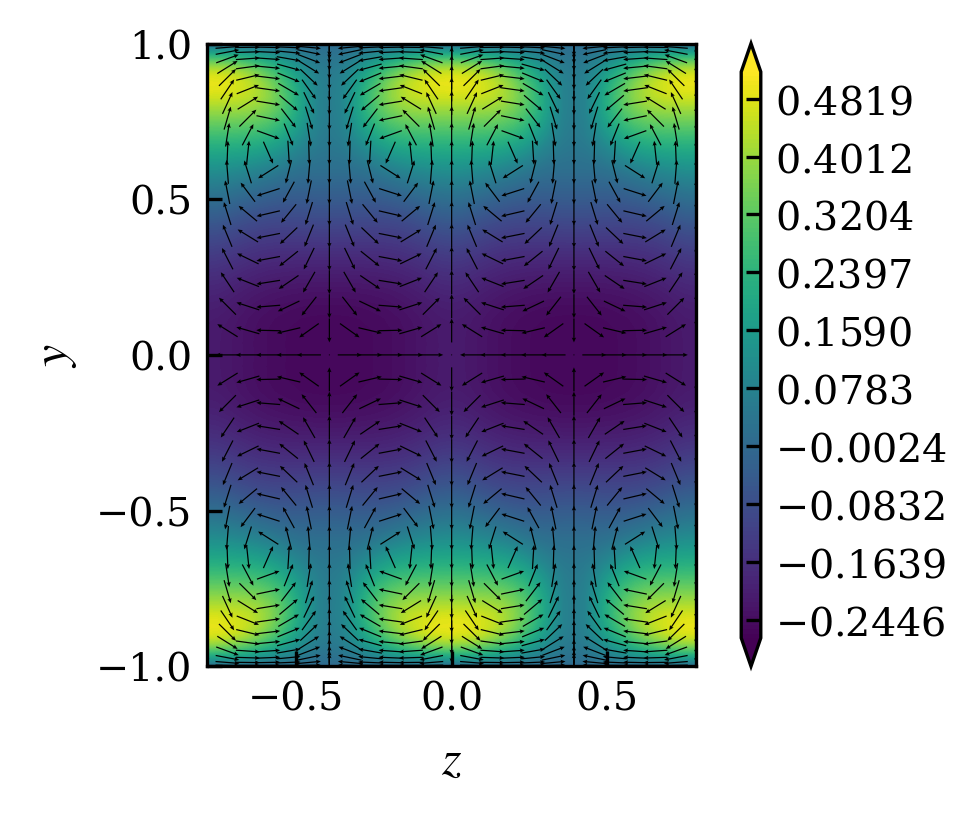}
    \caption{$(y,z)$ slice: $(v,w)$ quiver with $\langle u\rangle_x$ contours.}
    \label{fig:rpo2_yz_avg}
  \end{subfigure}
  \vspace{2mm}
  \begin{subfigure}[t]{\textwidth}
    \centering
    \includegraphics[width=0.7\linewidth]{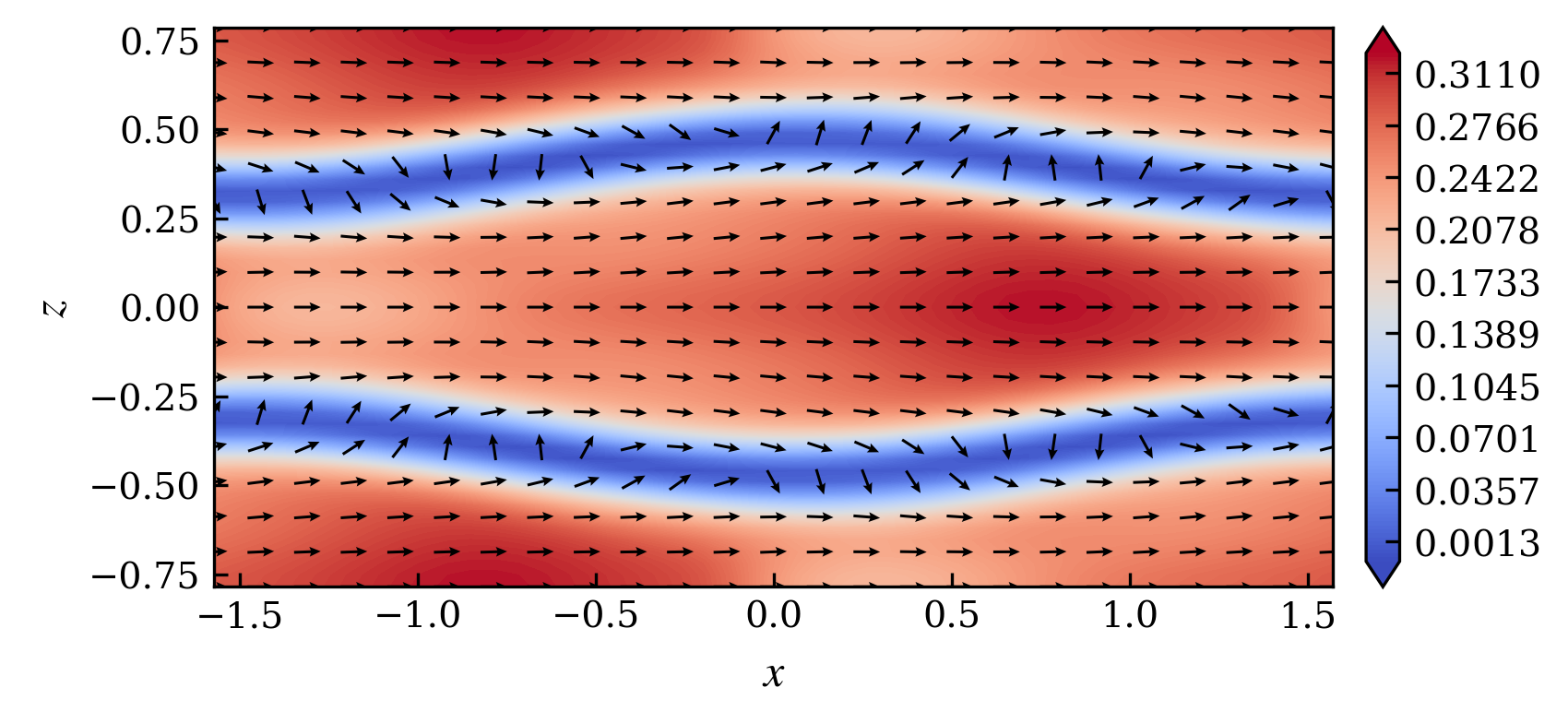}
    \caption{$(x,z)$ slice at $y=0.5$: $(u,w)$ quiver with $u$ contours.}
    \label{fig:rpo2_xz}
  \end{subfigure}
  \caption{Structure of RPO2 at $Re=1500$ shown via streak--roll projections.
  Background colour: streamwise velocity $u$ (or $\langle u\rangle_x$ in the
  middle panel). Arrows: in-plane velocity components in the plotted plane.}
  \label{fig:RPO2_quiver}
\end{figure}

Figure~\ref{fig:RPO2_quiver} summarises the spatial organisation of RPO2
through three complementary two-dimensional views, with the background colour
showing the streamwise velocity $u$ (or its streamwise average
$\langle u\rangle_x$) and the arrows indicating in-plane velocity components.
The cross-plane $(y,z)$ slice (figure~\ref{fig:rpo2_yz}) reveals
counter-rotating roll cells across the span, with alternating upwash and
downwash regions that produce strong near-wall streak modulation: high-speed
regions near the walls (warm colours) alternate with lower-speed regions
(cooler colours) in $z$, consistent with roll-driven lift-up. The
streamwise-averaged view (figure~\ref{fig:rpo2_yz_avg}) filters the
streamwise waviness and isolates the persistent mean streak/roll
organisation: the averaged $u$ field shows robust spanwise streak modulation
that is symmetric about the channel centreline, while the $(v,w)$ arrows
retain the multi-cell roll pattern, confirming that RPO2 contains a coherent,
repeatable roll--streak structure rather than a purely transient one. The
streamwise--spanwise $(x,z)$ slice at $y=0.5$ (figure~\ref{fig:rpo2_xz})
shows long streamwise-oriented streak bands with pronounced spanwise variation
and a weak streamwise waviness; the in-plane $(u,w)$ arrows indicate
alternating spanwise transport along the streak flanks, consistent with the
crossflow motions observed in the $(y,z)$ views.

\subsubsection{Floquet spectrum and linear stability}
\label{sec:rpo2_floquet}

\begin{figure}[htbp]
  \centering
  \includegraphics[width=\linewidth]{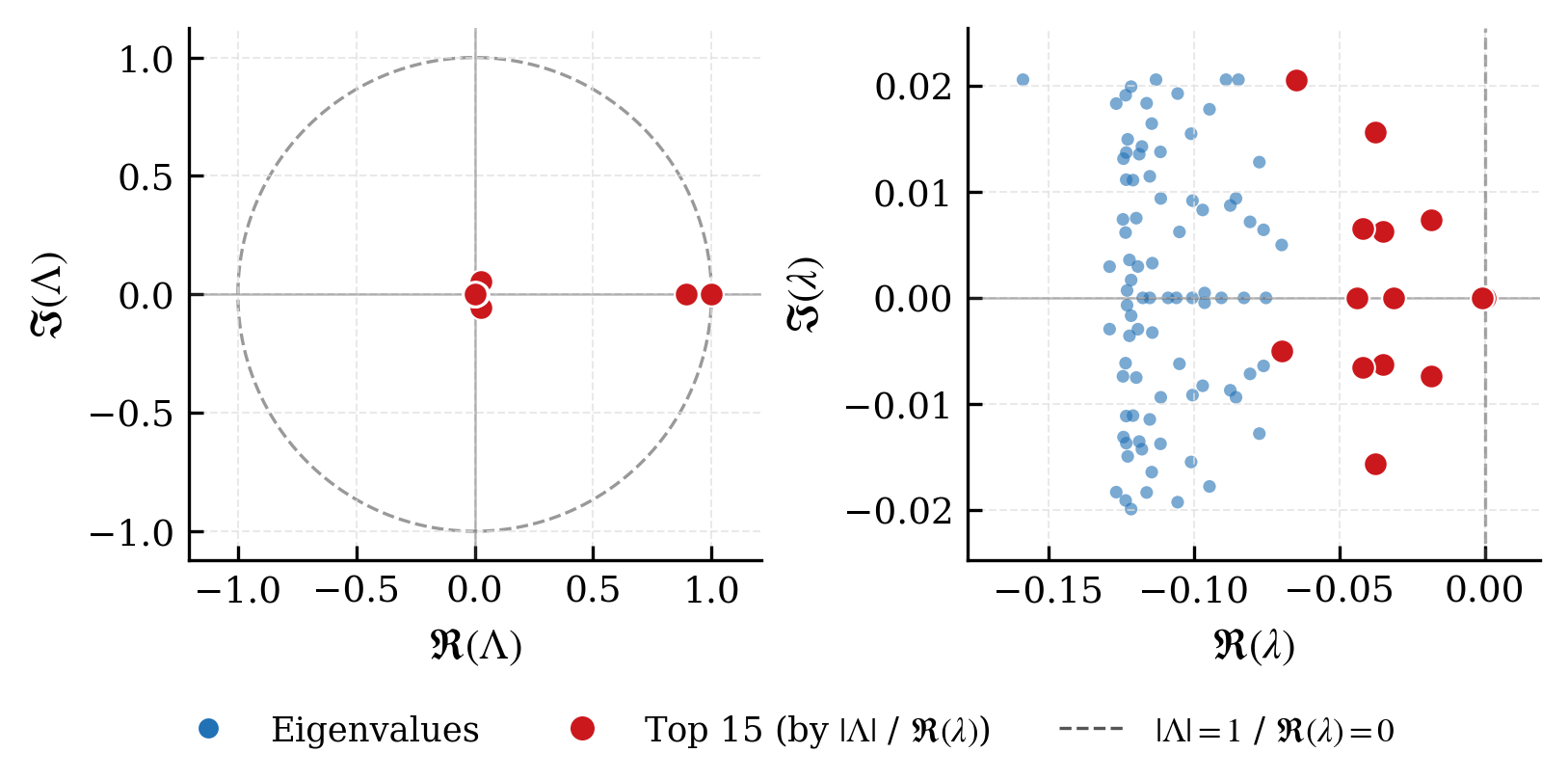}
  \caption{Floquet spectrum of RPO2 at $Re=1500$. (a)~Multipliers $\Lambda$
  in the complex plane; the dashed circle marks $|\Lambda|=1$.
  (b)~Corresponding exponents $\lambda=(1/T)\log\Lambda$; the dashed vertical
  line marks $\Re(\lambda)=0$. In both panels, blue markers show all computed
  eigenvalues and red markers highlight the 15 eigenvalues with largest
  $|\Lambda|$ (left) or largest $\Re(\lambda)$ (right).}
  \label{fig:rpo2_floquet}
\end{figure}

The linear stability of RPO2 is characterised by its Floquet multipliers
$\Lambda$ and corresponding exponents
$\lambda=(1/T)\log\Lambda$, computed over one period $T$.
Figure~\ref{fig:rpo2_floquet} shows the computed Floquet spectrum. In both
panels, blue markers denote the full set of computed eigenvalues, while the
red markers highlight the 15 eigenvalues of largest magnitude $|\Lambda|$
(left panel) or largest real part $\Re(\lambda)$ (right panel).

The left panel displays the multipliers in the complex plane together with
the unit circle $|\Lambda|=1$. All multipliers lie on or inside the unit
circle. Two multipliers are very close to unity on the real axis; these
correspond to neutral directions that are expected on general grounds because
RPO2, like any relative periodic orbit, possesses continuous symmetries that
leave the solution invariant, namely, a time shift along the orbit (which
maps one point on the orbit to another) and a streamwise translation
(reflecting the homogeneity of the flow direction). Since perturbations
aligned with these directions neither grow nor decay, the associated
multipliers must satisfy $|\Lambda|=1$.
All remaining multipliers lie well inside the unit circle, indicating that
perturbations along every non-neutral direction decay over one orbit period.

The right panel of fig. \ref{fig:rpo2_floquet} shows the corresponding exponents in the
$(\Re(\lambda),\Im(\lambda))$ plane, with the dashed vertical line marking
$\Re(\lambda)=0$. Consistent with the multiplier picture, the two
near-neutral modes appear with $\Re(\lambda)\approx 0$, while all other
exponents have strictly negative real parts, including several damped
oscillatory pairs with nonzero $\Im(\lambda)$. Taken together, these results
confirm that RPO2 is linearly stable within the symmetry subspace
$\langle \sigma_y,\,\sigma_z,\,\tau_{xz}\rangle$, aside from the neutral
directions generated by continuous symmetries. This stability is consistent
with the long near-recurrent windows observed in the $(D,I)$ projection,
which indicate that the turbulent trajectory spends extended intervals in
the neighbourhood of RPO2 before departing along weakly unstable directions
that lie outside the enforced symmetry subspace.

\subsection{TW1 at \texorpdfstring{$Re=1900$}{Re=1900}}
\label{sec:tw1}

\subsubsection{Overview and defining parameters}

TW1 is a travelling wave (relative equilibrium) computed at $Re=1900$.
In the results reported here, the converged solution lies in the
symmetry-invariant subspace
\[
\langle \sigma_y,\;\tau_x,\;\tau_z\rangle,
\qquad
\tau_x=\tau(L_x/2,0),\;\;\tau_z=\tau(0,L_z/2),
\]
i.e.\ TW1 is invariant under reflection about the channel centreplane and
independent half-box translations in $x$ and $z$ (hence also under
$\tau_{xz}=\tau_x\tau_z$, which is implied by $\tau_x$ and $\tau_z$).
As a travelling wave, TW1 is steady in a suitably translating frame: there exist
a time $T>0$ (the time horizon used to form the Newton residual) and drifts
$(a_x,a_z)$ such that the invariance condition is
\begin{equation}
\phi^{T}(\mathbf{u}) \;=\; \tau(a_x,a_z)\,\mathbf{u},
\end{equation}
\begin{equation}
\qquad \tau(a_x,a_z)=\tau(a_x,a_z)[u,v,w](x,y,z)=[u,v,w](x+a_x,y,z+a_z),
\label{eq:tw1_invariance}
\end{equation}
with the numerically converged drifts
\[
a_x \approx -0.011,\qquad a_z \approx 0.
\]
Thus, in the laboratory frame the dominant evolution of TW1 is a streamwise
translation, with negligible spanwise drift.

\subsubsection{Structure}
\label{sec:tw1_structure}

\begin{figure}[htbp]
  \centering
  \begin{subfigure}[t]{0.51\textwidth}
    \centering
    \includegraphics[width=1.2\linewidth]{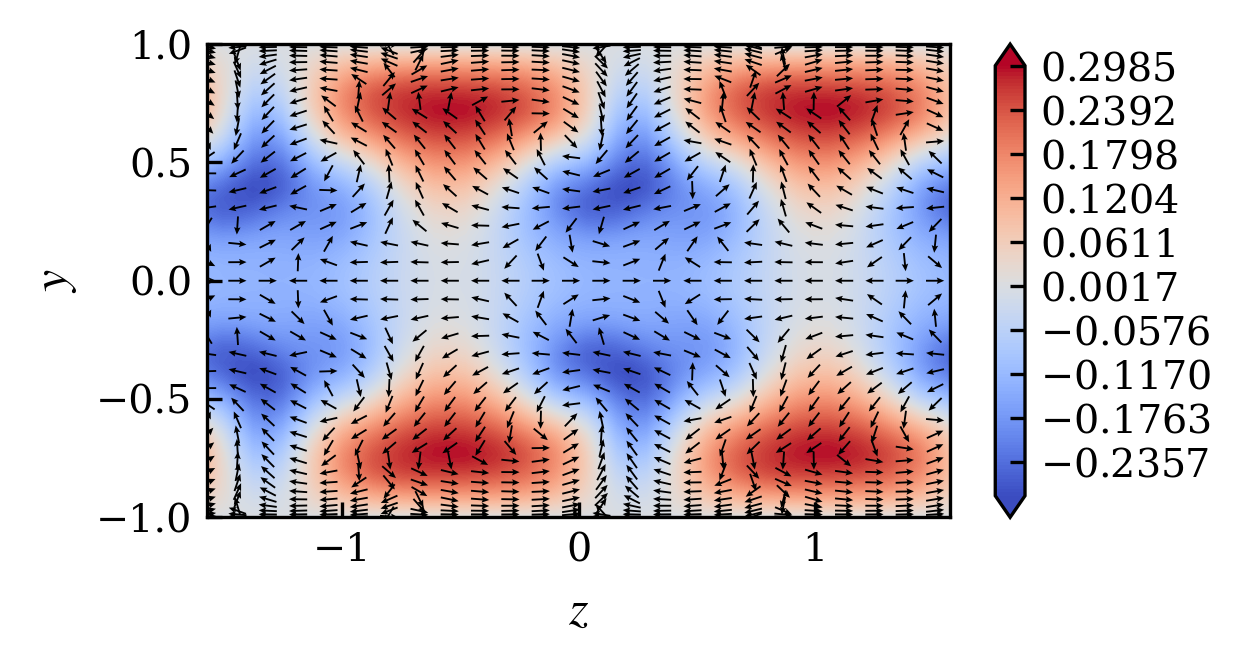}
    \caption{$(y,z)$ slice: $(v,w)$ quiver with $u$ contours.}
    \label{fig:tw1_yz}
  \end{subfigure}
  \hfill
  \begin{subfigure}[t]{0.51\textwidth}
    \centering
    \includegraphics[width=1.2\linewidth]{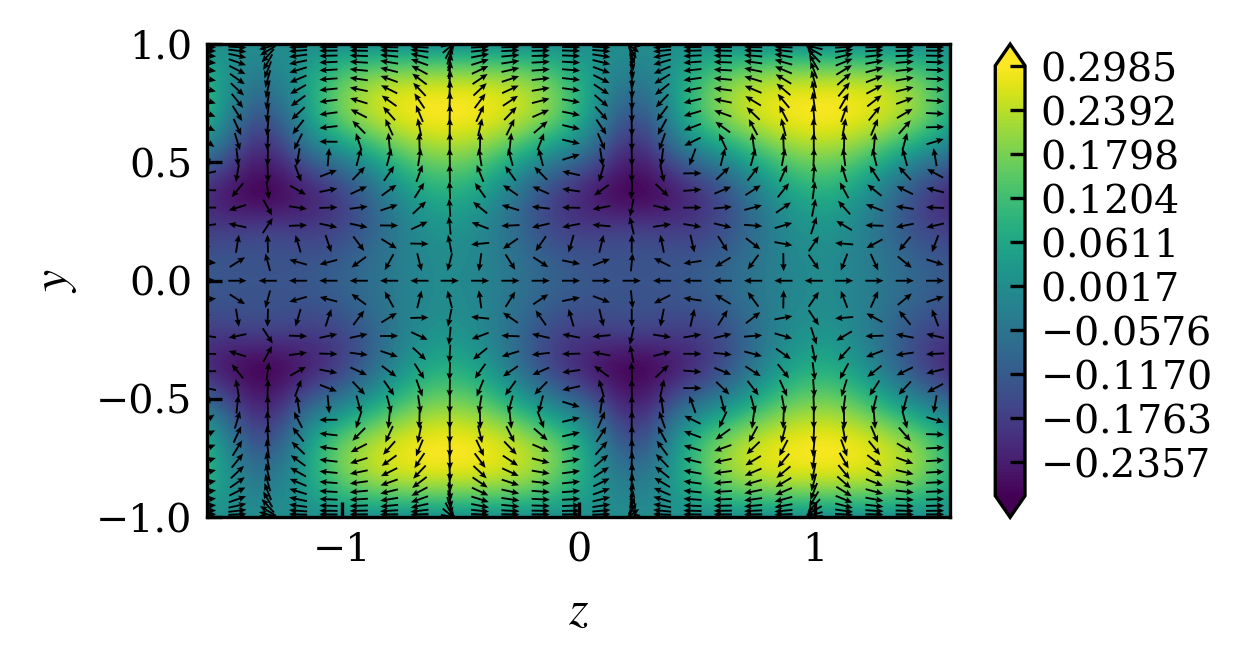}
    \caption{$(y,z)$ slice: $(v,w)$ quiver with $\langle u\rangle_x$ contours.}
    \label{fig:tw1_yz_avg}
  \end{subfigure}
  \vspace{2mm}
  \begin{subfigure}[t]{0.9\textwidth}
    \centering
    \includegraphics[width=0.9\linewidth]{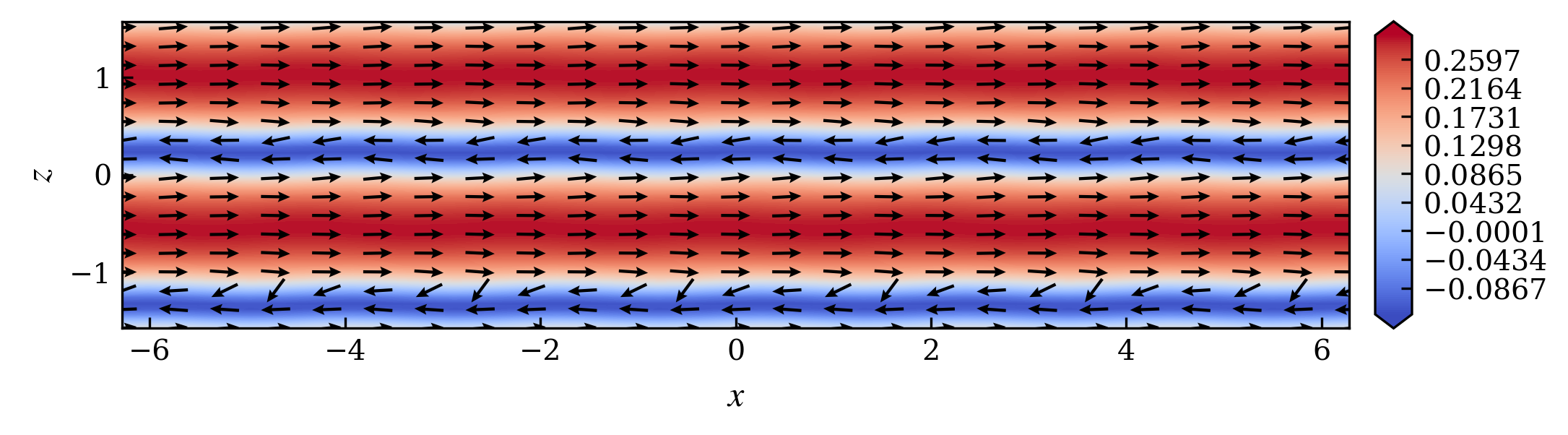}
    \caption{$(x,z)$ slice at $y=0.5$: $(u,w)$ quiver with $u$ contours.}
    \label{fig:tw1_xz}
  \end{subfigure}
  \caption{Structure of TW1 at $Re=1900$ shown via streak--roll projections.
  Background colour: streamwise velocity $u$ (or $\langle u\rangle_x$ in panel \emph{b}). Arrows: in-plane velocity components in the plotted plane.}
  \label{fig:TW1_quiver}
\end{figure}

Figure~\ref{fig:TW1_quiver} summarises the spatial organisation of TW1
through three complementary two-dimensional views, with the background colour
showing the streamwise velocity $u$ (or its streamwise average
$\langle u\rangle_x$) and the arrows indicating in-plane velocity components.
The cross-plane $(y,z)$ slice (figure~\ref{fig:tw1_yz}) reveals a roll--streak
organisation with multiple counter-rotating roll pairs spanning the channel that redistributes streamwise momentum into alternating
high- and low-speed streak regions; the strongest $u$ modulation is
concentrated near the walls, consistent with a lift-up mechanism, and the
$\sigma_y$ symmetry is visible in the mirrored circulation and streak pattern
between the upper and lower halves of the channel. The streamwise-averaged
view (figure~\ref{fig:tw1_yz_avg}) filters the streamwise waviness and
isolates the persistent backbone of the travelling wave: the roll cells remain
sharply defined and retain their symmetry about $y=0$, while the streak
pattern appears smoother, confirming that TW1 is organised around a robust
roll--streak scaffold. The streamwise--spanwise $(x,z)$ slice at $y=0.5$
(figure~\ref{fig:tw1_xz}) shows that $u$ forms
elongated spanwise bands with very weak streamwise modulation, and the
overlaid $(u,w)$ arrows indicate a predominantly streamwise motion with a
weaker spanwise component along the streak flanks, consistent with the
crossflow transport observed in the $(y,z)$ views.

\subsubsection{Floquet spectrum and linear stability}
\label{sec:tw1_floquet}

\begin{figure}[htbp]
  \centering
  \includegraphics[width=0.9\linewidth]{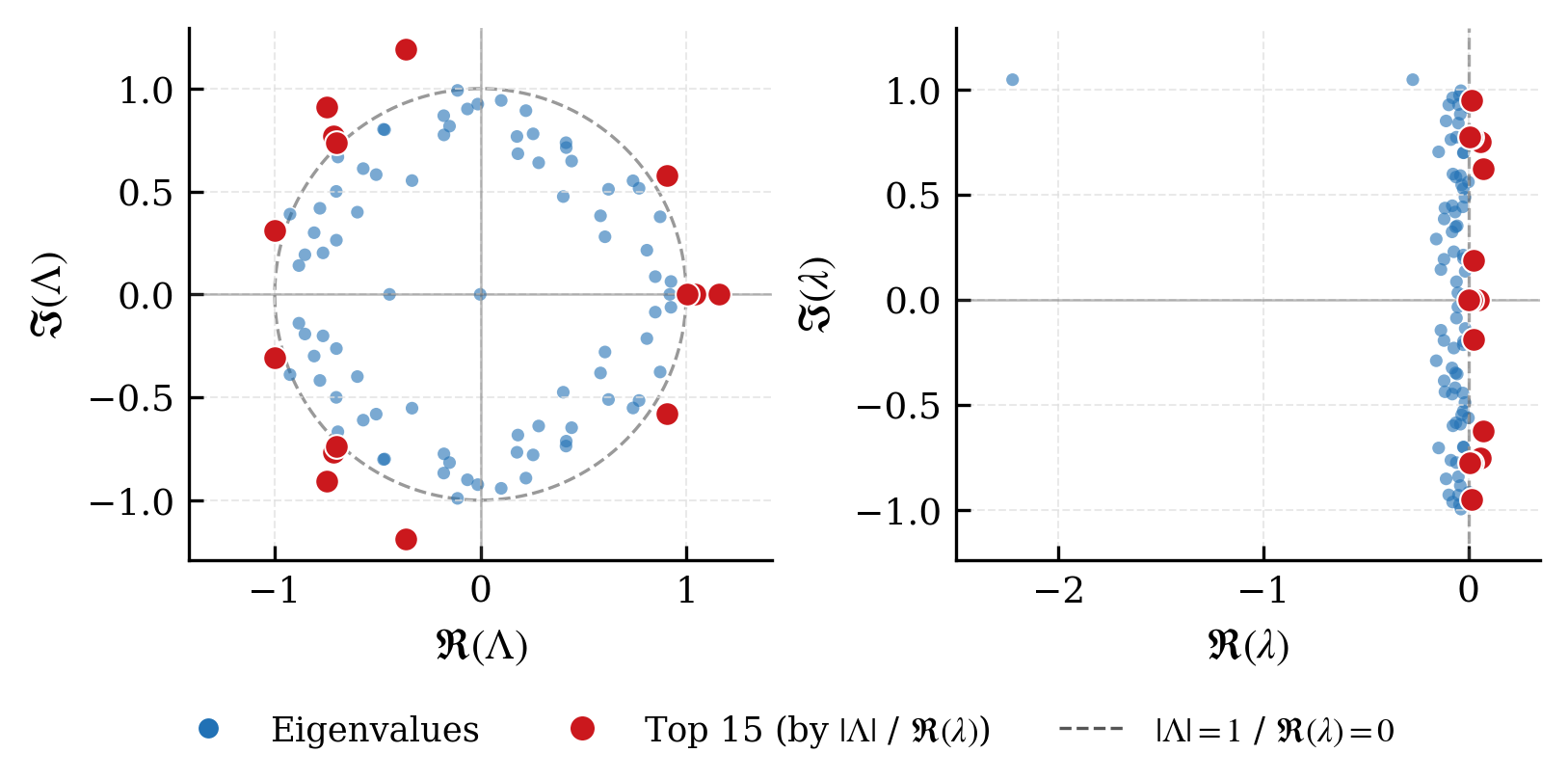}
  \caption{Floquet spectrum of TW1 at $Re=1900$. (a)~Multipliers $\Lambda$
  in the complex plane; the dashed circle marks $|\Lambda|=1$.
  (b)~Corresponding exponents $\lambda=(1/T)\log\Lambda$; the dashed vertical
  line marks $\Re(\lambda)=0$. In both panels, blue markers show all computed
  eigenvalues and red markers highlight the 15 eigenvalues with largest
  $|\Lambda|$ (left) or largest $\Re(\lambda)$ (right).}
  \label{fig:tw1_floquet}
\end{figure}

The linear stability of TW1 is characterised by its Floquet multipliers
$\Lambda$ and corresponding exponents $\lambda=(1/T)\log\Lambda$, computed
over the time horizon $T$ used in the Newton--Krylov formulation.
Figure~\ref{fig:tw1_floquet} shows the computed Floquet spectrum. In both
panels, blue markers denote the full set of computed eigenvalues, while the
red markers highlight the 15 eigenvalues of largest magnitude $|\Lambda|$
(left panel) or largest real part $\Re(\lambda)$ (right panel).

In contrast to the two RPOs presented earlier, which are linearly stable
within their symmetry subspace (all non-neutral multipliers inside the unit
circle), TW1 is \emph{unstable}: several leading multipliers lie outside the
unit circle. The left panel shows that the largest multipliers include
complex-conjugate pairs with $|\Lambda|>1$ as well as real multipliers with
$\Lambda > 1$. These two types of instability correspond to physically
distinct perturbation dynamics. A complex-conjugate pair with $|\Lambda|>1$
represents an \emph{oscillatory instability}: a perturbation aligned with
this mode grows exponentially in amplitude while oscillating at a frequency
set by $\Im(\lambda)$, so that the trajectory spirals away from TW1 in state
space. A real multiplier with $\Lambda > 1$, on the other hand, represents a
\emph{monotone instability}: the corresponding perturbation grows
exponentially without oscillation, driving the trajectory directly away from
TW1 along a fixed direction. In exponent form (right panel), these appear as
a small number of modes with $\Re(\lambda)>0$ the oscillatory pairs are the ones with nonzero $\Im(\lambda)$ and the monotone modes have $\Im(\lambda)=0$ while
all remaining exponents have $\Re(\lambda)<0$.

Despite possessing these unstable directions, TW1 has strongly contracting
dynamics along the majority of resolved directions, as indicated by the
concentration of eigenvalues well inside the unit circle. The total number of
unstable modes remains small, so that TW1 acts as a saddle-type invariant
solution in state space: nearby trajectories are attracted toward its
neighbourhood along the many stable directions, but are ultimately deflected
away along the low-dimensional unstable manifold. This saddle character
distinguishes the travelling waves from the RPOs in the present study and is
consistent with the general observation that travelling waves in wall-bounded
flows typically possess a small number of unstable directions while organising
the surrounding turbulent dynamics through their stable manifolds
\citep{gibson2009equilibrium,park2015exact,graham2021exact}.

\subsection{TW2 at \texorpdfstring{$Re=2000$}{Re=2000}}
\label{sec:tw2}

\subsubsection{Overview and defining parameters}

TW2 is a travelling-wave (relative equilibrium) solution at $Re=2000$. Although the
DNS trajectory used to initialize the search was evolved with only $\sigma_y$ enforced, the
\emph{converged} travelling wave is found to lie in the smaller symmetry-invariant subspace
\[
\langle \sigma_y,\tau_{xz}\rangle,
\qquad \tau_{xz}=\tau(L_x/2,L_z/2),
\]
i.e. ,\ it is invariant under reflection about the channel centreplane and the discrete
half-box shift in $(x,z)$.
As a travelling wave, TW2 is steady in a translating frame: there exist drifts
$(a_x,a_z)$ (equivalently, wave speeds $(c_x,c_z)$) such that the time-$T$ flow map closes
up to a continuous translation,
\begin{equation}
\phi^{T}(\mathbf{u}) \;=\; \tau(a_x,a_z)\,\mathbf{u},
\qquad (a_x,a_z)=(c_xT,c_zT),
\label{eq:tw2_invariance}
\end{equation}
with the numerically converged drifts
\[
a_x \approx 0.076,\qquad
a_z \approx 0.
\]
Thus the laboratory-frame evolution is dominated by streamwise drift, with negligible net
spanwise drift.

\subsubsection{Structure}
\label{sec:tw2_structure}

\begin{figure}[htbp]
  \centering
  \begin{subfigure}[t]{0.48\textwidth}
    \centering
    \includegraphics[width=\linewidth]{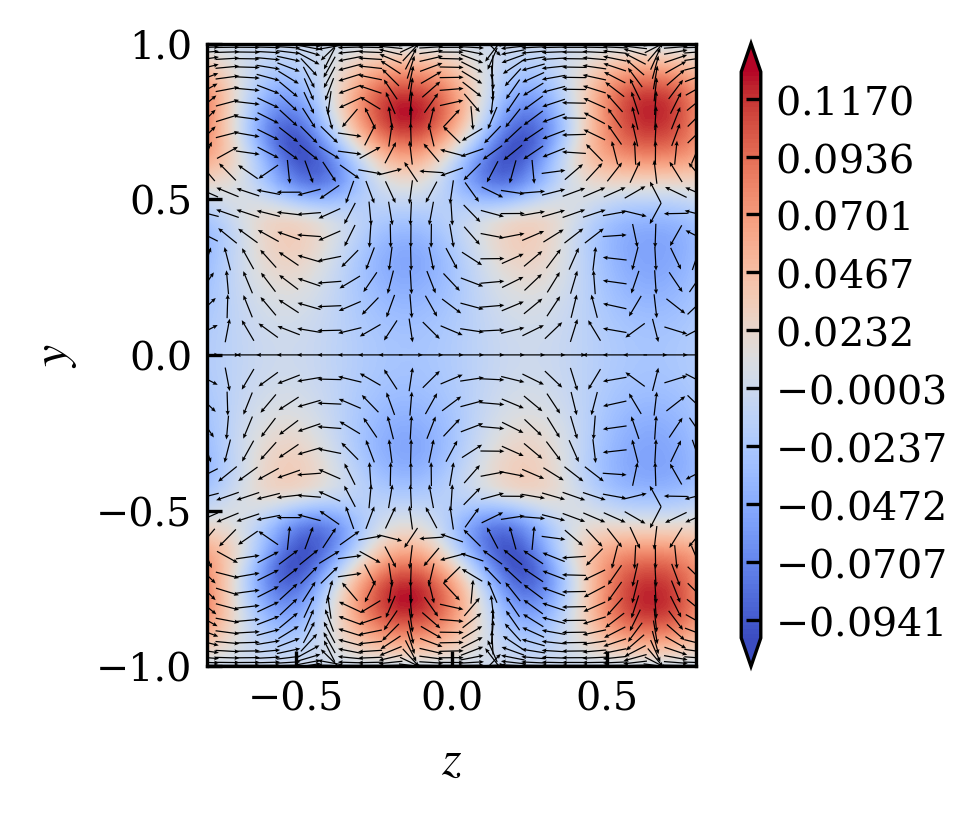}
    \caption{$(y,z)$ slice: $(v,w)$ quiver with $u$ contours.}
    \label{fig:tw2_yz}
  \end{subfigure}
  \begin{subfigure}[t]{0.48\textwidth}
    \centering
    \includegraphics[width=\linewidth]{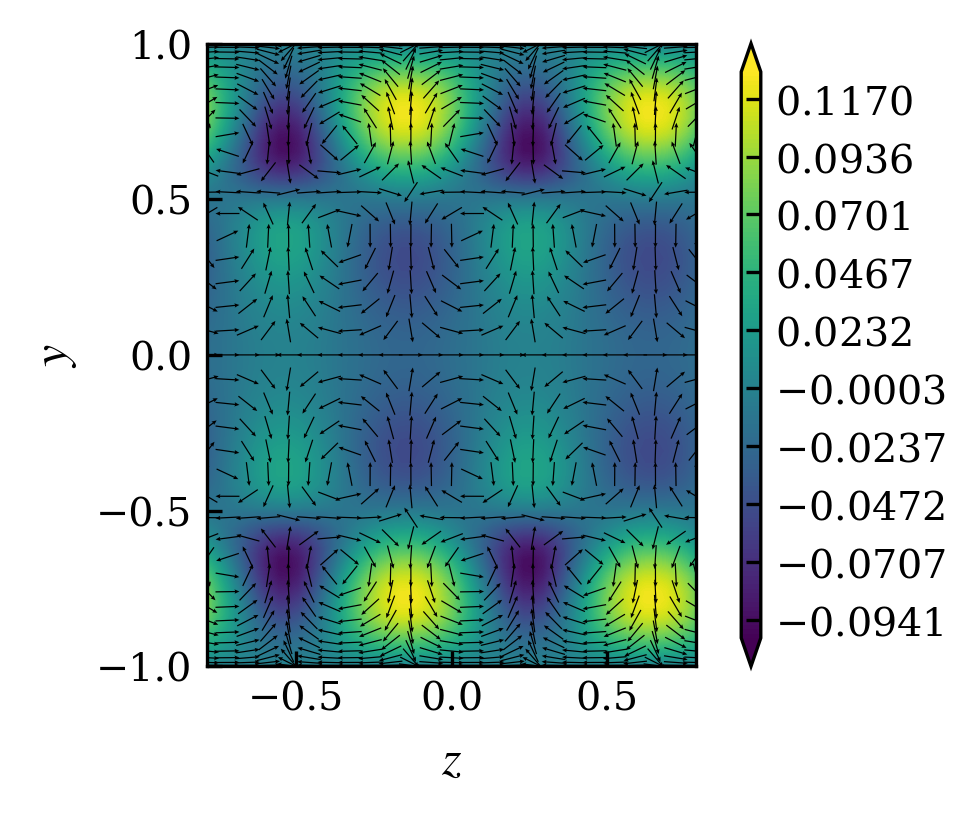}
    \caption{$(y,z)$ slice: $(v,w)$ quiver with $\langle u\rangle_x$ contours.}
    \label{fig:tw2_yz_avg}
  \end{subfigure}
  \vspace{2mm}
  \begin{subfigure}[t]{0.7\textwidth}
    \centering
    \includegraphics[width=\linewidth]{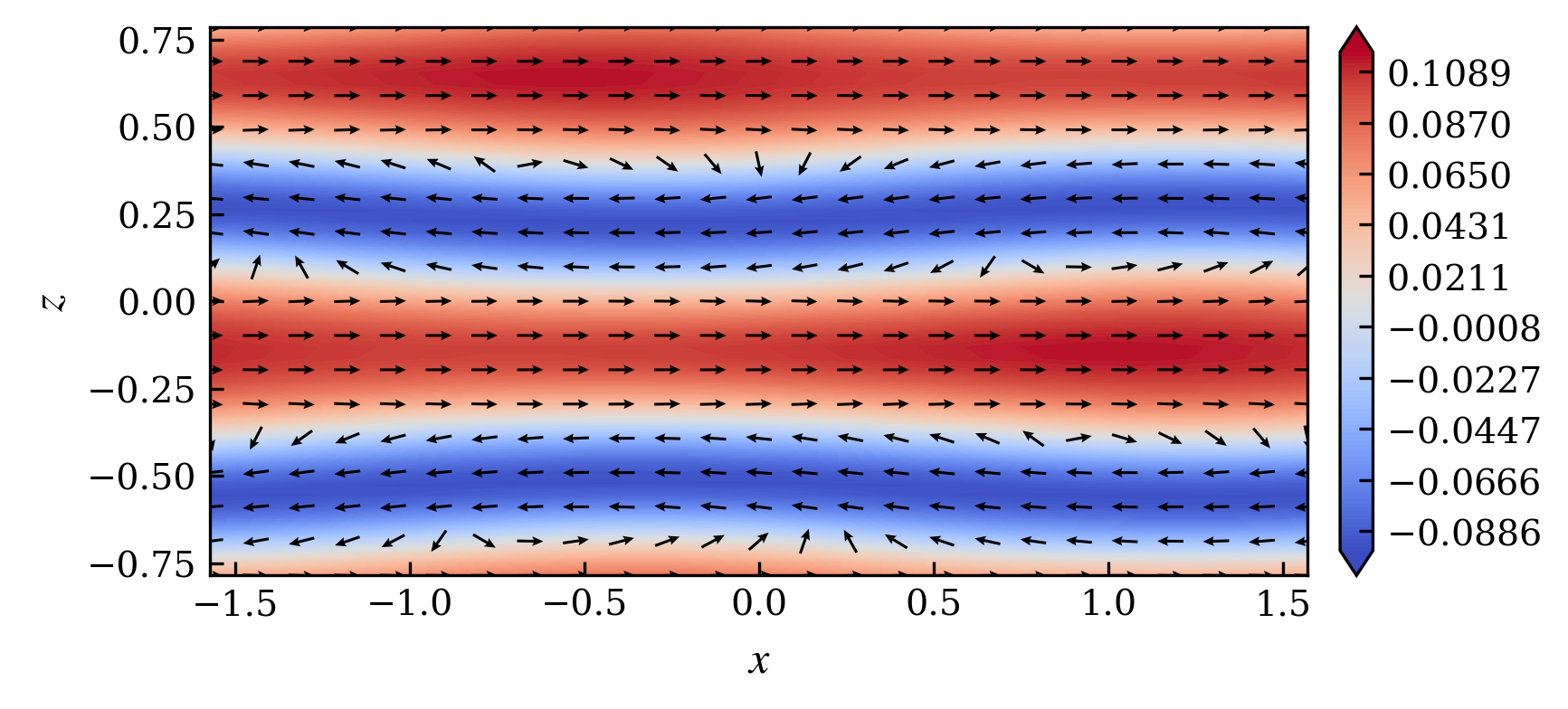}
    \caption{$(x,z)$ slice at $y=0.5$: $(u,w)$ quiver with $u$ contours.}
    \label{fig:tw2_xz}
  \end{subfigure}
  \caption{Structure of TW2 at $Re=2000$ shown via streak--roll projections.
  Background colour: streamwise velocity $u$ (or $\langle u\rangle_x$ in the
  middle panel). Arrows: in-plane velocity components in the plotted plane.}
  \label{fig:TW2_quiver}
\end{figure}

Figure~\ref{fig:TW2_quiver} summarises the spatial organisation of TW2
through three complementary two-dimensional views, with the background colour
showing the streamwise velocity $u$ (or its streamwise average
$\langle u\rangle_x$) and the arrows indicating in-plane velocity components.
The cross-plane $(y,z)$ slice (figure~\ref{fig:tw2_yz}) shows a compact
roll--streak arrangement with four prominent recirculation cells (two across
the span and two stacked across the channel height), accompanied by strong
vertical exchange near the spanwise boundaries; the resulting $u$ field has
localised high-speed patches near the walls and reduced velocity in the core,
and the $\sigma_y$ symmetry is evident in the mirrored pattern about $y=0$.
Notably, the streamwise velocity amplitudes are substantially weaker than
those of TW1 (peak $|u| \approx 0.13$ compared with $\approx 0.29$ for TW1),
consistent with TW2 having been identified via edge tracking and therefore
residing closer to the laminar--turbulent boundary, where invariant solutions
are known to carry weaker velocity fluctuations than their turbulent
counterparts
\citep{schneider2008laminar,kerswell2018nonlinear}. The streamwise-averaged
view (figure~\ref{fig:tw2_yz_avg}) reveals a more regular array of roll cells
and well-ordered streak bands whose strongest modulation is concentrated near
the walls; the periodicity of this pattern over half-box offsets confirms the
presence of the discrete shift symmetry $\tau_{xz}$ in the converged state.
The streamwise--spanwise $(x,z)$ slice at $y=0.5$
(figure~\ref{fig:tw2_xz}) shows broad streamwise-elongated bands with
spanwise alternation, and the overlaid $(u,w)$ arrows indicate structured but
comparatively weak spanwise transport concentrated along the interfaces
between high- and low-speed bands, consistent with the crossflow motions
observed in the $(y,z)$ views.

\subsubsection{Floquet spectrum and linear stability}
\label{sec:tw2_floquet}

\begin{figure}[htbp]
  \centering
  \includegraphics[width=0.9\linewidth]{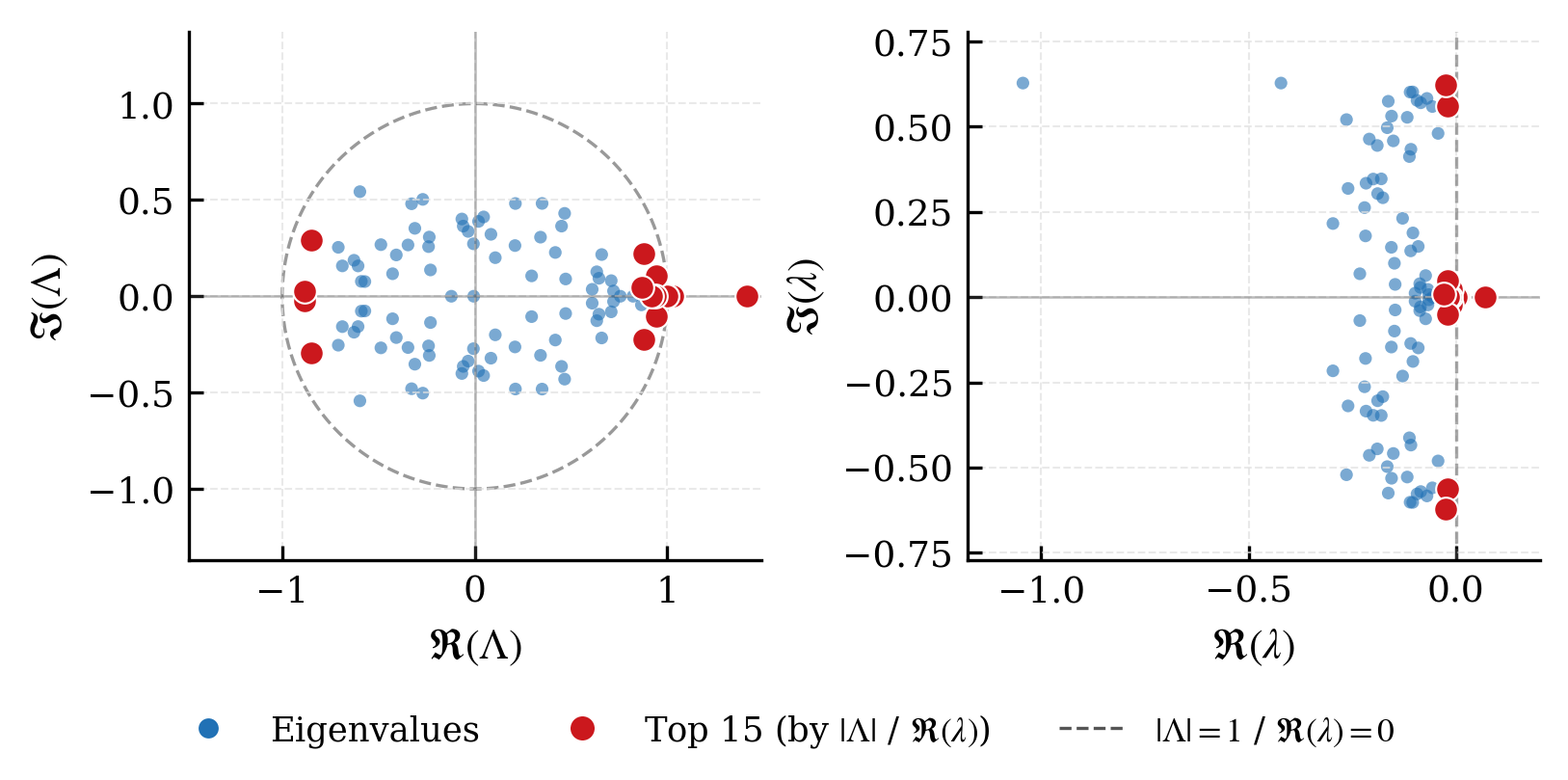}
  \caption{Floquet spectrum of TW2 at $Re=2000$. (a)~Multipliers $\Lambda$
  in the complex plane; the dashed circle marks $|\Lambda|=1$.
  (b)~Corresponding exponents $\lambda=(1/T)\log\Lambda$; the dashed vertical
  line marks $\Re(\lambda)=0$. In both panels, blue markers show all computed
  eigenvalues and red markers highlight the 15 eigenvalues with largest
  $|\Lambda|$ (left) or largest $\Re(\lambda)$ (right).}
  \label{fig:tw2_floquet}
\end{figure}

The linear stability of TW2 is characterised by its Floquet multipliers
$\Lambda$ and corresponding exponents $\lambda=(1/T)\log\Lambda$, computed
over the time horizon $T$ used in the Newton--Krylov formulation.
Figure~\ref{fig:tw2_floquet} shows the computed Floquet spectrum. In both
panels, blue markers denote the full set of computed eigenvalues, while the
red markers highlight the 15 eigenvalues of largest magnitude $|\Lambda|$
(left panel) or largest real part $\Re(\lambda)$ (right panel).

Like TW1, TW2 is unstable within its symmetry subspace, but its instability
structure is qualitatively different. Whereas TW1 possesses both oscillatory
and monotone unstable directions, the leading multipliers of TW2 are
dominated by \emph{two real instabilities}:
\[
\Lambda_1 \approx 1.414,\qquad \Lambda_2 \approx 1.030,
\]
corresponding to positive growth rates, both with
$\Im(\lambda)=0$. Thus the unstable subspace of TW2 is two-dimensional and
purely non-oscillatory: perturbations depart monotonically from the
travelling wave along these real eigendirections. The spectrum also contains
two near-neutral multipliers close to unity ($\Lambda\approx 1.000026$ and
$\Lambda\approx 0.999680$). These are expected on general grounds: a
travelling wave is a relative equilibrium, so translating it in the
streamwise direction or shifting the time origin of the Newton formulation
produces an equivalent solution. Perturbations aligned with these continuous
symmetry directions neither grow nor decay, and the corresponding multipliers
must therefore satisfy $|\Lambda|=1$; their marginal departure from exactly
unity reflects the finite numerical precision of the computation.

All remaining multipliers lie well inside the unit circle, and the
corresponding exponents have strictly negative real parts. This strong
contraction reflects the viscous damping that acts on all perturbation
components not sustained by the nonlinear self-interaction of the travelling
wave. Because TW2 resides close to the laminar--turbulent boundary (having
been identified via edge tracking), its velocity fluctuations are
comparatively weak, and only a small number of perturbation directions
receive sufficient energy from the mean shear to overcome viscous decay---a
feature that is characteristic of edge states and lower-branch solutions more
generally
\citep{wang2007lower,schneider2008laminar,kerswell2018nonlinear}. The most
prominent stable modes include a complex-conjugate pair with
$|\Lambda|\approx 0.947$ that represents a damped oscillatory deformation,
confirming that oscillatory perturbations also decay back toward the
solution.

Overall, TW2 is a saddle-type travelling wave with two real unstable
directions and strong transverse contraction. Compared with TW1, which has
both oscillatory and monotone instabilities and stronger velocity
fluctuations, TW2 has a simpler and lower-dimensional unstable manifold.
This is consistent with its proximity to the laminar--turbulent boundary,
where invariant solutions are expected to have fewer unstable directions
\citep{skufca2006edge,schneider2008laminar,kerswell2018nonlinear}.
Trajectories in the symmetry subspace can therefore approach and transiently
shadow TW2, but ultimately depart along the two preferred real unstable
directions.

\subsection{TW3 at \texorpdfstring{$Re=2000$}{Re=2000}}
\label{sec:tw3}

\subsubsection{Overview and defining parameters}

TW3 is a travelling-wave (relative equilibrium) solution at $Re=2000$.
It was computed within the $\langle \sigma_y,\,\sigma_z\rangle$-invariant subspace, and the
converged state additionally satisfies the discrete half-box shift $\tau_{xz}=\tau(L_x/2,L_z/2)$,
so that TW3 lies in the larger subgroup $\langle \sigma_y,\,\sigma_z,\,\tau_{xz}\rangle$.
In a suitably translating frame TW3 is steady, while in the laboratory frame it drifts primarily
in the streamwise direction. The converged drifts are
\[
a_x \approx 0.12,\qquad a_z = 0.
\]
Here $a_z=0$ indicates no measurable spanwise drift for this solution.
For a travelling wave, the defining relative-equilibrium condition is that the time-$T$ flow map
returns the state to itself up to a translation,
\begin{equation}
\phi^{T}(\mathbf{u}) \;=\; \tau(a_x,a_z)\,\mathbf{u},
\label{eq:tw3_invariance}
\end{equation}
with $\tau(a_x,a_z)$ the continuous translation in the $(x,z)$-periodic directions. In our case,
$a_z=0$, so the drift is purely streamwise.

\subsubsection{Structure}
\label{sec:tw3_structure}

\begin{figure}[htbp]
  \centering
  \begin{subfigure}[t]{0.51\textwidth}
    \centering
    \includegraphics[width=1.2\linewidth]{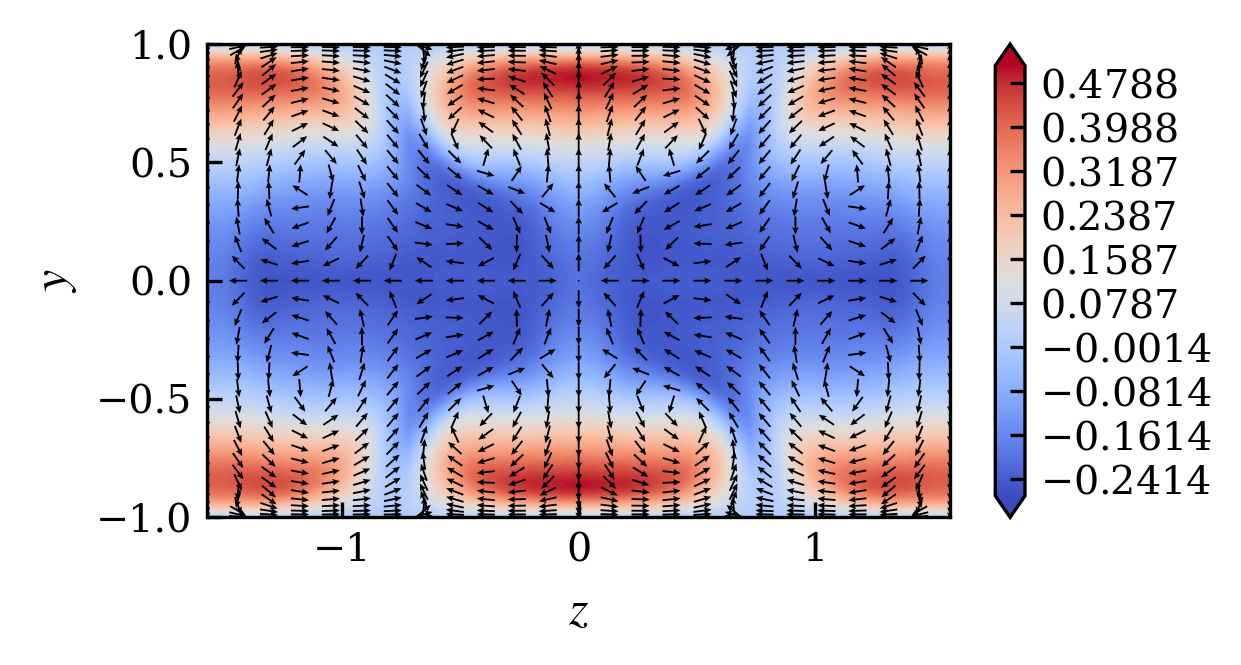}
    \caption{$(y,z)$ slice: $(v,w)$ quiver with $u$ contours.}
    \label{fig:tw3_yz}
  \end{subfigure}
  \begin{subfigure}[t]{0.51\textwidth}
    \centering
    \includegraphics[width=1.2\linewidth]{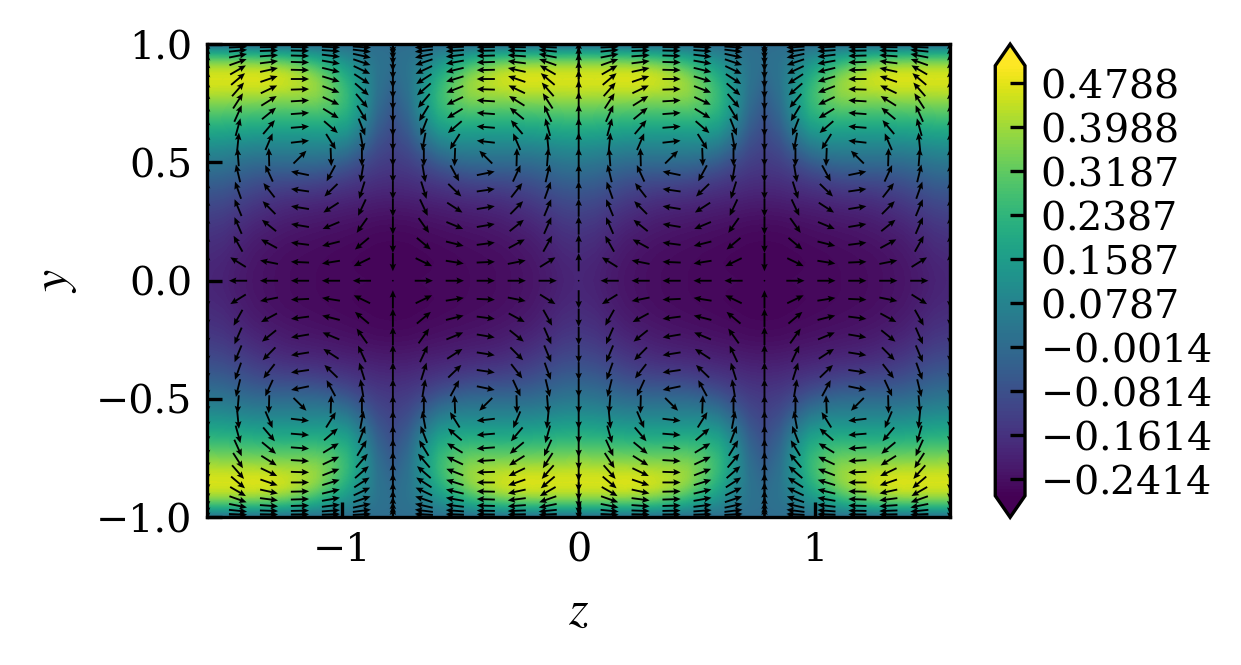}
    \caption{$(y,z)$ slice: $(v,w)$ quiver with $\langle u\rangle_x$ contours.}
    \label{fig:tw3_yz_avg}
  \end{subfigure}
  \vspace{2mm}
  \begin{subfigure}[t]{0.51\textwidth}
    \centering
    \includegraphics[width=1.2\linewidth]{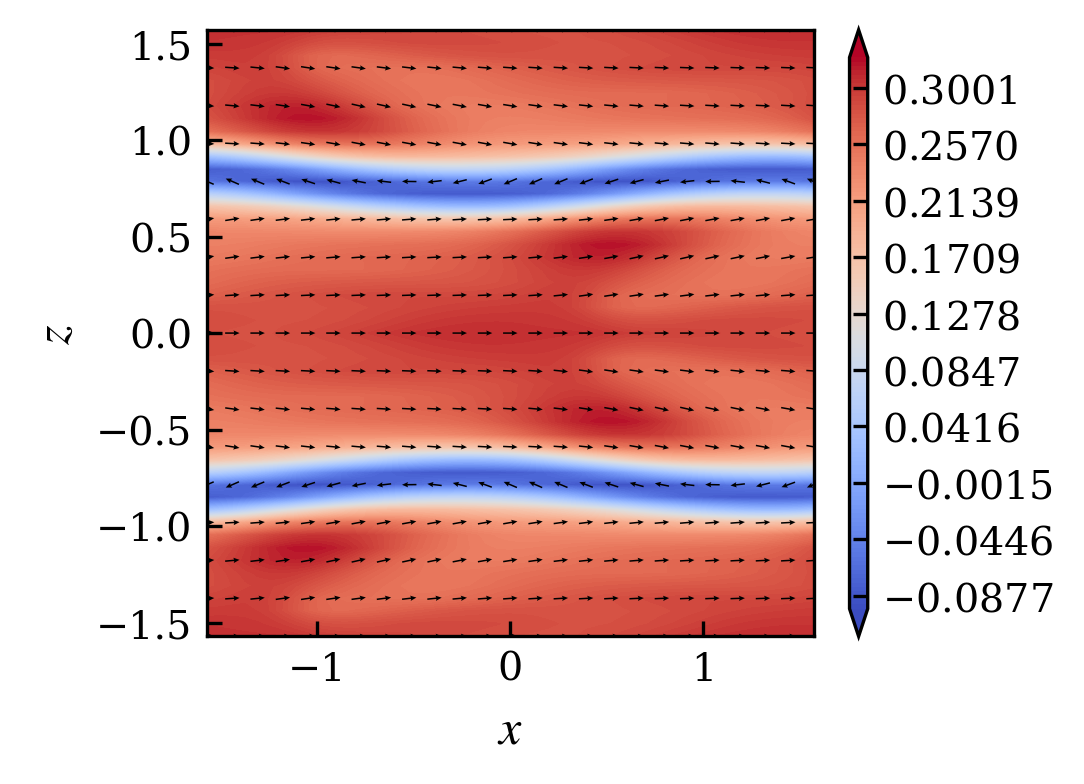}
    \caption{$(x,z)$ slice at $y=0.5$: $(u,w)$ quiver with $u$ contours.}
    \label{fig:tw3_xz}
  \end{subfigure}
  \caption{Structure of TW3 at $Re=2000$ shown via streak--roll projections.
  Background colour: streamwise velocity $u$ (or $\langle u\rangle_x$ in the
  middle panel). Arrows: in-plane velocity components in the plotted plane.}
  \label{fig:TW3_quiver}
\end{figure}

Figure~\ref{fig:TW3_quiver} summarises the spatial organisation of TW3
through three complementary two-dimensional views, with the background colour
showing the streamwise velocity $u$ (or its streamwise average
$\langle u\rangle_x$) and the arrows indicating in-plane velocity components.
The cross-plane $(y,z)$ slice (figure~\ref{fig:tw3_yz}) reveals a multi-cell
pattern of counter-rotating rolls across the span that drive alternating
upwash and downwash, reorganising the streamwise momentum into strongly
spanwise-modulated streak with peak amplitudes (${|u| \approx 0.50}$)
comparable to those of the RPOs and considerably larger than TW2; the pattern
is mirrored about both the channel centreplane ($\sigma_y$) and the spanwise
midplane ($\sigma_z$), reflecting the full set of imposed symmetries. The
streamwise-averaged view (figure~\ref{fig:tw3_yz_avg}) filters residual
streamwise waviness of TW3: the roll
cells and the near-wall streak bands remain sharply organised, confirming
that the cross-plane circulation is the primary sustaining mechanism of the
wave rather than a transient feature. The half-box periodicity in the
averaged field is consistent with the additional $\tau_{xz}$ symmetry of the
converged state. The streamwise--spanwise $(x,z)$ slice at $y=0.5$
(figure~\ref{fig:tw3_xz}) shows streamwise-elongated bands whose dominant
variation is spanwise, with a pronounced low-speed streak running the full
streamwise extent near $z\approx 0.8$; the overlaid $(u,w)$ arrows indicate
structured spanwise transport aligned with the streak gradients, consistent
with the crossflow motions observed in the $(y,z)$ views.

\subsubsection{Floquet spectrum and linear stability}
\label{sec:tw3_floquet}

The linear stability of TW3 is characterised by its Floquet multipliers
$\Lambda$ and corresponding exponents $\lambda=(1/T)\log\Lambda$, computed
over the time horizon $T$ used in the Newton--Krylov formulation.
Figure~\ref{fig:tw3_floquet} shows the computed Floquet spectrum. In both
panels, blue markers denote the full set of computed eigenvalues, while the
red markers highlight the 15 eigenvalues of largest magnitude $|\Lambda|$
(left panel) or largest real part $\Re(\lambda)$ (right panel).

As with TW1 and TW2, TW3 is unstable within its symmetry subspace, but its
instability character differs from both. The left panel shows that several
leading multipliers lie outside the unit circle, and these are arranged as
complex-conjugate pairs with $|\Lambda|>1$ at nonzero $\Im(\Lambda)$. In the
exponent plane (right panel), the corresponding modes appear with small
positive $\Re(\lambda)$ and nonzero $\Im(\lambda)$, confirming that the
dominant instabilities are \emph{oscillatory}: perturbations aligned with
these modes grow while oscillating, causing the trajectory to spiral away
from TW3. This contrasts with TW1, whose instabilities include both oscillatory and monotone (real) modes with comparatively larger growth rates, and with TW2, whose two unstable directions are purely monotone. The spectrum also contains near-neutral multipliers close to unity on the real axis, corresponding to the expected neutral directions associated with continuous symmetries (streamwise translation and the phase-fixing freedom).

All remaining multipliers lie inside the unit circle, and the corresponding
exponents have strictly negative real parts. As with TW2, this strong
contraction along the majority of resolved directions reflects the viscous
damping of perturbation components that are not sustained by the nonlinear
self-interaction of the travelling wave. The combination of a small number of
weakly unstable oscillatory modes and strong transverse contraction gives TW3
the character of a saddle-type invariant solution: nearby trajectories in the
symmetry subspace $\langle \sigma_y,\,\sigma_z,\,\tau_{xz}\rangle$ can
approach and transiently shadow TW3, but ultimately depart along the
low-dimensional unstable manifold via slowly growing oscillatory modulations.

\begin{figure}[htbp]
  \centering
  \includegraphics[width=0.9\linewidth]{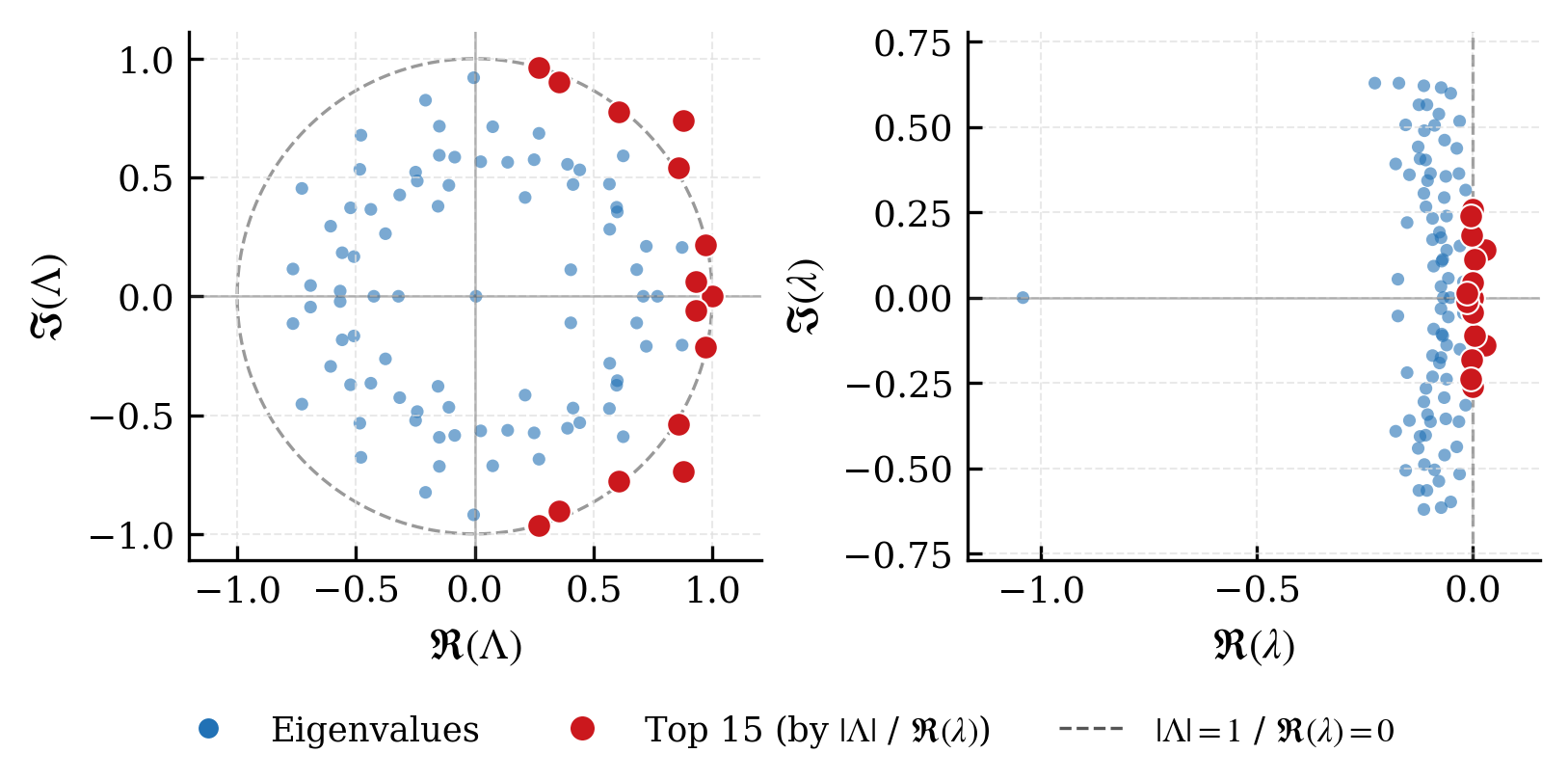}
  \caption{Floquet spectrum of TW3 at $Re=2000$. (a)~Multipliers $\Lambda$
  in the complex plane; the dashed circle marks $|\Lambda|=1$.
  (b)~Corresponding exponents $\lambda=(1/T)\log\Lambda$; the dashed vertical
  line marks $\Re(\lambda)=0$. In both panels, blue markers show all computed
  eigenvalues and red markers highlight the 15 eigenvalues with largest
  $|\Lambda|$ (left) or largest $\Re(\lambda)$ (right).}
  \label{fig:tw3_floquet}
\end{figure}

\section{Continuation in Reynolds number and spanwise domain length}
\label{sec:continuation}

We now examine how each of the five exact coherent structures (RPO1, RPO2,
TW1--TW3) extends into one-parameter families under two complementary
continuation procedures: (i) varying the Reynolds number $Re$ at fixed domain
size and symmetry constraints, and (ii) varying the spanwise period $L_z$ at
fixed $Re$ (and fixed $L_x$), again within the same symmetry-invariant
subspace. In both cases, a converged solution at a reference parameter value
is used as an initial guess for a Newton--Krylov continuation step, generating
a branch of nearby invariant solutions.

Along each continuation curve we track the dissipation $D$, the temporal
period $T$ (for RPOs), and the streamwise drift $a_x$. These quantities are
presented as bifurcation diagrams that reveal the creation, evolution and
possible annihilation of each solution family through saddle-node folds. We
also inspect representative flow visualisations at selected points along each
branch to identify qualitative structural changes such as modifications to
the roll--streak organisation or changes in spanwise wavelength as the
parameters $(Re, L_z)$ are varied. In addition, we compute the Floquet
spectrum at selected points along the continuation branches (fold/turning
points and opposite-branch states) to determine how the linear stability
evolves with the control parameter and, in particular, whether the number or
character of unstable directions changes across the folds.

The following subsections treat each ECS in turn. For each, we present the
$Re$-continuation and $L_z$-continuation diagrams and summarise the
associated evolution of spatial structure and linear stability along the
computed branches.

\subsection{RPO1: continuation in $Re$ and $L_z$}
\label{sec:rpo1_continuation}

We continue RPO1 in two independent parameters: the Reynolds number $Re$ at
fixed domain size, and the spanwise period $L_z$ at fixed $Re$ and $L_x$.
Along each branch we track the dissipation $D$, the temporal period $T$ and
the relative streamwise shift $a_x$.

The $Re$-continuation (figure~\ref{fig:rpo1_bif}\emph{a--c}) yields a
non-monotone family in dissipation with a clear saddle-node fold: the branch
turns back at a lower-$Re$ turning point near $Re \approx 986$, forming a
closed loop in the $(Re, D)$ plane, and develops small secondary loops at the
high-$Re$ end near $Re \approx 1010$--$1015$. Away from these turning
regions, the diagnostics vary smoothly: $T$ increases approximately linearly
with $Re$ (from $\approx 40.8$ to $\approx 41.3$), and $a_x$ likewise
increases near-linearly toward zero.

The $L_z$-continuation (figure~\ref{fig:rpo1_bif}\emph{d--f}) also exhibits
fold-type behaviour: the dissipation forms a prominent loop in the $(L_z, D)$
plane with turning points near $L_z \approx 1.507$ and $L_z \approx 1.508$,
followed by a broader fold near $L_z \approx 1.55$. As in the $Re$ case, $T$
increases monotonically over the computed range (from $\approx 37$ to
$\approx 41$) and $a_x$ varies smoothly with $L_z$.

\begin{figure}[htbp]
  \centering
  \begin{subfigure}[t]{\textwidth}
    \centering
    \includegraphics[width=\linewidth]{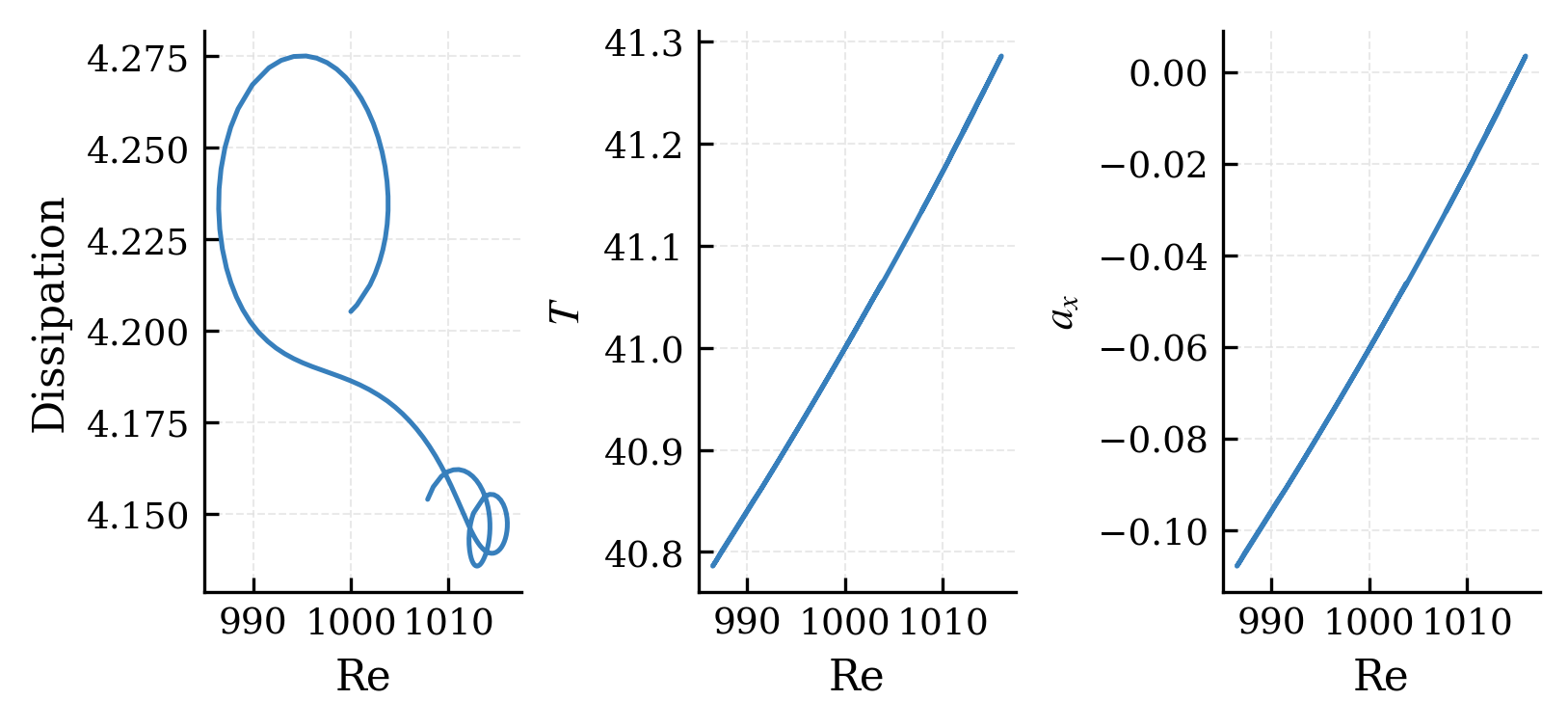}
    \caption{$Re$-continuation: (a)~$D$ vs $Re$, (b)~$T$ vs $Re$,
    (c)~$a_x$ vs $Re$.}
    \label{fig:rpo1_bif_Re}
  \end{subfigure}
  \vspace{2mm}
  \begin{subfigure}[t]{\textwidth}
    \centering
    \includegraphics[width=\linewidth]{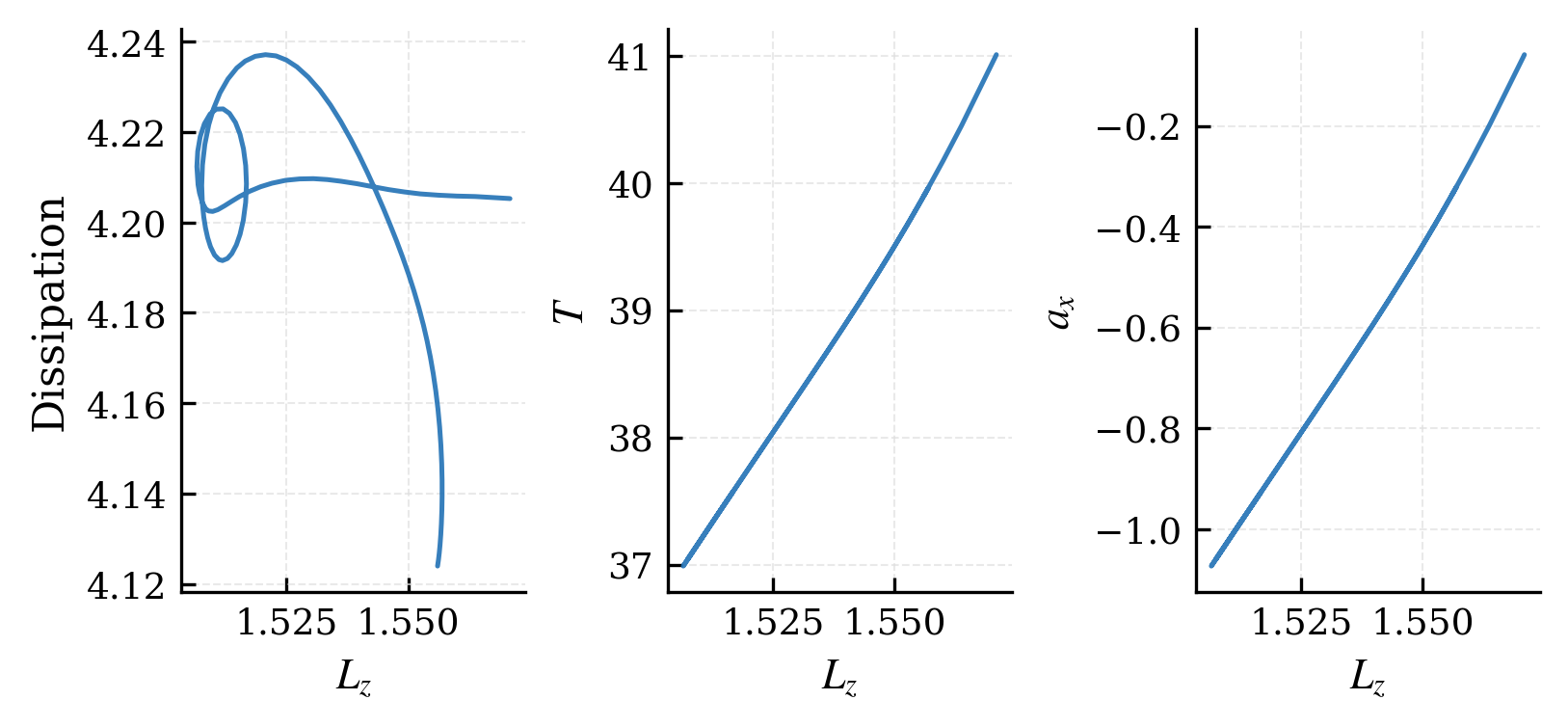}
    \caption{$L_z$-continuation: (d)~$D$ vs $L_z$, (e)~$T$ vs $L_z$,
    (f)~$a_x$ vs $L_z$.}
    \label{fig:rpo1_bif_Lz}
  \end{subfigure}
  \caption{Continuation and bifurcation structure of RPO1.
  Top row: $Re$-continuation at fixed domain size.
  Bottom row: $L_z$-continuation at fixed $Re$ and $L_x$.}
  \label{fig:rpo1_bif}
\end{figure}

To assess whether the folds correspond to qualitative changes in the flow, we
compare representative solutions at the reference parameter value, at the
fold/turning point and on the opposite branch. For the $Re$-continuation, the
streak--roll organisation remains topologically unchanged in that the number and arrangement of roll cells, the alternating
pattern of high- and low-speed streaks, and the symmetry of the flow field are
all preserved, with only minor amplitude variations (peak $u$ changes by less
than $4\%$). The same conclusion holds in the $(y,z)$ cross-plane views and
for the $L_z$-continuation, where the structural changes are similarly limited
to gentle intensity variations consistent with a smooth geometric rescaling in
$z$. All structural comparisons for both continuations are provided in the
supplementary material (figures~S2--S5).

Floquet spectra computed at the fold and opposite-branch states confirm that
RPO1 remains linearly stable throughout both continuations. Along the
$Re$-continuation, the spectra at the fold ($Re \approx 986$) and on the
opposite branch are nearly indistinguishable from the reference spectrum
(figure~\ref{fig:rpo1_floquet}): all non-neutral multipliers remain well
inside the unit circle with the largest at $|\Lambda| \approx 0.5$, and no
multiplier approaches or crosses $|\Lambda| = 1$. Along the
$L_z$-continuation, the leading eigenvalue structure reorganises modestly
near the fold: a complex-conjugate pair becomes the most prominent
non-neutral mode (reaching $|\Lambda| \approx 0.63$ at the $L_z$ fold), but
it remains inside the unit circle and all Floquet exponents retain strictly
negative real parts. Thus, the folds in both continuations are purely
geometric features of the solution branch---they do not coincide with any
change in stability character, and RPO1 persists as a linearly stable
relative periodic orbit across the entire computed parameter range. The
Floquet spectra along both continuation branches are provided in the
supplementary material (figure~S6).

\subsection{RPO2: continuation in $Re$ and $L_z$}
\label{sec:rpo2_continuation}

We continue RPO2 in two independent parameters: the Reynolds number $Re$ at
fixed domain size, and the spanwise period $L_z$ at fixed $Re$ and $L_x$.
Along each branch we track the dissipation $D$, the temporal period $T$ and
the relative streamwise shift $a_x$.

The $Re$-continuation (figure~\ref{fig:rpo2_bif}\emph{a--c}) yields a
non-monotone family in dissipation with a saddle-node fold near
$Re \approx 1499$: the branch turns back at this lower-$Re$ turning point
and reaches a broad dissipation maximum near $Re \approx 1502$ before
decreasing again toward higher $Re$. The period $T$ decreases near-linearly
with $Re$ (from $\approx 152.9$ to $\approx 152.7$), and $a_x$ likewise
decreases approximately linearly, becoming more negative with increasing
$Re$.

The $L_z$-continuation (figure~\ref{fig:rpo2_bif}\emph{d--f}) shows more
pronounced fold-type behaviour: the dissipation forms a closed loop in the
$(L_z, D)$ plane with turning points near $L_z \approx 1.550$ and
$L_z \approx 1.570$. Along this branch, $a_x$ varies strongly with $L_z$
(from $\approx 0.4$ down to $\approx -0.1$), while $T$ decreases
non-monotonically with a minimum near $L_z \approx 1.565$. Overall,
varying $L_z$ has a more pronounced effect on the continuation geometry and
orbit diagnostics than varying $Re$ over the explored intervals.

\begin{figure}[htbp]
  \centering
  \begin{subfigure}[t]{\textwidth}
    \centering
    \includegraphics[width=\linewidth]{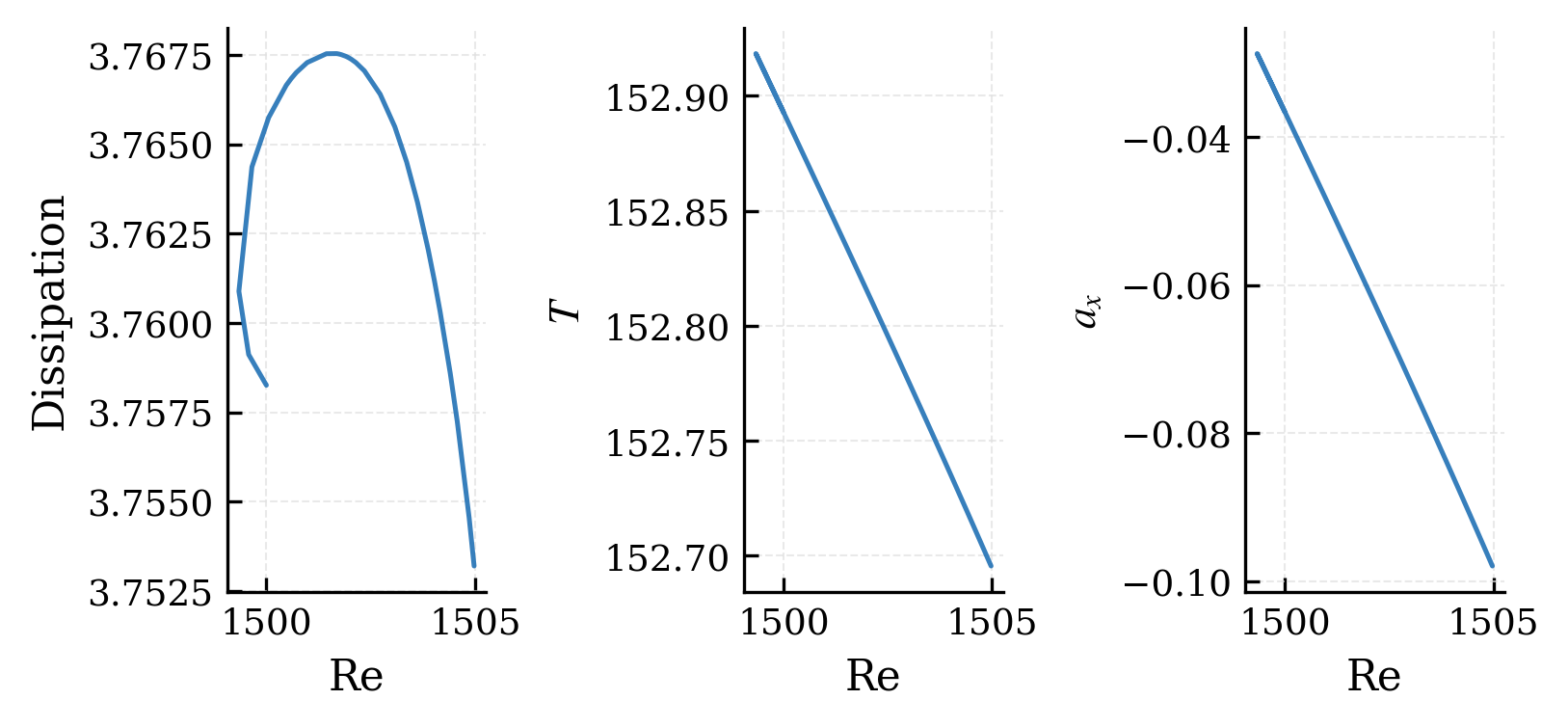}
    \caption{$Re$-continuation: (a)~$D$ vs $Re$, (b)~$T$ vs $Re$,
    (c)~$a_x$ vs $Re$.}
    \label{fig:rpo2_bif_Re}
  \end{subfigure}
  \vspace{2mm}
  \begin{subfigure}[t]{\textwidth}
    \centering
    \includegraphics[width=\linewidth]{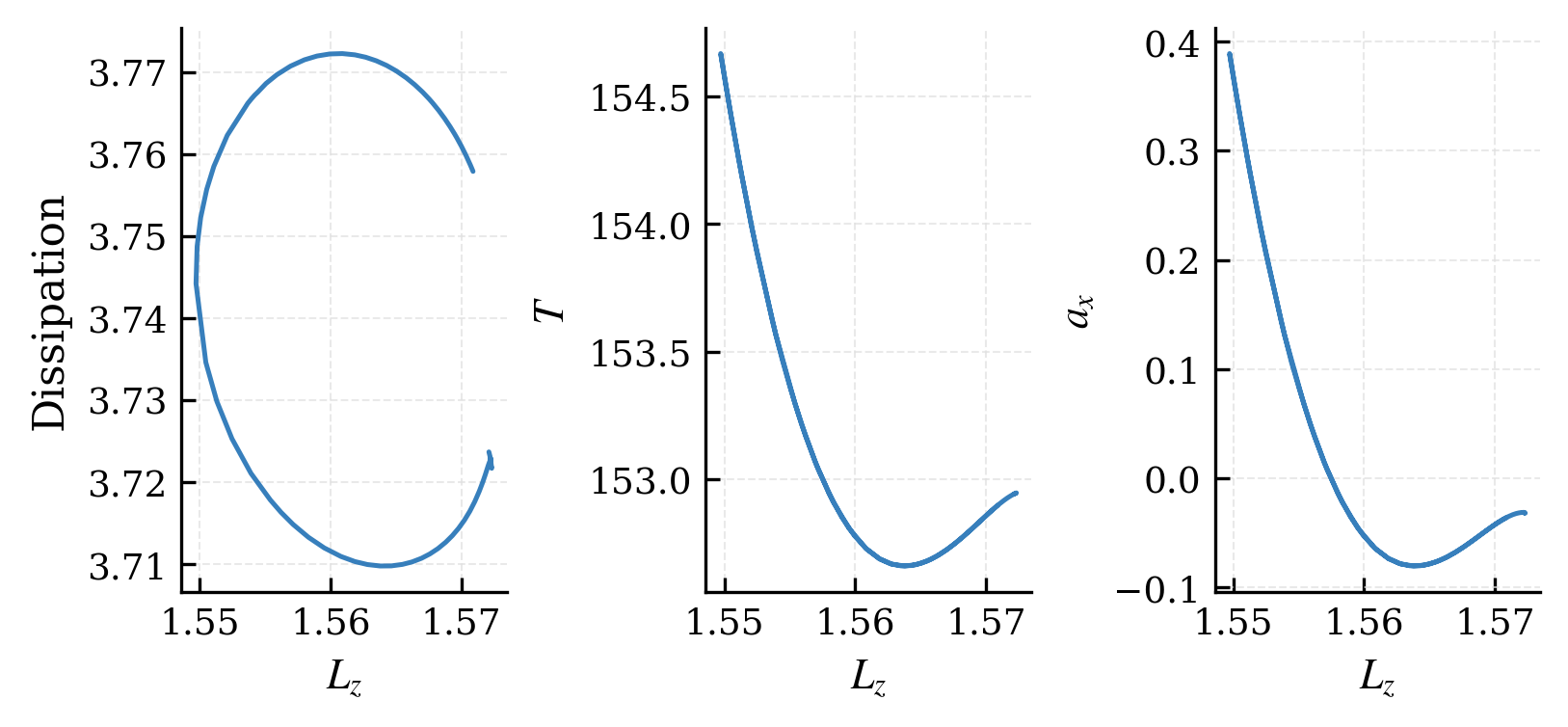}
    \caption{$L_z$-continuation: (d)~$D$ vs $L_z$, (e)~$T$ vs $L_z$,
    (f)~$a_x$ vs $L_z$.}
    \label{fig:rpo2_bif_Lz}
  \end{subfigure}
  \caption{Continuation and bifurcation structure of RPO2.
  Top row: $Re$-continuation at fixed domain size.
  Bottom row: $L_z$-continuation at fixed $Re$ and $L_x$.}
  \label{fig:rpo2_bif}
\end{figure}

To assess whether the folds correspond to qualitative changes in the flow, we
compare representative solutions at the reference parameter value, at the
fold/turning point and on the opposite branch. For the $Re$-continuation, the
streak--roll organisation remains unchanged: the number and
arrangement of streak bands, the alternating high- and low-speed pattern, and
the symmetry of the flow field are all preserved, with only minor amplitude
variations. The same conclusion holds in the $(y,z)$ cross-plane views and
for the $L_z$-continuation, where the structural changes are similarly limited
to gentle intensity variations. All structural comparisons for both
continuations are provided in the supplementary material (figures~S7--S10).

As with RPO1, Floquet spectra computed at the fold and opposite-branch states
confirm that RPO2 remains linearly stable throughout both continuations. Along
the $Re$-continuation, the spectra at the fold ($Re \approx 1499$) and on the
upper branch are virtually indistinguishable from the reference spectrum
(figure~\ref{fig:rpo2_floquet}): the leading non-neutral multiplier remains a
real eigenvalue at $|\Lambda| \approx 0.85$, well inside the unit circle, and
all Floquet exponents retain strictly negative real parts with no appreciable
shift toward the stability boundary. Along the $L_z$-continuation, the
spectrum undergoes a more noticeable reorganisation: the leading non-neutral
multiplier contracts from $|\Lambda| \approx 0.85$ at the reference to
$\approx 0.5$ at the fold and $\approx 0.65$ on the opposite branch,
indicating that the orbit becomes more strongly attracting as $L_z$ is
varied, consistent with the more pronounced effect of $L_z$ on the
continuation geometry noted above. Despite this reorganisation, no multiplier
approaches or crosses the unit circle, and RPO2 remains linearly stable
across the entire computed parameter range. The Floquet spectra along both
continuation branches are provided in the supplementary material
(figure~S11).

\subsection{TW1: continuation in $Re$ and $L_z$}
\label{sec:tw1_continuation}

We continue TW1 in two independent parameters: the Reynolds number $Re$ at
fixed domain size, and the spanwise period $L_z$ at fixed $Re$ and $L_x$.
Along each branch we track the dissipation $D$ and the streamwise drift
$a_x$.

The $Re$-continuation (figure~\ref{fig:tw1_bif}\emph{a,\,b}) exhibits a
clear saddle-node fold: the branch turns back near $Re \approx 1400$,
creating an upper branch that extends to $Re \approx 2600$ at substantially
higher dissipation ($D \approx 3.0$) and a lower branch at smaller
dissipation ($D \approx 2.0$) that terminates within the plotted range. The
reference TW1 at $Re = 1900$ lies on the lower branch. The streamwise drift
$a_x$ remains near $-0.010$ on the upper branch and decreases to
$\approx -0.035$ at higher $Re$, indicating a progressively faster
downstream translation as $Re$ increases.

The $L_z$-continuation (figure~\ref{fig:tw1_bif}\emph{c,\,d}) is likewise
non-monotone, with a pronounced fold at the small-$L_z$ end near
$L_z \approx 2.1$. The upper branch rises steeply to $D \approx 2.5$ over a
narrow $L_z$ interval, while the lower branch varies only weakly with $L_z$
(a shallow rise from $D \approx 1.9$ to $\approx 2.0$). The drift $a_x$
peaks near $L_z \approx 2.15$ and then decreases monotonically, becoming
more negative with increasing $L_z$.

\begin{figure}[htbp]
  \centering
  \begin{subfigure}[t]{\textwidth}
    \centering
    \includegraphics[width=0.8\linewidth]{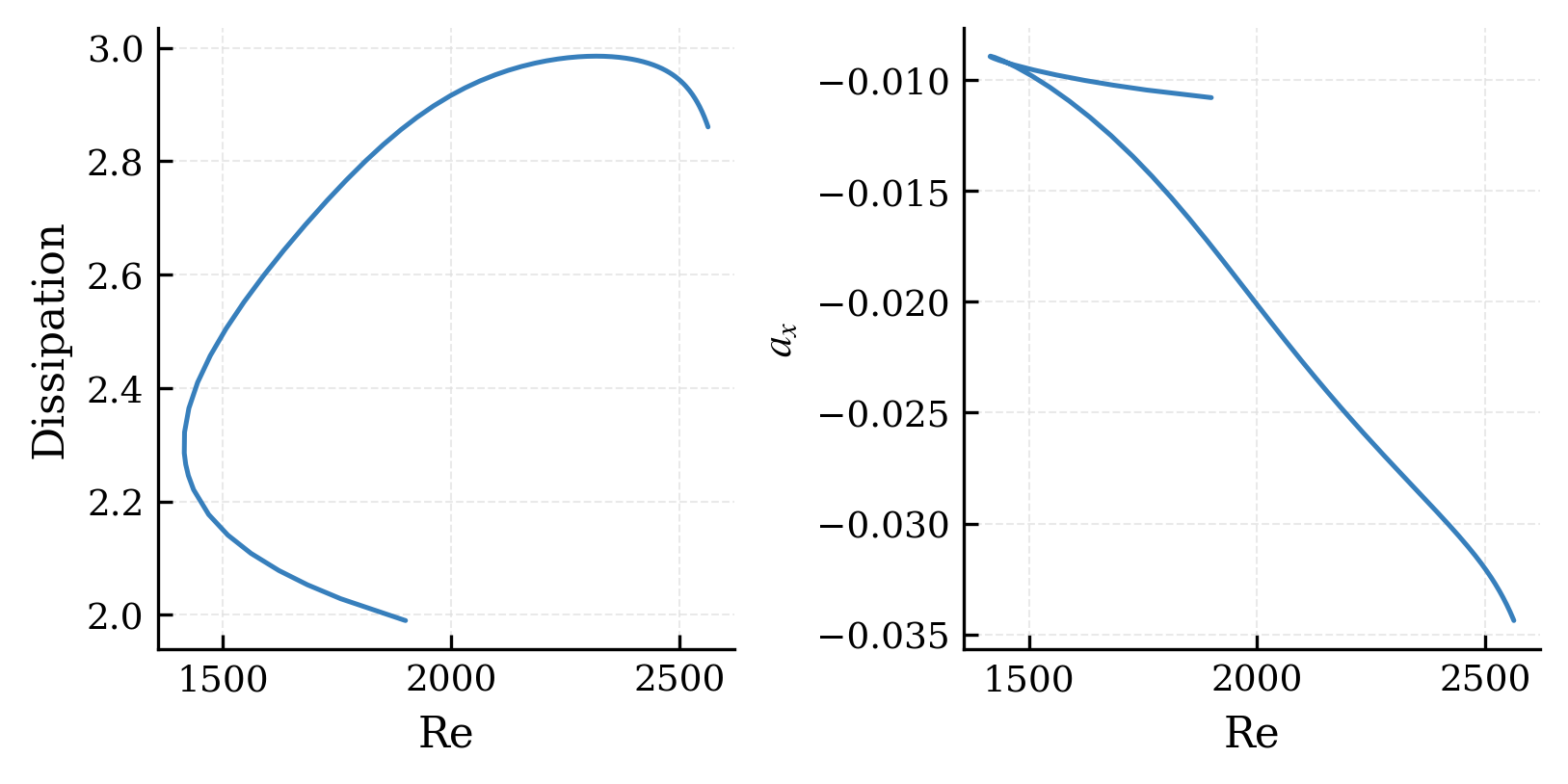}
    \caption{$Re$-continuation: (a)~$D$ vs $Re$, (b)~$a_x$ vs $Re$.}
    \label{fig:tw1_bif_Re}
  \end{subfigure}
  \vspace{2mm}
  \begin{subfigure}[t]{\textwidth}
    \centering
    \includegraphics[width=0.8\linewidth]{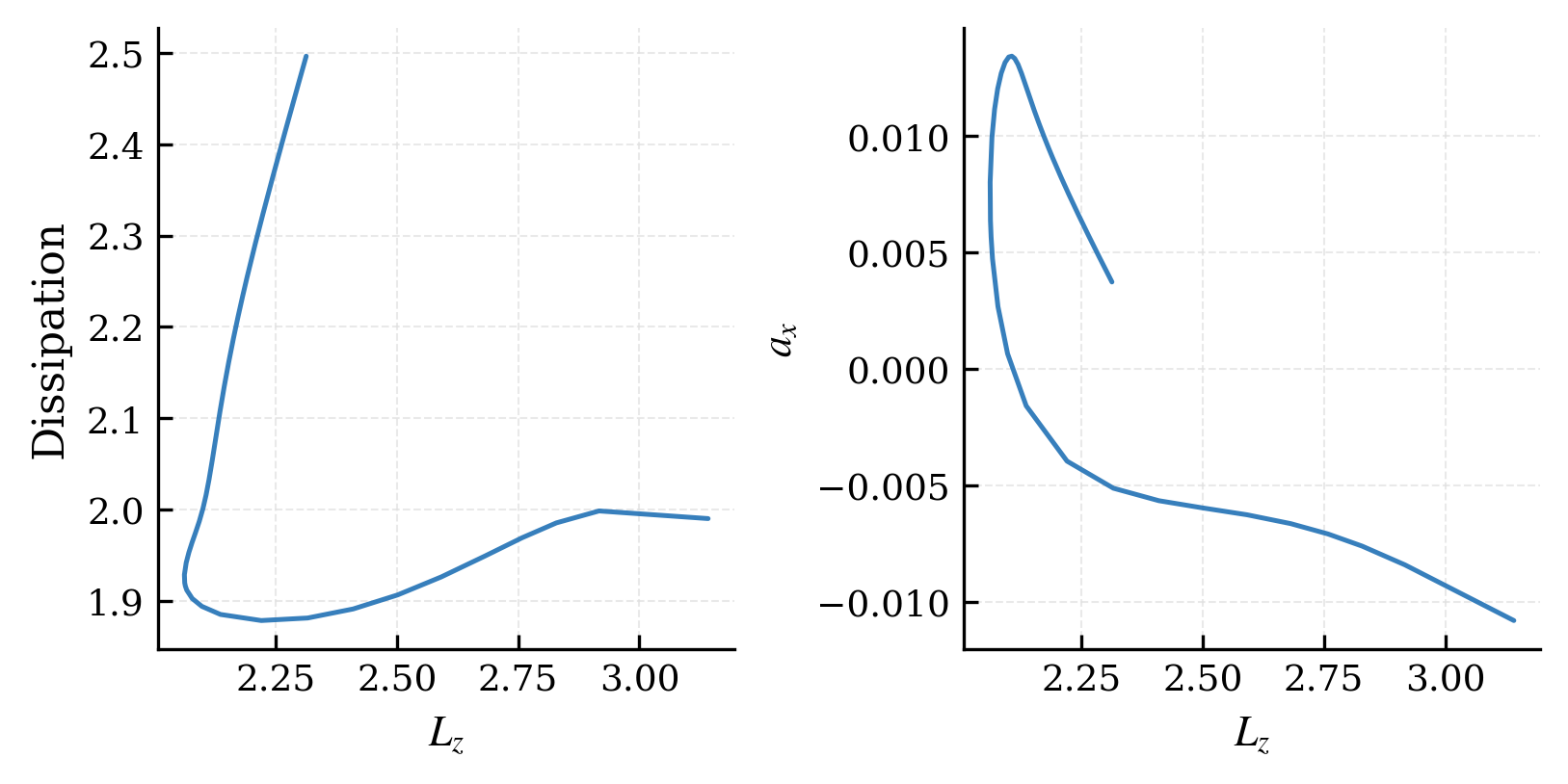}
    \caption{$L_z$-continuation: (c)~$D$ vs $L_z$, (d)~$a_x$ vs $L_z$.}
    \label{fig:tw1_bif_Lz}
  \end{subfigure}
  \caption{Continuation and bifurcation structure of TW1.
  Top row: $Re$-continuation at fixed domain size.
  Bottom row: $L_z$-continuation at fixed $Re$ and $L_x$.}
  \label{fig:tw1_bif}
\end{figure}

To assess whether the folds correspond to qualitative changes in the flow, we
compare representative solutions at the reference parameter value, at the
fold/turning point and on the opposite branch.
Figure~\ref{fig:tw1_Re_struct_yz} shows this comparison for the
$Re$-continuation in the $(y,z)$ cross-plane: the counter-rotating roll-cell
pattern and the alternating high- and low-speed streak arrangement are
preserved across all three states---that is, the number and arrangement of
roll cells, the streak pattern and the symmetry of the flow field are all
retained. However, the amplitude increases markedly along the branch, from peak
$|u| \approx 0.29$ at the reference state to $\approx 0.35$ at the fold and
$\approx 0.39$ on the upper branch, indicating that the upper branch
corresponds to a more energetic version of the same roll--streak structure.
For the $L_z$-continuation, the same topological preservation is observed:
the fold solution at $L_z \approx 2.1$ retains essentially the same
amplitude as the reference (peak $|u| \approx 0.28$ versus $0.29$), while
the upper-branch solution shows a substantial increase to peak
$|u| \approx 0.38$, with the roll cells adapting to the narrower spanwise
domain. In neither continuation does the number of roll cells or the basic
streak arrangement change. The $(x,z)$ midplane views and the
$L_z$-continuation structural comparisons confirm these conclusions and are
provided in the supplementary material (figures~S12--S14).

\begin{figure}[htbp]
  \centering
  \includegraphics[width=0.65\linewidth]{Quiver_TW1_YZ.png}
  \includegraphics[width=0.65\linewidth]{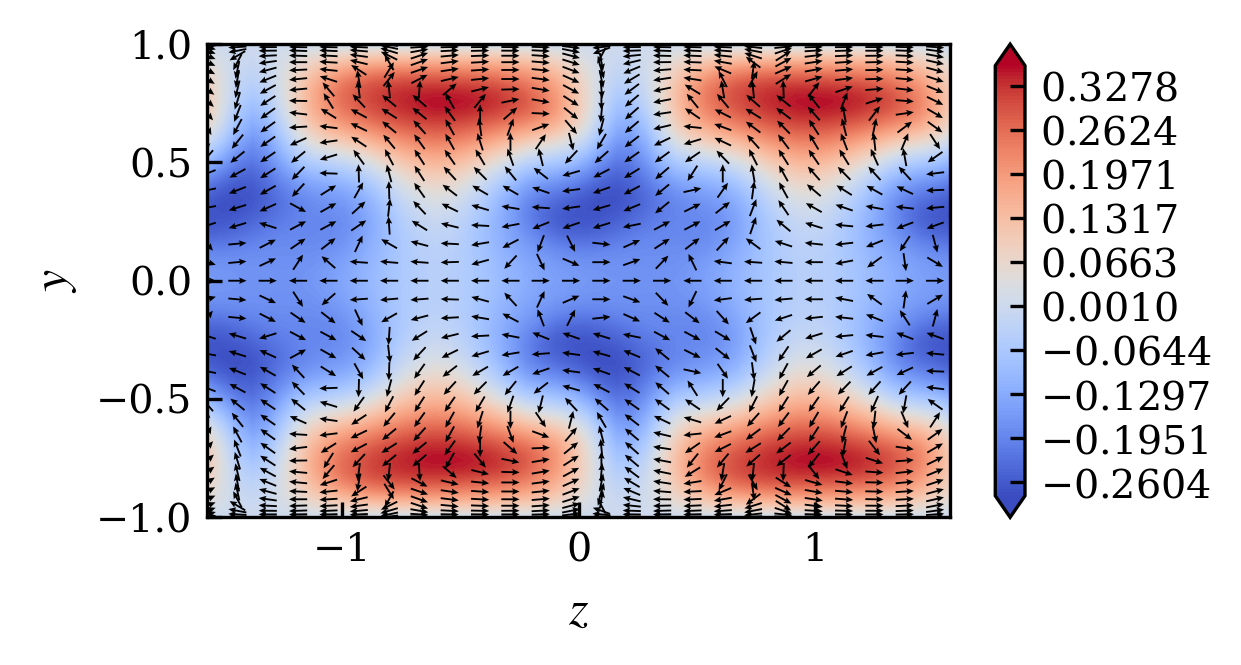}
\includegraphics[width=0.65\linewidth]{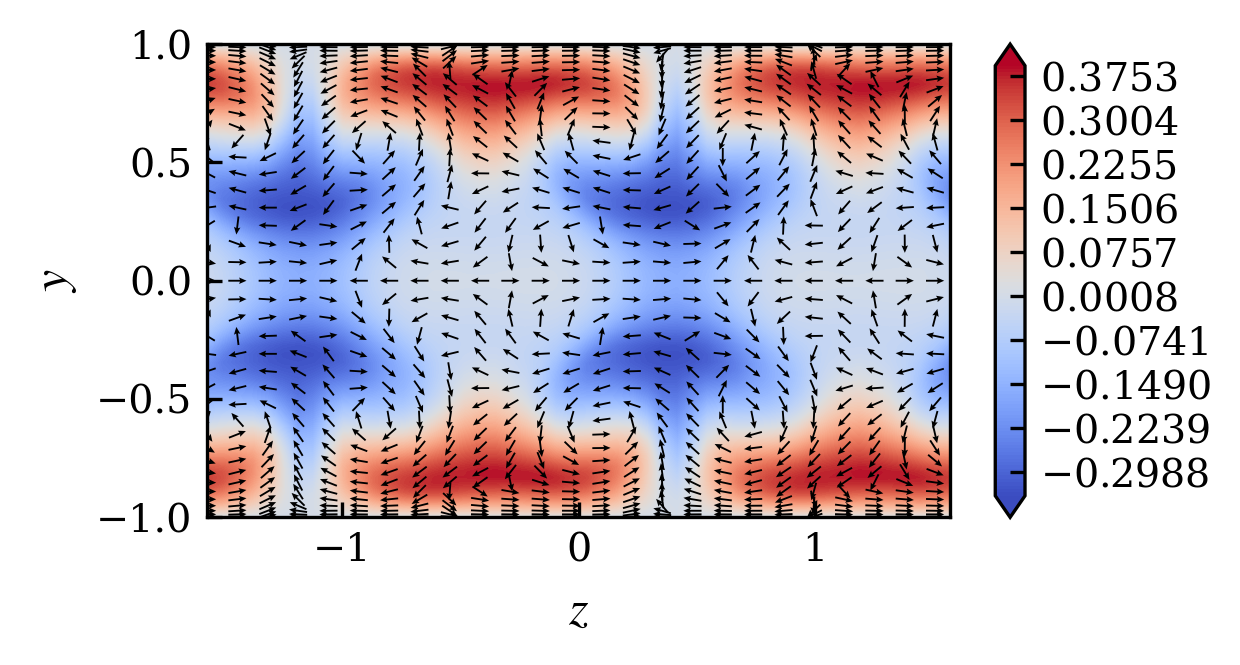}
  \caption{$Re$-continuation structural comparison for TW1 in the $(y,z)$
  plane: (left) reference TW1 at $Re=1900$ (top), (middle)
  fold/turning-point near $Re \approx 1400$, (bottom) upper-branch solution.
  Background colour: streamwise velocity $u$; arrows: in-plane $(v,w)$
  components. The roll--streak topology is preserved, with amplitude
  increasing from left to right.}
  \label{fig:tw1_Re_struct_yz}
\end{figure}

Unlike the two RPOs, whose stability is unaffected by continuation, the
Floquet spectrum of TW1 evolves substantially along the $Re$-continuation.
At the fold ($Re \approx 1400$), the instability weakens relative to the
reference: the leading multiplier magnitude decreases from
$|\Lambda| \approx 1.22$ to $\approx 1.09$, and fewer multipliers lie
outside the unit circle, indicating that the unstable manifold contracts as
the fold is approached. On the upper branch, this trend reverses sharply:
the leading multiplier grows to $|\Lambda| \approx 1.45$, additional
complex-conjugate pairs move outside the unit circle, and the unstable
subspace becomes higher-dimensional with stronger growth rates. Thus, the
upper branch carries a stronger instability than the reference state,
consistent with its higher dissipation and stronger streak amplitudes, and
the fold marks a local minimum in the degree of instability along the
$Re$-continuation. Along the $L_z$-continuation, the evolution is more
moderate: the fold solution is only weakly unstable (leading
$|\Lambda| \approx 1.03$), while the upper-branch state recovers a
stronger instability ($|\Lambda| \approx 1.1$) comparable to but slightly
weaker than the reference. Throughout both continuations, the instability
retains the mixed oscillatory-and-monotone character identified at the
reference state (section~\ref{sec:tw1_floquet}), with no qualitative change
in the type of unstable modes. The Floquet spectra along both continuation
branches are provided in the supplementary material (figure~S15).

\subsection{TW2: continuation in $Re$ and $L_z$}
\label{sec:tw2_continuation}

We continue TW2 in two independent parameters: the Reynolds number $Re$ at
fixed domain size, and the spanwise period $L_z$ at fixed $Re$ and $L_x$.
Along each branch we track the dissipation $D$ and the streamwise drift
$a_x$.

The $Re$-continuation (figure~\ref{fig:tw2_bif}\emph{a,\,b}) exhibits a
clear saddle-node fold near $Re \approx 700$: the branch turns back at this
lower-$Re$ turning point, creating an upper branch that extends to higher
$Re$ at substantially larger dissipation ($D \approx 2.6$ near the fold)
and a lower branch at smaller dissipation that decreases gradually toward
$D \approx 1.54$ at $Re = 2000$. The reference TW2 at $Re = 2000$ lies on
the lower branch. The drift $a_x$ displays the same fold signature: it is
multi-valued near the turning region but varies smoothly and
near-monotonically with $Re$ away from the fold, increasing from negative
values on the lower branch to $a_x \approx 0.076$ at the reference state.

The $L_z$-continuation (figure~\ref{fig:tw2_bif}\emph{c,\,d}) is likewise
non-monotone, with a compact fold region at the smallest $L_z$ values near
$L_z \approx 1.05$. Beyond the fold, the branches separate: one varies
only weakly with $L_z$ (a broad minimum near $L_z \approx 1.4$), while the
other remains at higher dissipation over a shorter $L_z$ interval. The
drift $a_x$ mirrors this structure, with a much stronger variation on the
high-dissipation branch ($a_x$ increasing from $\approx -0.06$ to
$\approx 0.30$).

\begin{figure}[htbp]
  \centering
  \begin{subfigure}[t]{\textwidth}
    \centering
    \includegraphics[width=0.8\linewidth]{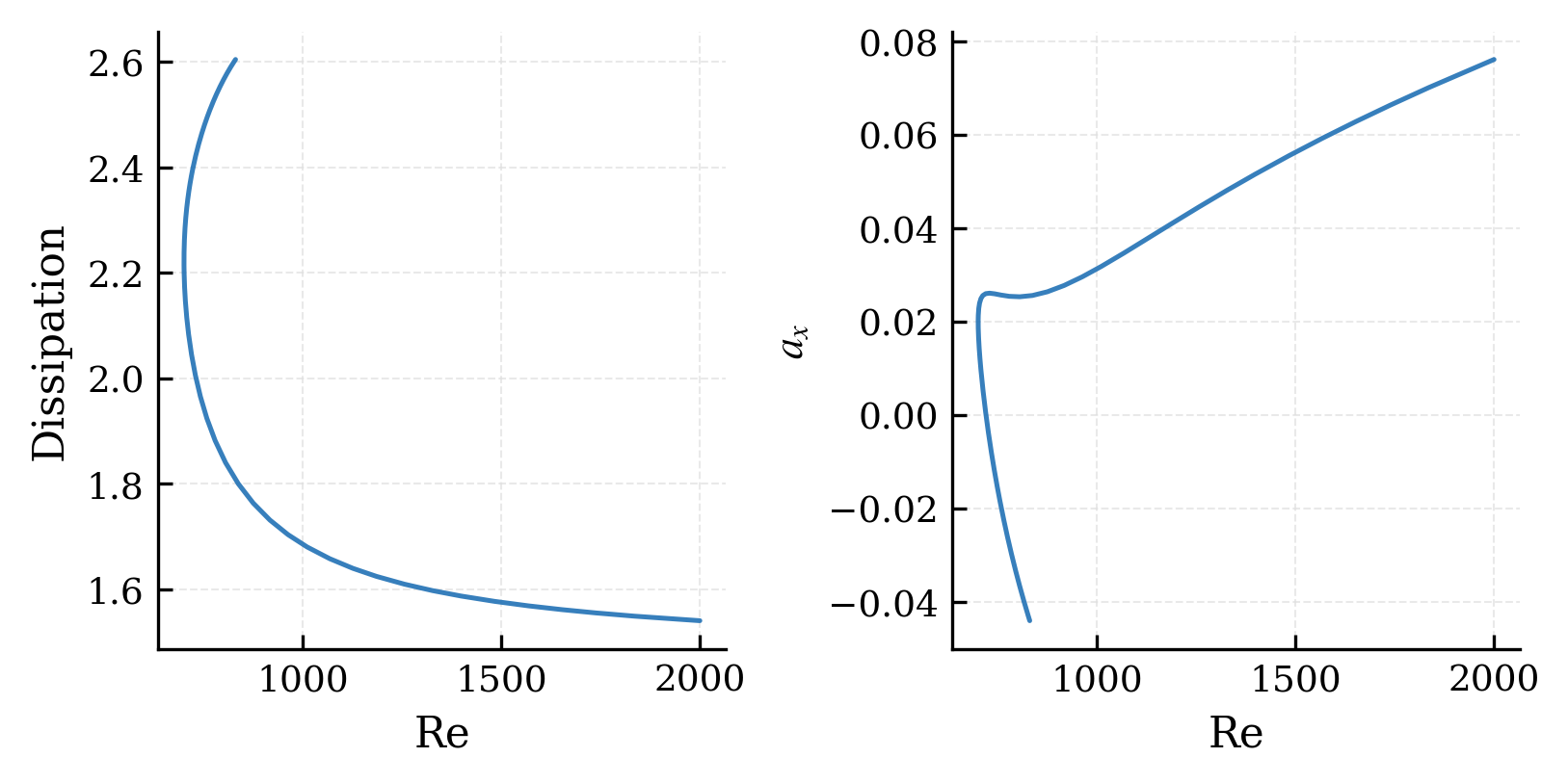}
    \caption{$Re$-continuation: (a)~$D$ vs $Re$, (b)~$a_x$ vs $Re$.}
    \label{fig:tw2_bif_Re}
  \end{subfigure}
  \vspace{2mm}
  \begin{subfigure}[t]{\textwidth}
    \centering
    \includegraphics[width=0.8\linewidth]{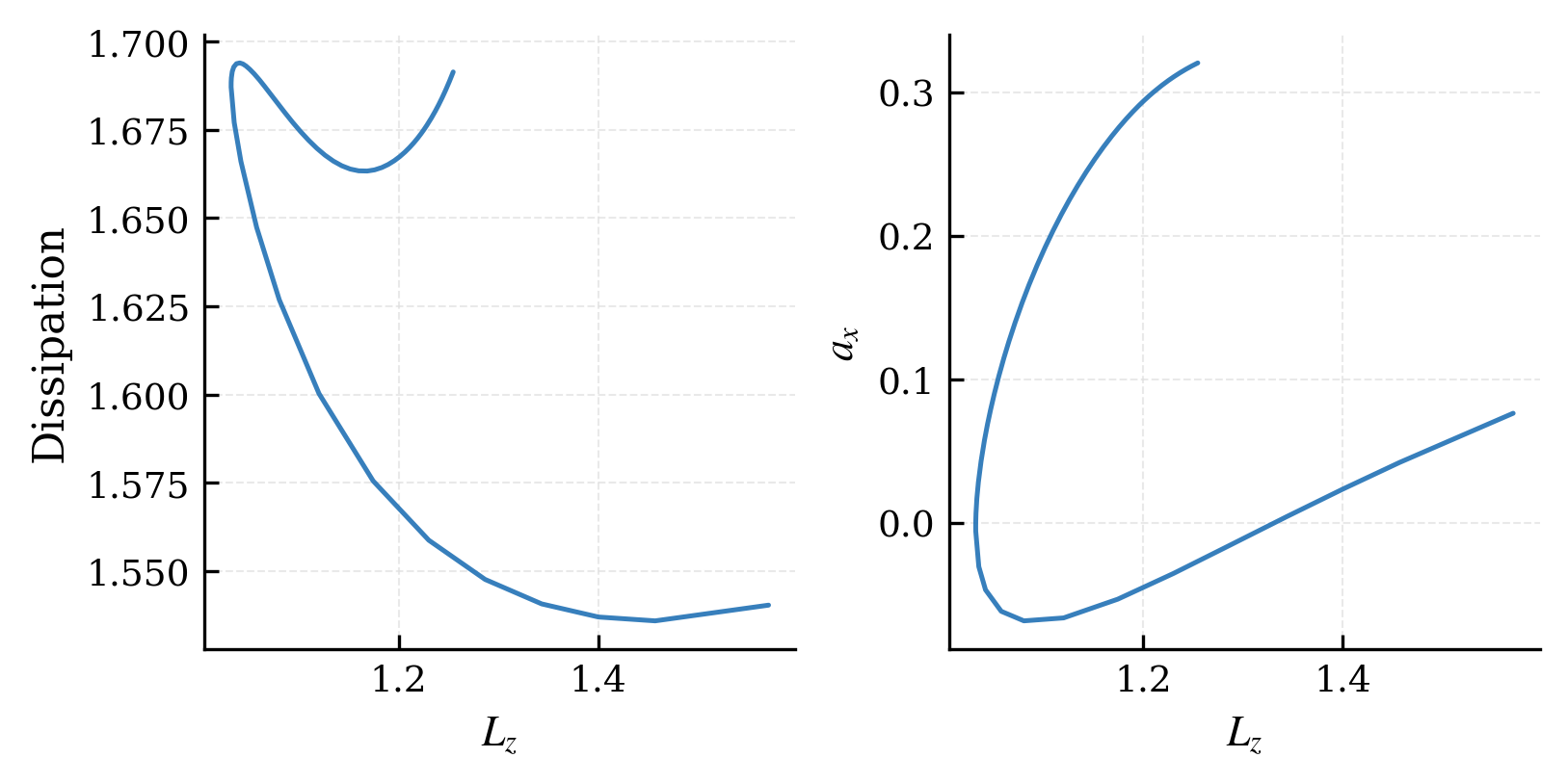}
    \caption{$L_z$-continuation: (c)~$D$ vs $L_z$, (d)~$a_x$ vs $L_z$.}
    \label{fig:tw2_bif_Lz}
  \end{subfigure}
  \caption{Continuation and bifurcation structure of TW2.
  Top row: $Re$-continuation at fixed domain size.
  Bottom row: $L_z$-continuation at fixed $Re$ and $L_x$.}
  \label{fig:tw2_bif}
\end{figure}

To assess whether the folds correspond to qualitative changes in the flow,
we compare representative solutions at the reference parameter value, at the
fold/turning point and on the opposite branch.
Figure~\ref{fig:tw2_Re_struct_yz} shows this comparison for the
$Re$-continuation in the $(y,z)$ cross-plane: the counter-rotating roll-cell
pattern is preserved across all three states. The same four-cell arrangement
persists with no cell creation or annihilation but the amplitude increases
markedly along the branch, from peak $|u| \approx 0.13$ at the reference
state to $\approx 0.32$ at the fold and $\approx 0.35$ on the upper branch.
This is consistent with the large dissipation contrast between the two branches. The
upper-branch state additionally shows a spanwise rephasing of the roll--streak
pattern (cell centres and near-wall extrema shift in $z$, consistent with
translational invariance) and more pronounced wall-normal exchange at the cell
interfaces. The $(x,z)$ midplane views confirm these trends: the reference
state displays weak, nearly uniform streaks, whereas the fold and
upper-branch solutions show increasingly pronounced streamwise modulation and
streak tilting. For the $L_z$-continuation, the structural evolution is
considerably milder: the roll--streak topology is again preserved, with the
peak amplitude increasing only modestly from $|u| \approx 0.13$ at the
reference state to $\approx 0.15$ at the fold and $\approx 0.19$ on the
upper branch; the dominant change is a rescaling of the spanwise wavelength
to accommodate the new periodicity, and the $(x,z)$ slices show that streaks
become progressively straighter and more $z$-periodic on the upper branch,
consistent with the travelling wave relaxing toward a more purely
spanwise-periodic roll--streak balance as the domain widens. The $(x,z)$
comparisons for both continuations and the $L_z$-continuation $(y,z)$ views
are provided in the supplementary material (figures~S16--S19).

\begin{figure}[htbp]
  \centering
  \includegraphics[width=0.48\linewidth]{Quiver_TW2_YZ.png}
  \includegraphics[width=0.48\linewidth]{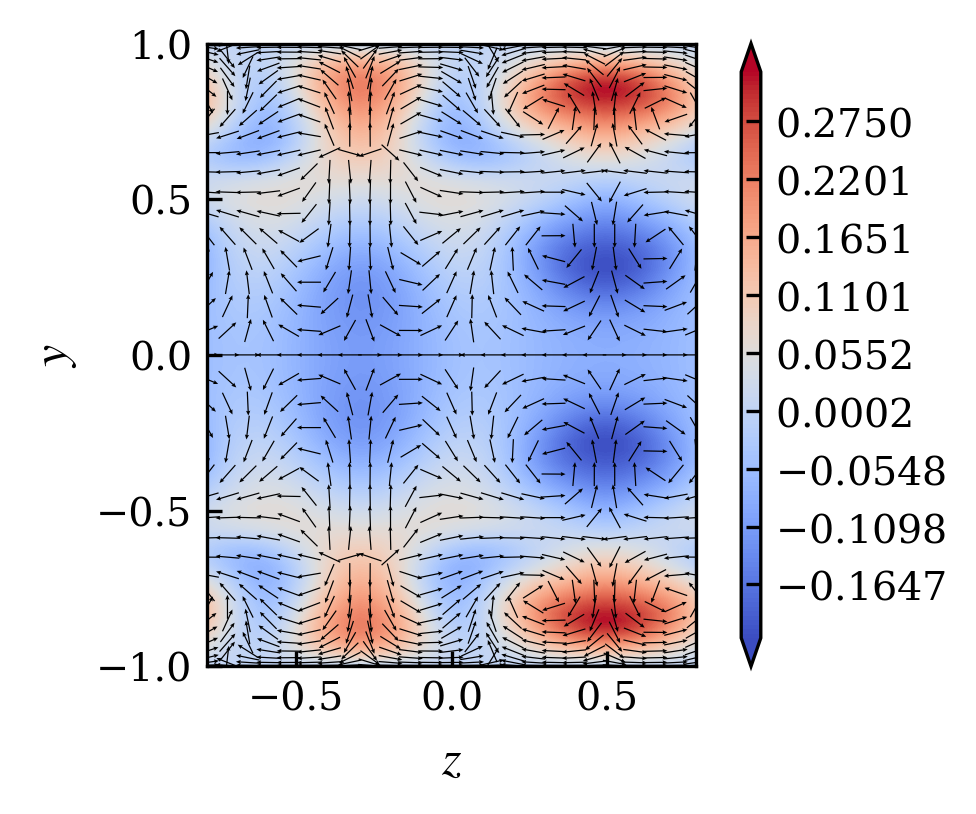}
  \includegraphics[width=0.48\linewidth]{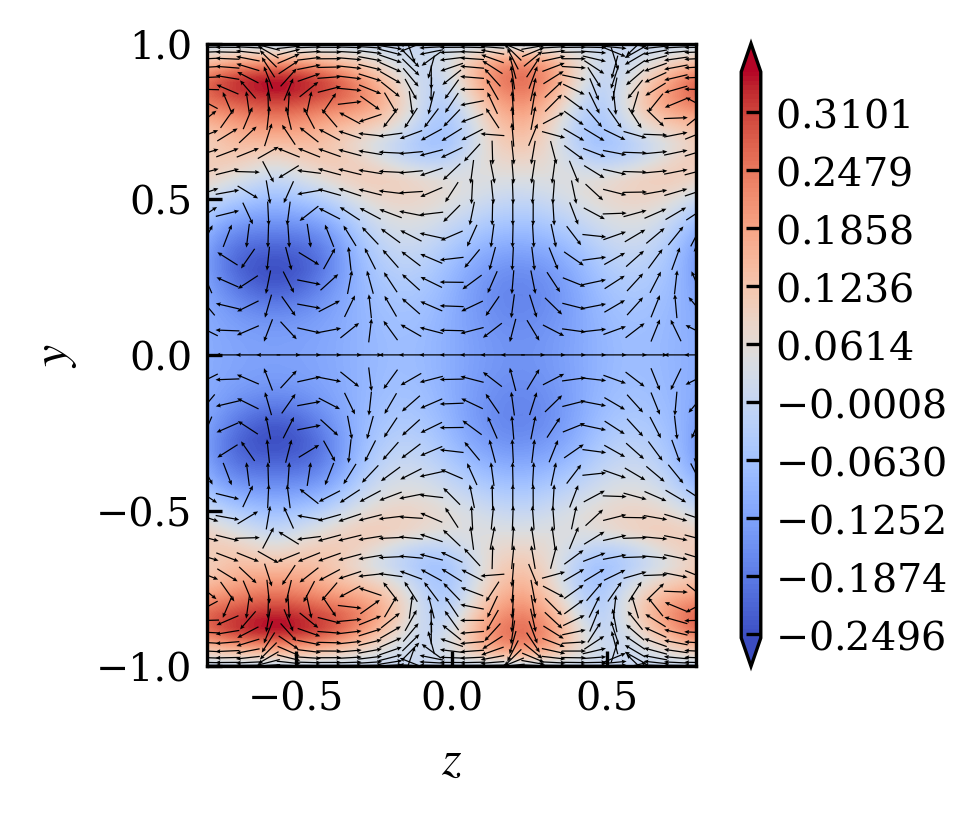}
  \caption{$Re$-continuation structural comparison for TW2 in the $(y,z)$
  plane: (left) reference TW2 at $Re=2000$ (lower branch), (right)
  fold/turning-point near $Re \approx 700$, (bottom) upper-branch solution.
  Background colour: streamwise velocity $u$; arrows: in-plane $(v,w)$
  components. The roll--streak topology is preserved, with amplitude
  increasing markedly from left to right.}
  \label{fig:tw2_Re_struct_yz}
\end{figure}

The Floquet spectrum of TW2 evolves markedly along the $Re$-continuation,
including a qualitative change in the character of the instability. At the
reference state (section~\ref{sec:tw2_floquet}), the unstable subspace is
two-dimensional and purely monotone ($\Lambda_1 \approx 1.414$,
$\Lambda_2 \approx 1.030$, both real). At the fold ($Re \approx 700$), the
instability weakens and the leading multipliers contract to
$|\Lambda| \approx 1$, placing the solution close to marginal stability, so
that the fold coincides with a near-stabilisation of the travelling wave.
On the upper branch, the instability re-emerges but with a qualitatively
different structure: complex-conjugate pairs with $|\Lambda| > 1$ appear
alongside real unstable modes, indicating that the upper branch develops
oscillatory instabilities that are absent at the reference state. This
transition from a purely monotone to a mixed oscillatory-and-monotone
unstable manifold distinguishes the upper and lower branches dynamically,
not merely in amplitude, and suggests that trajectories departing the
upper-branch TW2 spiral away in state space rather than departing along
fixed directions as they do near the reference state.

Along the $L_z$-continuation, a similar weakening occurs at the fold
($L_z \approx 1.05$), where the leading multipliers again approach the unit
circle and the solution becomes only marginally unstable. On the
upper $L_z$ branch, the instability strengthens modestly relative to the
fold, with the leading multiplier recovering to $|\Lambda|$ slightly above
unity, though the overall instability remains weaker than at the reference
state. The Floquet spectra along both continuation branches are provided in
the supplementary material (figure~S20).


\subsection{TW3: continuation in $Re$ and $L_z$}
\label{sec:tw3_continuation}

We continue TW3 in two independent parameters: the Reynolds number $Re$ at
fixed domain size, and the spanwise period $L_z$ at fixed $Re$ and $L_x$.
Along each branch we track the dissipation $D$ and the streamwise drift
$a_x$.

The $Re$-continuation (figure~\ref{fig:tw3_bif}\emph{a,\,b}) exhibits a
pronounced S-shaped, multi-branch structure with multiple saddle-node folds
concentrated near $Re \approx 1200$--$1350$. Three distinct dissipation
levels are visible over the computed range: a lower branch that decreases
from $D \approx 1.9$ to a weakly $Re$-dependent plateau near
$D \approx 1.55$ at the high-$Re$ end, an intermediate branch at
$D \approx 2.2$--$2.5$, and an upper branch that increases steadily to
$D \approx 3.5$ near $Re = 2000$. These levels are connected through narrow
fold regions at the low-$Re$ end, creating an interval of coexisting TW3
states a hysteretic character. This is analogous to subcritical bifurcation
scenarios \citep{rosas2016globally}, although here we trace travelling-wave
branches rather than a single symmetry-breaking normal form. The reference
TW3 at $Re = 2000$ lies on the lower branch. The drift $a_x$
(figure~\ref{fig:tw3_bif}\emph{b}) splits into three well-separated bands
that track the dissipation levels: $a_x \approx 0.12$--$0.13$ on the lower
branch, $\approx 0.25$ on the intermediate branch, and
$\approx 0.30$--$0.45$ on the upper branch, with only mild variation along
each band away from the folds.

The $L_z$-continuation (figure~\ref{fig:tw3_bif}\emph{c,\,d}) is also
non-monotone, with a compact fold/loop region at small
$L_z \approx 2.83$--$2.90$ where the dissipation reaches $D \approx 4.3$.
Beyond this turning region, the dissipation decreases markedly with
increasing $L_z$ (to $D \approx 3.5$ at $L_z \approx 3.15$), indicating
that widening the spanwise domain progressively weakens TW3 in this scalar
measure. The drift $a_x$ forms a corresponding loop over the same
small-$L_z$ interval and then varies only weakly with $L_z$
($a_x \approx 0.116$--$0.119$), so the main dynamical change is
concentrated near the fold, after which the continuation primarily retunes
amplitude and drift gradually.

\begin{figure}[htbp]
  \centering
  \begin{subfigure}[t]{\textwidth}
    \centering
    \includegraphics[width=0.8\linewidth]{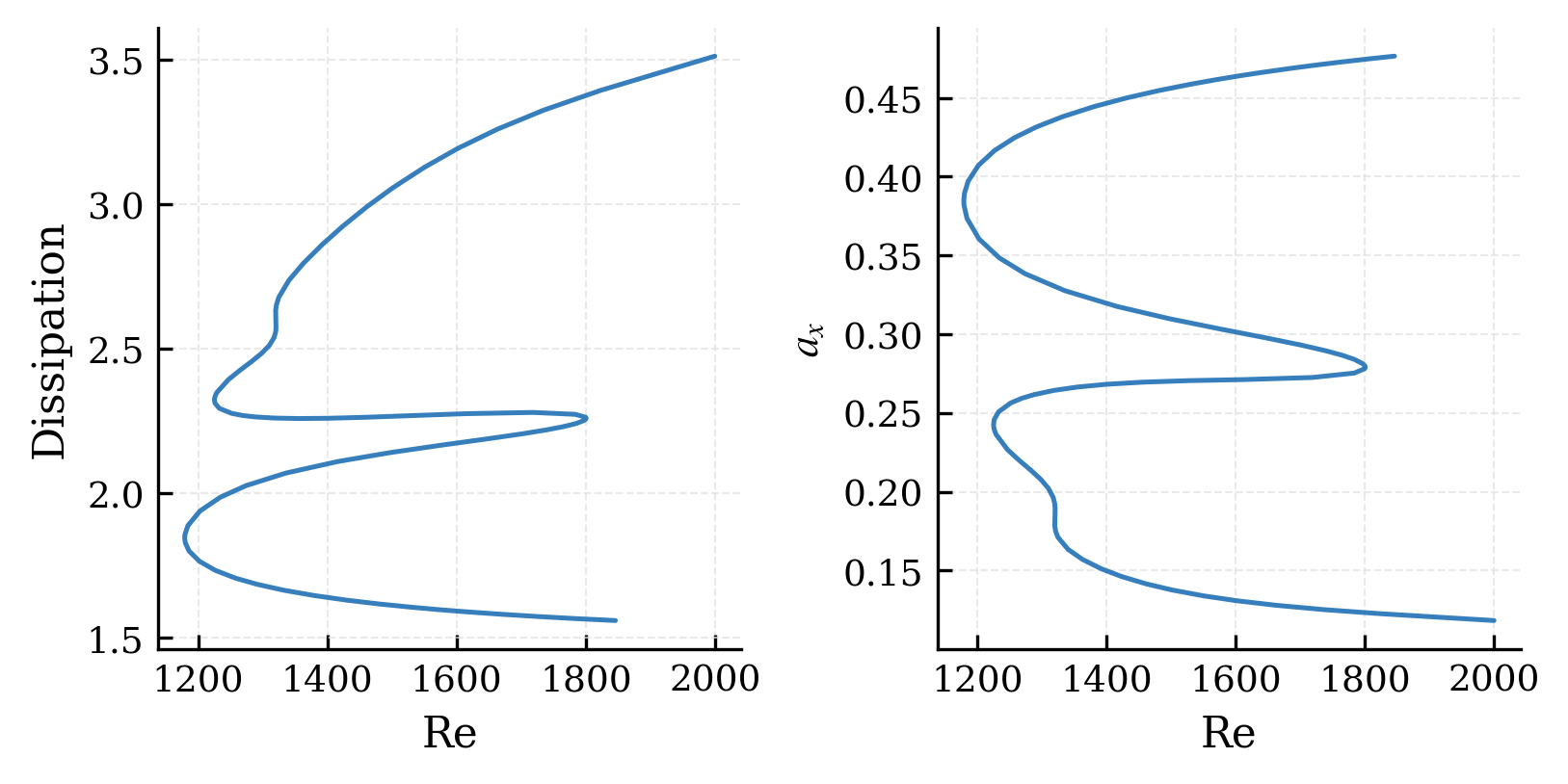}
    \caption{$Re$-continuation: (a)~$D$ vs $Re$, (b)~$a_x$ vs $Re$.}
    \label{fig:tw3_bif_Re}
  \end{subfigure}
  \vspace{2mm}
  \begin{subfigure}[t]{\textwidth}
    \centering
    \includegraphics[width=0.8\linewidth]{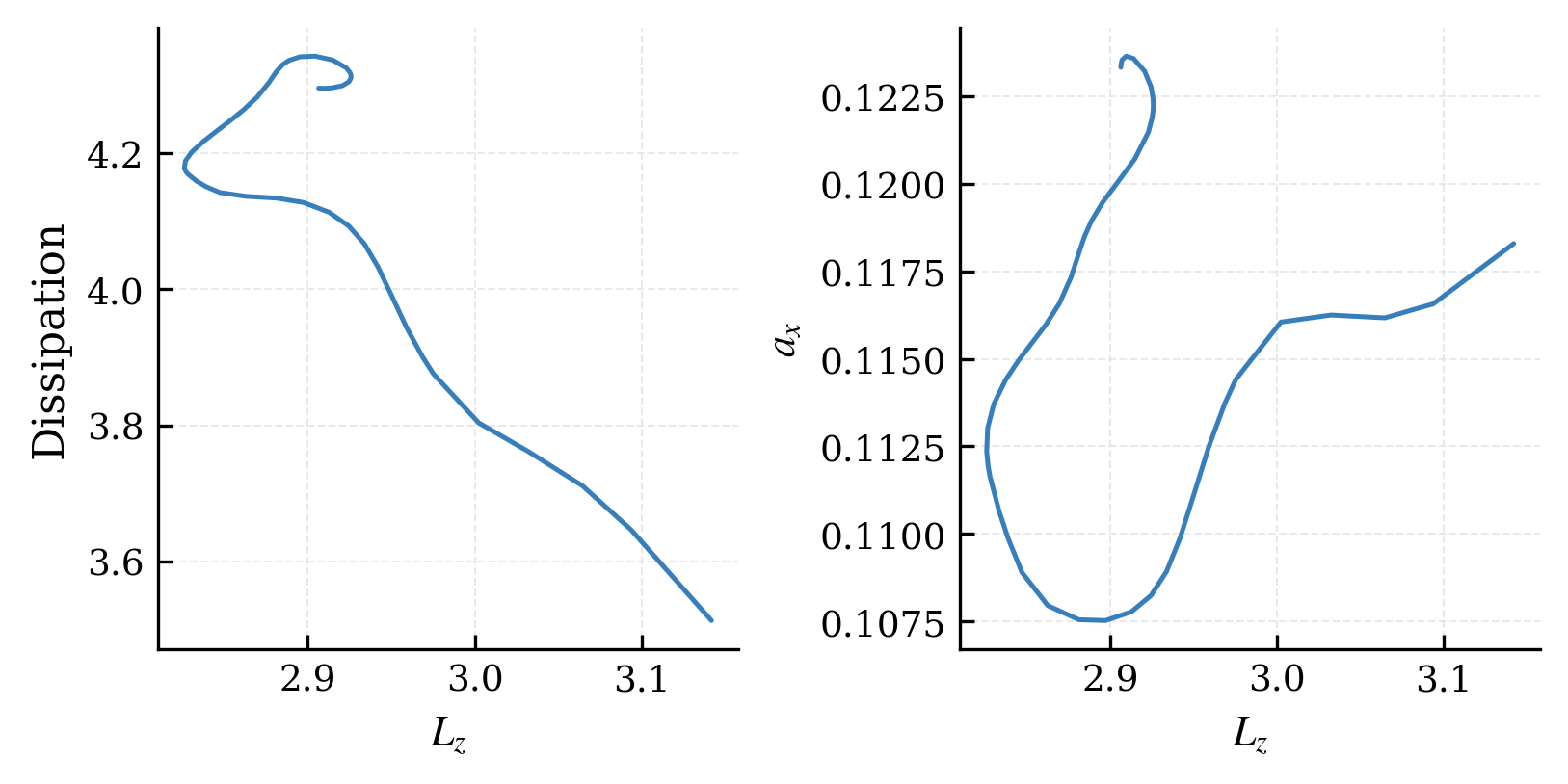}
    \caption{$L_z$-continuation: (c)~$D$ vs $L_z$, (d)~$a_x$ vs $L_z$.}
    \label{fig:tw3_bif_Lz}
  \end{subfigure}
  \caption{Continuation and bifurcation structure of TW3.
  Top row: $Re$-continuation at fixed domain size.
  Bottom row: $L_z$-continuation at fixed $Re$ and $L_x$.}
  \label{fig:tw3_bif}
\end{figure}

\begin{figure}[htbp]
  \centering
  \includegraphics[width=0.6\linewidth]{Quiver_TW3r_YZ.png}
  \includegraphics[width=0.6\linewidth]{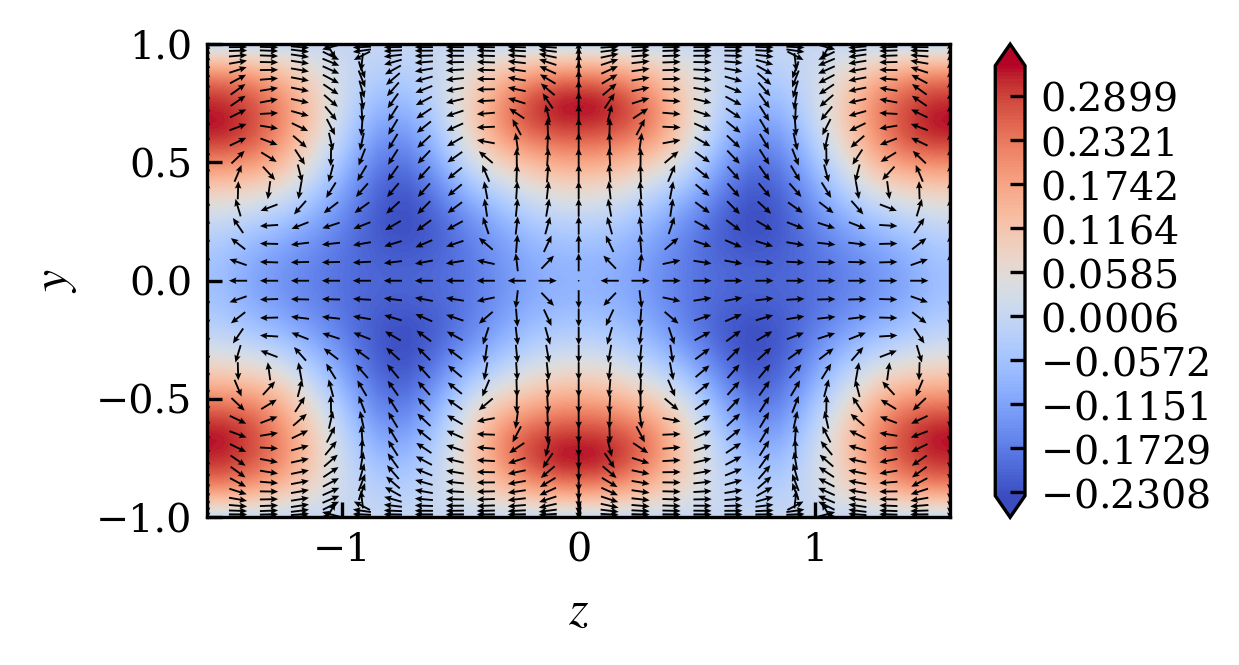}
  \includegraphics[width=0.6\linewidth]{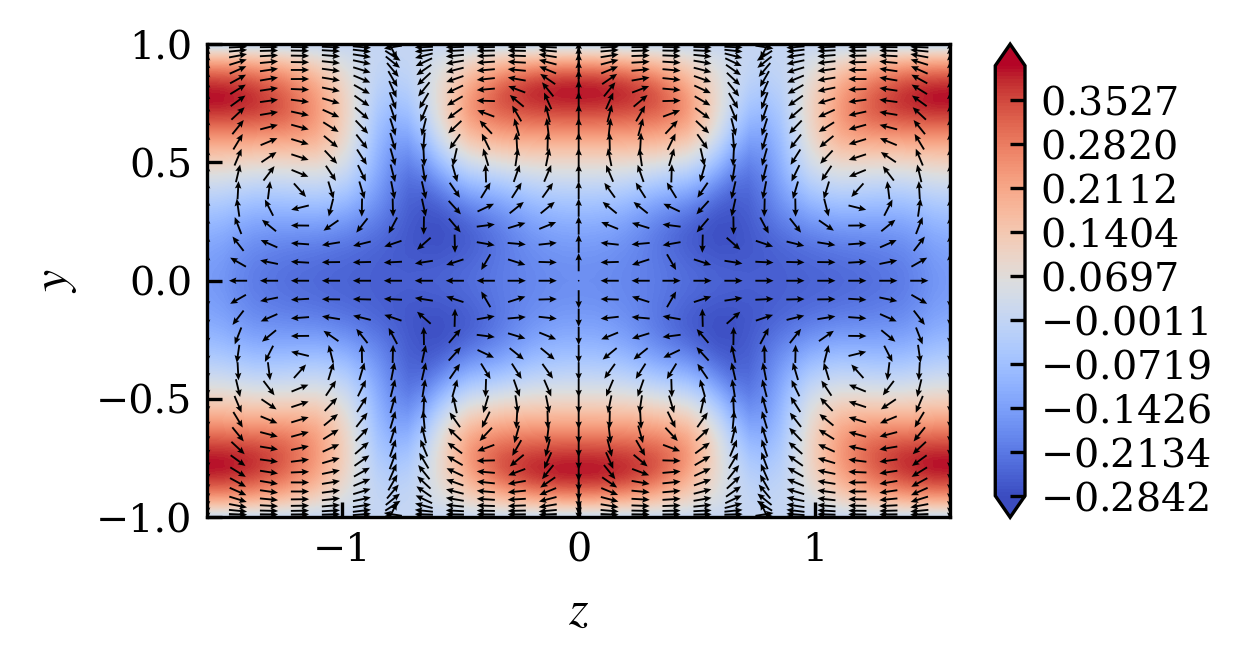}
  \caption{$Re$-continuation structural comparison for TW3 in the $(y,z)$
  plane: (top) reference TW3 at $Re=2000$ (lower branch), (middle)
  lower fold/turning-point near $Re \approx 1200$, (bottom) upper-branch
  solution. Background colour: streamwise velocity $u$; arrows: in-plane
  $(v,w)$ components. The roll--streak topology is preserved; peak $|u|$
  decreases from the reference ($\approx 0.50$) to the lower fold
  ($\approx 0.32$) and partially recovers on the upper branch
  ($\approx 0.40$).}
  \label{fig:tw3_Re_struct_yz}
\end{figure}

To assess whether the folds correspond to qualitative changes in the flow,
we compare representative solutions at the reference parameter value, at
fold/turning points and on the upper branch.
Figure~\ref{fig:tw3_Re_struct_yz} shows this comparison for the
$Re$-continuation in the $(y,z)$ cross-plane, sampling three states that
span the full S-curve: the reference TW3 on the lower branch, the lower
fold (connecting the lower and intermediate branches), and a state on the
upper branch. The multi-cell counter-rotating roll pattern is preserved
across all three but the peak
streamwise velocity varies non-monotonically across the branches: it
decreases from $|u| \approx 0.50$ at the reference state to $\approx 0.32$
at the lower fold, and then recovers to $\approx 0.40$ on the upper branch.
A similar non-monotonic trend holds on the intermediate branch, where the
peak amplitude drops further to $|u| \approx 0.25$. Despite the reduction in
peak streak amplitude relative to the reference, the upper-branch state
resides at substantially higher dissipation ($D \approx 3.5$ versus
$\approx 1.55$), indicating that the velocity gradients steepen and the
small-scale structure intensifies as the branch rises. The roll--streak
cell count, spacing, and symmetry about $y = 0$ and
$z = L_z/2$ remains unchanged throughout, confirming that the folds mark
coexisting amplitude/gradient regimes within the same topological family
rather than transitions to a qualitatively different coherent structure. For
the $L_z$-continuation, the structural evolution differs from the $Re$ case:
the peak amplitude increases modestly from $|u| \approx 0.50$ at the
reference to $\approx 0.57$ at the fold and $\approx 0.58$ on the upper
branch, while the
roll--streak topology is again preserved and the dominant change is a
rescaling of the spanwise wavelength to accommodate the narrower periodicity.
The $(x,z)$ midplane views for both continuations, the $L_z$-continuation
$(y,z)$ views, and the intermediate-branch snapshots are provided in the
supplementary material (figures~S21--S26).

The Floquet spectrum of TW3 undergoes the most complex evolution of any ECS
in this study, reflecting the S-shaped multi-branch structure of the
$Re$-continuation. At the reference state
(section~\ref{sec:tw3_floquet}), the instability is purely oscillatory, with
weakly unstable complex-conjugate pairs near the unit circle. Strikingly,
the Floquet analysis at multiple points along the S-curve reveals that both
the upper fold and the lower fold coincide with near-stabilisation: at these
turning points, all non-neutral multipliers lie on or inside the unit circle,
so that TW3 is linearly stable (or at most marginally unstable) precisely
where the branch reverses direction. Moreover, both the intermediate and
upper branches contain segments that are nearly stable. This alternation
between stable and weakly unstable segments on the same continuation branch
is consistent with the narrow fold geometry as the closely spaced turning
points create short parameter intervals over which eigenvalues cross the unit
circle, producing a patchwork of stability along the S-curve.

Along the $L_z$-continuation, the instability evolves differently: at the
fold ($L_z \approx 2.85$), the leading multiplier magnitude increases to
$|\Lambda| \approx 1.15$, making the fold solution more unstable than the
reference rather than less, in contrast to the $Re$-continuation behaviour.
On the upper $L_z$ branch, the instability weakens back to a level
comparable to the reference ($|\Lambda| \approx 1.05$). The oscillatory
character is maintained throughout. The Floquet spectra along both
continuation branches are provided in the supplementary material
(figure~S27).

\section{Conclusions}
\label{sec:conclusions}

We have computed five new exact coherent states of the incompressible
Navier--Stokes equations in plane Poiseuille flow. The two relative periodic
orbits (RPO1 at $Re=1000$, RPO2 at $Re=1500$) and three travelling waves
(TW1 at $Re=1900$, TW2 at $Re=2000$, TW3 at $Re=2000$) reside in
distinct symmetry-invariant subspaces. All solutions were obtained by
applying a Newton--Krylov--hookstep solver
\citep{viswanath2007recurrent,Aghor_Gibson_2025} to near-recurrent or
near-steady episodes identified in direct numerical simulations, and were
subsequently continued in both Reynolds number $Re$ and spanwise period
$L_z$ to map out their bifurcation structure and stability evolution.

Across all five states, the dominant physical organisation is the roll--streak
mechanism familiar from the self-sustaining process
\citep{hamilton1995regeneration,waleffe1997self}: streamwise velocity
streaks are sustained by counter-rotating cross-plane rolls, with near-wall
high- and low-speed patches aligned with the boundaries between adjacent
roll cells. The peak streak amplitudes span a wide range, from
$|u| \approx 0.13$ for TW2 (an edge-proximal state identified via edge
tracking) to $|u| \approx 0.50$ for TW3, with the two RPOs and TW1 at
intermediate levels ($|u| \approx 0.23$--$0.29$). Despite this amplitude
variation, the cell count, arrangement, and symmetry of the roll--streak pattern are robust along each continuation branch. The changes are expressed primarily as intensification or weakening of streak contrast, spanwise rephasing consistent with translational neutrality, and modulation of streamwise waviness. This separation between robust topology and tunable amplitude is a recurring theme in both the RPO and TW families, and mirrors
the structural persistence observed in the upper- and lower-branch equilibria
of plane Couette flow. For instance, \cite{gibson2009equilibrium} found that
the Nagata equilibria EQ$_1$/EQ$_2$ and the EQ$_3$/EQ$_4$ pairs retain the
same roll--streak skeleton across their entire $Re$ and $L_z$ continuation
ranges, with amplitude as the primary distinguishing feature between the
branches, precisely as we observe here for the Poiseuille-flow ECS.

The Floquet analysis at the reference parameter values reveals a clear
dichotomy in linear stability: the two RPOs are linearly stable within their
shared symmetry subspace
$\langle \sigma_y, \sigma_z, \tau_{xz} \rangle$, with all non-neutral
multipliers lying inside the unit circle, whereas all three travelling waves
are unstable saddle-type solutions. The character of the instability differs
among the TWs: TW1 possesses both oscillatory and monotone unstable
directions, TW2 has a purely two-dimensional monotone unstable subspace
(consistent with its proximity to the laminar--turbulent boundary, where fewer unstable directions are expected
\citep{skufca2006edge,schneider2008laminar,kerswell2018nonlinear}), and TW3
is destabilised by weakly growing oscillatory modes alone. In all three cases, the number of unstable directions remains small while the majority of
resolved perturbation directions contract strongly, giving each TW the
saddle character that is typical of travelling waves and equilibria in
wall-bounded flows
\citep{gibson2009equilibrium,park2015exact,graham2021exact}. In plane
Couette flow, the Nagata lower-branch equilibrium EQ$_1$ similarly possesses
a single unstable direction and serves as a gatekeeper between laminar and
turbulent dynamics \citep{gibson2009equilibrium}, a role that TW2, with
its two-dimensional monotone unstable manifold and edge-proximal
character, may play in the Poiseuille-flow symmetry subspace. The stable
RPOs, by contrast, act as attractors within their symmetry subspace and are
consistent with the long near-recurrent episodes observed in the parent DNS
trajectories, analogous to the role played by the periodic orbit of
\cite{kawahara2001periodic} in plane Couette flow, which organises quiescent
phases of the turbulent dynamics.

The Floquet spectra computed along the continuation branches reveal that the
stability landscape evolves substantially with the control parameters and
differs qualitatively between the RPOs and TWs. Both RPOs remain linearly
stable across the entire computed $Re$ and $L_z$ ranges, with the folds
acting as purely geometric features that do not coincide with any stability
transition. The travelling waves, by contrast, exhibit different spectral
evolution. TW1 shows a local minimum in instability at the fold
($|\Lambda|$ decreasing from $\approx 1.22$ to $\approx 1.09$), followed
by a marked intensification on the upper branch ($|\Lambda| \approx 1.45$)
with additional unstable modes. TW2 undergoes a qualitative change: its
purely monotone instability at the reference state is replaced by a mixed
oscillatory-and-monotone unstable manifold on the upper branch, while the
fold itself coincides with near-stabilisation. TW3 displays the most
complex behaviour, with both folds of its S-shaped $Re$-continuation acting
as stabilisation points, and both the intermediate and upper branches
containing linearly stable segments interspersed with weakly unstable ones
($|\Lambda| \approx 1.03$). Together, these results suggest that the fold
is generically a site of reduced instability for the travelling waves, while
the upper branches carry stronger and sometimes qualitatively different
instabilities. This pattern is consistent with the observation in Couette
flow that upper-branch equilibria (e.g.\ EQ$_2$ with its 8-dimensional
unstable manifold) are substantially more unstable than the lower-branch
ones (e.g.\ EQ$_1$ with a single unstable direction)
\citep{gibson2009equilibrium}.

The continuation diagrams show the bifurcation geometry. RPO1 and RPO2
exhibit localised saddle-node folds with narrow coexistence intervals (near
$Re \approx 986$ and $Re \approx 1499$, respectively) across which the
solution properties change only quantitatively. TW1 and TW2 each display a
single prominent fold (near $Re \approx 1400$ and $Re \approx 700$),
creating well-separated upper and lower branches with substantial dissipation
contrast. These fold Reynolds numbers are comparable to the saddle-node
thresholds reported for pipe-flow travelling waves
\citep{faisst2003traveling,wedin2004exact} and for the lower-branch channel
ECS of \cite{waleffe2001exact,waleffe2003homotopy}. In plane Couette flow,
\cite{gibson2009equilibrium} similarly found that all equilibrium families
originate at saddle-node bifurcations at critical $Re$ (e.g.\ EQ$_1$/EQ$_2$
at $Re \approx 218.5$, EQ$_3$/EQ$_4$ at $Re \approx 364$), reinforcing the
view that fold-type creation at moderate $Re$ is a generic feature of
wall-bounded ECS families across different flow geometries. TW3 stands out
with a pronounced S-shaped, multi-branch structure containing three distinct
dissipation levels connected by multiple folds near
$Re \approx 1200$--$1350$, producing an extended interval of coexisting
states with a hysteretic character. In all cases, the relative streamwise
drift $a_x$ provides an independent diagnostic that tracks the branch
structure. It becomes multi-valued near folds in step with the dissipation,
while separating into distinct bands where multiple branches coexist,
indicating that amplitude changes are accompanied by changes in
wave speed rather than being purely energetic changes.

Continuation in $L_z$ complements the $Re$ picture by acting primarily as a
spanwise-wavenumber constraint. Changing $L_z$ changes the admissible
spanwise organisation without forcing topological reorganisation. As $L_z$
increases, states generally weaken in dissipation and become less
streamwise-modulated, while near fold regions, sharp transitions between coexisting amplitudes can occur. The $L_z$ folds are typically more
localised than their $Re$ counterparts (e.g.\ the RPO1 fold spans
$\Delta L_z \approx 0.001$, compared with $\Delta Re \approx 30$ in the
$Re$ continuation), reflecting the stricter geometric constraint imposed by
the spanwise periodicity on the roll--streak tiling. A similar sensitivity to spanwise cell width was observed by \cite{gibson2009equilibrium} in
plane Couette flow, where several equilibria (including the Nagata pair)
terminate in saddle-node or pitchfork bifurcations under $L_z$ variation, and only a few (EQ$_4$, EQ$_7$, EQ$_9$) could be continued across a wide
range of spanwise periods. Overall, $Re$ controls the energetic level and
waviness of each state, while $L_z$ selects and reshapes the spanwise mode
content.

The present results extend the known catalogue of invariant solutions in
plane Poiseuille flow and provide new evidence for the central role of
discrete symmetries in structuring the existence, stability and bifurcation
of exact coherent states. The five new ECS, together with their continuation
branches and stability properties, give a set of dynamically relevant
landmarks in state space that can suggest further investigations into the organisation, transition, and suppression of turbulence.

\begin{bmhead}[Acknowledgements and Funding disclosure]
We sincerely thank Pratik P Aghor for extensive discussions. AN's graduate research is funded by the Indian Institute of Technology (IIT) Delhi. RT acknowledges support provided by the seed grant of IIT Delhi and the Prime Minister Early Career Research Grant (Project No. RP05041G\_SN) of the Anusandhan National Research Foundation (ANRF) of India. The computational facilities provided by IIT Delhi are greatly appreciated.
\end{bmhead}

\begin{bmhead}[Conflict disclosure]
The authors report no conflict of interest.
\end{bmhead}


\bibliographystyle{jfm}
\bibliography{jfm}

@inproceedings{landau1944problem,
  title={On the problem of turbulence},
  author={Landau, Lev D},
  booktitle={Dokl. Akad. Nauk USSR},
  volume={44},
  pages={311},
  year={1944}
}

@article{hamilton1995regeneration,
  title={Regeneration mechanisms of near-wall turbulence structures},
  author={Hamilton, James M and Kim, John and Waleffe, Fabian},
  journal={Journal of Fluid Mechanics},
  volume={287},
  pages={317--348},
  year={1995},
  publisher={Cambridge University Press}
}

@article{waleffe1997self,
  title={On a self-sustaining process in shear flows},
  author={Waleffe, Fabian},
  journal={Physics of Fluids},
  volume={9},
  number={4},
  pages={883--900},
  year={1997},
  publisher={American Institute of Physics}
}

@article{waleffe2001exact,
  title={Exact coherent structures in channel flow},
  author={Waleffe, Fabian},
  journal={Journal of Fluid Mechanics},
  volume={435},
  pages={93--102},
  year={2001},
  publisher={Cambridge University Press}
}

@article{graham2021exact,
  title={Exact coherent states and the nonlinear dynamics of wall-bounded turbulent flows},
  author={Graham, Michael D and Floryan, Daniel},
  journal={Annual Review of Fluid Mechanics},
  volume={53},
  number={1},
  pages={227--253},
  year={2021},
  publisher={Annual Reviews}
}

@article{nagata1990three,
  title={Three-dimensional finite-amplitude solutions in plane Couette flow: bifurcation from infinity},
  author={Nagata, Masato},
  journal={Journal of Fluid Mechanics},
  volume={217},
  pages={519--527},
  year={1990},
  publisher={Cambridge University Press}
}

@article{clever1992three,
  title={Three-dimensional convection in a horizontal fluid layer subjected to a constant shear},
  author={Clever, Richard M and Busse, Friedrich H},
  journal={Journal of Fluid Mechanics},
  volume={234},
  pages={511--527},
  year={1992},
  publisher={Cambridge University Press}
}

@article{faisst2003traveling,
  title={Traveling waves in pipe flow},
  author={Faisst, Holger and Eckhardt, Bruno},
  journal={Physical Review Letters},
  volume={91},
  number={22},
  pages={224502},
  year={2003},
  publisher={APS}
}

@article{wedin2004exact,
  title={Exact coherent structures in pipe flow: travelling wave solutions},
  author={Wedin, Hakan and Kerswell, Rich R},
  journal={Journal of Fluid Mechanics},
  volume={508},
  pages={333--371},
  year={2004},
  publisher={Cambridge University Press}
}

@article{gibson2009equilibrium,
  title={Equilibrium and travelling-wave solutions of plane Couette flow},
  author={Gibson, John F and Halcrow, Jonathan and Cvitanovi{\'c}, Predrag},
  journal={Journal of Fluid Mechanics},
  volume={638},
  pages={243--266},
  year={2009},
  publisher={Cambridge University Press}
}

@article{park2015exact,
  title={Exact coherent states and connections to turbulent dynamics in minimal channel flow},
  author={Park, Jae Sung and Graham, Michael D},
  journal={Journal of Fluid Mechanics},
  volume={782},
  pages={430--454},
  year={2015},
  publisher={Cambridge University Press}
}

@article{gibson2014spanwise,
  title={Spanwise-localized solutions of planar shear flows},
  author={Gibson, John F and Brand, Evan},
  journal={Journal of fluid mechanics},
  volume={745},
  pages={25--61},
  year={2014},
  publisher={Cambridge University Press}
}

@article{zammert2015crisis,
  title={Crisis bifurcations in plane Poiseuille flow},
  author={Zammert, Stefan and Eckhardt, Bruno},
  journal={Physical Review E},
  volume={91},
  number={4},
  pages={041003},
  year={2015},
  publisher={APS}
}

@article{Aghor_Gibson_2025, 
title={Symmetry groups and invariant solutions of plane Poiseuille flow}, author={Aghor, Pratik P. and Gibson, John F.}, 
year={2025},
volume={1019}, 
DOI={10.1017/jfm.2025.10576}, 
journal={Journal of Fluid Mechanics}, 
pages={A36}

}

@article{viswanath2007recurrent,
  title={Recurrent motions within plane Couette turbulence},
  author={Viswanath, Divakar},
  journal={Journal of Fluid Mechanics},
  volume={580},
  pages={339--358},
  year={2007},
  publisher={Cambridge University Press}
}

@article{Gibson2008,
  author  = {Gibson, John F. and Halcrow, Jonathan and Cvitanovi{\'c}, Predrag},
  title   = {Visualizing the geometry of state space in plane Couette flow},
  journal = {Journal of Fluid Mechanics},
  year    = {2008},
  volume  = {611},
  pages   = {107--130},
  doi     = {10.1017/S002211200800267X}
}

@article{kim1987turbulence,
  title={Turbulence statistics in fully developed channel flow at low Reynolds number},
  author={Kim, John and Moin, Parviz and Moser, Robert},
  journal={Journal of fluid mechanics},
  volume={177},
  pages={133--166},
  year={1987},
  publisher={Cambridge University Press}
}

@article{rosas2016globally,
  title={Globally coupled stochastic two-state oscillators: synchronization of infinite and finite arrays},
  author={Rosas, Alexandre and Escaff, Daniel and Pinto, Italo’Ivo Lima Dias and Lindenberg, Katja},
  journal={Journal of Physics A: Mathematical and Theoretical},
  volume={49},
  number={9},
  pages={095001},
  year={2016},
  publisher={IOP Publishing}
}

@article{hopf1948mathematical,
   author  = {Hopf, Eberhard},
   title   = {A mathematical example displaying features of turbulence},
   journal = {Communications on Pure and Applied Mathematics},
   volume  = {1},
   pages   = {303--322},
   year    = {1948}
 }

@book{doering1995applied,
   author    = {Doering, Charles R. and Gibbon, John D.},
  title     = {Applied Analysis of the {Navier--Stokes} Equations},
   publisher = {Cambridge University Press},
  year      = {1995}
 }

@article{clever1997tertiary,
   author  = {Clever, R. M. and Busse, F. H.},
  title   = {Tertiary and quaternary solutions for plane {Couette} flow},
  journal = {Journal of Fluid Mechanics},
   volume  = {344},
   pages   = {137--153},
  year    = {1997}
}

@article{waleffe1998three,
    author  = {Waleffe, Fabian},
    title   = {Three-dimensional coherent states in plane shear flows},
    journal = {Physical Review Letters},
    volume  = {81},
    pages   = {4140--4143},
    year    = {1998}
  }

@article{waleffe2003homotopy,
    author  = {Waleffe, Fabian},
    title   = {Homotopy of exact coherent structures in plane shear flows},
    journal = {Physics of Fluids},
    volume  = {15},
    pages   = {1517--1534},
    year    = {2003}
  }

@article{jimenez1999autonomous,
    author  = {Jim\'{e}nez, Javier and Pinelli, Alfredo},
    title   = {The autonomous cycle of near-wall turbulence},
    journal = {Journal of Fluid Mechanics},
    volume  = {389},
    pages   = {335--359},
    year    = {1999}
  }

@article{kawahara2001periodic,
    author  = {Kawahara, Genta and Kida, Shigeo},
    title   = {Periodic motion embedded in plane {Couette} turbulence: regeneration cycle and burst},
    journal = {Journal of Fluid Mechanics},
    volume  = {449},
    pages   = {291--300},
    year    = {2001}
  }

@article{toh2003periodic,
    author  = {Toh, Sadayoshi and Itano, Tomoaki},
    title   = {A periodic-like solution in channel flow},
    journal = {Journal of Fluid Mechanics},
    volume  = {481},
    pages   = {67--76},
    year    = {2003}
  }

@article{halcrow2009heteroclinic,
    author  = {Halcrow, Jonathan and Gibson, John F. and Cvitanovi\'{c}, Predrag and Viswanath, Divakar},
    title   = {Heteroclinic connections in plane {Couette} flow},
    journal = {Journal of Fluid Mechanics},
    volume  = {621},
    pages   = {365--376},
    year    = {2009}
  }

@article{budanur2017relative,
    author  = {Budanur, Nazmi Burak and Short, Keith Y. and Farazmand, Mohammad and Willis, Ashley P. and Cvitanovi\'{c}, Predrag},
    title   = {Relative periodic orbits form the backbone of turbulent pipe flow},
    journal = {Journal of Fluid Mechanics},
    volume  = {833},
    pages   = {274--301},
    year    = {2017}
  }

@article{van2011homoclinic,
    author  = {van Veen, Lennaert and Kawahara, Genta},
    title   = {Homoclinic tangle on the edge of shear turbulence},
    journal = {Physical Review Letters},
    volume  = {107},
    pages   = {114501},
    year    = {2011}
  }

@article{lustro2019onset,
    author  = {Lustro, Julius R. T. and Kawahara, Genta and van Veen, Lennaert and Shimizu, Masaki and Kokubu, Hiroshi},
    title   = {The onset of transient turbulence in minimal plane {Couette} flow},
    journal = {Journal of Fluid Mechanics},
    volume  = {862},
    pages   = {R2},
    year    = {2019}
  }

@article{park2018bursting,
    author  = {Park, Jae Sung and Shekar, Ashwin and Graham, Michael D.},
    title   = {Bursting and critical layer frequencies in minimal turbulent dynamics and connections to exact coherent states},
    journal = {Physical Review Fluids},
    volume  = {3},
    pages   = {014611},
    year    = {2018}
  }

@article{skufca2006edge,
    author  = {Skufca, Joseph D. and Yorke, James A. and Eckhardt, Bruno},
    title   = {Edge of chaos in a parallel shear flow},
    journal = {Physical Review Letters},
    volume  = {96},
    pages   = {174101},
    year    = {2006}
  }

@article{schneider2008laminar,
    author  = {Schneider, Tobias M. and Gibson, John F. and Lagha, Maher and de Lillo, Filippo and Eckhardt, Bruno},
    title   = {Laminar-turbulent boundary in plane {Couette} flow},
    journal = {Physical Review E},
    volume  = {78},
    pages   = {037301},
    year    = {2008}
  }

@article{schneider2010localized,
    author  = {Schneider, Tobias M. and Marinc, Daniel and Eckhardt, Bruno},
    title   = {Localized edge states nucleate turbulence in extended plane {Couette} cells},
    journal = {Journal of Fluid Mechanics},
    volume  = {646},
    pages   = {441--451},
    year    = {2010}
  }

@article{duguet2009localized,
    author  = {Duguet, Yohann and Schlatter, Philipp and Henningson, Dan S.},
    title   = {Localized edge states in plane {Couette} flow},
    journal = {Physics of Fluids},
    volume  = {21},
    pages   = {111701},
    year    = {2009}
  }

@article{schneider2009edge,
    author  = {Schneider, Tobias M. and Eckhardt, Bruno},
    title   = {Edge states intermediate between laminar and turbulent dynamics in pipe flow},
    journal = {Philosophical Transactions of the Royal Society A},
    volume  = {367},
    pages   = {577--587},
    year    = {2009}
  }

@article{willis2009turbulent,
    author  = {Willis, Ashley P. and Kerswell, Richard R.},
    title   = {Turbulent dynamics of pipe flow captured in a reduced model: puff relaminarization and localized `edge' states},
    journal = {Journal of Fluid Mechanics},
    volume  = {619},
    pages   = {213--233},
    year    = {2009}
  }

@article{xi2012dynamics,
    author  = {Xi, Li and Graham, Michael D.},
    title   = {Dynamics on the laminar-turbulent boundary and the origin of the maximum drag reduction asymptote},
    journal = {Physical Review Letters},
    volume  = {108},
    pages   = {028301},
    year    = {2012}
  }

@article{kerswell2018nonlinear,
    author  = {Kerswell, Rich R.},
    title   = {Nonlinear nonmodal stability theory},
    journal = {Annual Review of Fluid Mechanics},
    volume  = {50},
    pages   = {319--345},
    year    = {2018}
  }

@incollection{holmes2015dynamical,
    author    = {Holmes, Philip},
    title     = {Dynamical systems},
    booktitle = {Princeton Companion to Applied Mathematics},
    editor    = {Higham, Nicholas J.},
    pages     = {383--393},
    publisher = {Princeton University Press},
    year      = {2015}
  }

@misc{willis2019computing,
    author       = {Willis, Ashley P.},
    title        = {Equilibria, periodic orbits and computing them},
    year         = {2019},
    eprint       = {1908.06730},
    archivePrefix= {arXiv},
    primaryClass = {physics.flu-dyn}
  }

@article{schneider2010snakes,
    author  = {Schneider, Tobias M. and Gibson, John F. and Burke, John},
    title   = {Snakes and ladders: localized solutions of plane {Couette} flow},
    journal = {Physical Review Letters},
    volume  = {104},
    pages   = {104501},
    year    = {2010}
  }

@article{brand2014doubly,
    author  = {Brand, Evan and Gibson, John F.},
    title   = {A doubly-localized equilibrium solution of plane {Couette} flow},
    journal = {Journal of Fluid Mechanics},
    volume  = {750},
    pages   = {R3},
    year    = {2014}
  }

@article{viswanath2009critical,
    author  = {Viswanath, Divakar},
    title   = {The critical layer in pipe flow at high {Reynolds} number},
    journal = {Philosophical Transactions of the Royal Society A},
    volume  = {367},
    pages   = {561--576},
    year    = {2009}
  }

@article{hussain1986coherent,
  title={Coherent structures and turbulence},
  author={Hussain, A. K. M. F.},
  journal={Journal of Fluid Mechanics},
  volume={173},
  pages={303--356},
  year={1986},
  publisher={Cambridge University Press}
}

@article{robinson1991coherent,
  title={Coherent motions in the turbulent boundary layer},
  author={Robinson, Stephen K.},
  journal={Annual Review of Fluid Mechanics},
  volume={23},
  pages={601--639},
  year={1991},
  publisher={Annual Reviews}
}

@article{mckeon2010critical,
  title={A critical-layer framework for turbulent pipe flow},
  author={McKeon, Beverley J. and Sharma, Ati S.},
  journal={Journal of Fluid Mechanics},
  volume={658},
  pages={336--382},
  year={2010},
  publisher={Cambridge University Press}
}

@article{mckeon2017engine,
  title={The engine behind (wall) turbulence: perspectives on scale interactions},
  author={McKeon, Beverley J.},
  journal={Journal of Fluid Mechanics},
  volume={817},
  pages={P1},
  year={2017},
  publisher={Cambridge University Press}
}

@article{sharma2013coherent,
  title={On coherent structure in wall turbulence},
  author={Sharma, Ati S. and McKeon, Beverley J.},
  journal={Journal of Fluid Mechanics},
  volume={728},
  pages={196--238},
  year={2013},
  publisher={Cambridge University Press}
}

@article{suri2024predictive,
  title={Predictive framework for flow reversals and excursions in turbulence},
  author={Suri, Balachandra},
  journal={Physical Review Letters},
  volume={133},
  number={15},
  pages={154002},
  year={2024},
  publisher={American Physical Society}
}

@article{wang2007lower,
  title={Lower branch coherent states in shear flows: transition and control},
  author={Wang, Jue and Gibson, John and Waleffe, Fabian},
  journal={Physical review letters},
  volume={98},
  number={20},
  pages={204501},
  year={2007},
  publisher={APS}
}

\end{document}